\documentclass[letterpaper,12pt]{report}

\pdfoutput=1

\usepackage[latin1]{inputenc}
\usepackage{geometry}
\usepackage{graphicx}
\usepackage{latexsym}
\usepackage{makeidx}
\usepackage{cuthesis}
\usepackage{named}
\usepackage{url}

\newcommand{\xf}[1]{Figure~\ref{#1}}
\newcommand{\xp}[1]{page~\pageref{#1}}
\newcommand{\xs}[1]{Section~\ref{#1}}
\newcommand{\xa}[1]{Appendix~\ref{#1}}
\newcommand{\xc}[1]{Chapter~\ref{#1}}
\newcommand{\xt}[1]{Table~\ref{#1}}

\newcommand{\rpc}{{RPC\index{RPC}}}
\newcommand{\rmi}{{RMI\index{RMI}}}
\newcommand{\clp}{{CLP\index{CLP}}}
\newcommand{\tlp}{{TLP\index{TLP}}}
\newcommand{\slp}{{SLP\index{SLP}}}
\newcommand{\complus}{{DCOM+\index{DCOM+}}}
\newcommand{\corba}{{CORBA\index{CORBA}}}
\newcommand{\jini}{{Jini\index{Jini}}}
\newcommand{\dotnet}{{.NET Remoting\index{.NET Remoting}}}
\newcommand{\gnu}{{GNU\index{GNU}}}
\newcommand{\tcpip}{{TCP/IP\index{TCP/IP}}}
\newcommand{\AST}{{AST\index{AST}}}

\newcommand{\gipc}{{GIPC\index{GIPC}}}
\newcommand{\gicf}{{GICF\index{GICF}}}
\newcommand{\iplcf}{{IPLCF\index{IPLCF}}}
\newcommand{\gee}{{GEE\index{GEE}}}
\newcommand{\geer}{{GEER\index{GEER}}}
\newcommand{\gipsy}{{GIPSY\index{GIPSY}}}
\newcommand{\ripe}{{RIPE\index{RIPE}}}
\newcommand{\dpr}{{DPR\index{DPR}}}
\newcommand{\dms}{{DMS\index{DMS}}}
\newcommand{\dfg}{{DFG\index{DFG}}}

\newcommand{\glu}{{GLU\index{GLU}}}
\newcommand{\glusharp}{{GLU\#\index{GLU\#}}}
\newcommand{\gipl}{{GIPL\index{GIPL}}}
\newcommand{\sipl}{{SIPL\index{SIPL}}}
\newcommand{\lucid}{{Lucid\index{Lucid}}}
\newcommand{\ilucid}{{Indexical Lucid\index{Indexical Lucid}}}
\newcommand{\jlucid}{{JLucid\index{JLucid}}}
\newcommand{\olucid}{{Objective Lucid\index{Tensor Lucid}}}
\newcommand{\tlucid}{{Tensor Lucid\index{Tensor Lucid}}}
\newcommand{\plucid}{{Partial Lucid\index{Partial Lucid}}}
\newcommand{\onyx}{{Onyx\index{Onyx}}}
\newcommand{\lucx}{{Lucx\index{Lucx}}}

\newcommand{\C}{{C\index{C}}}
\newcommand{\cpp}{{C++\index{C++}}}
\newcommand{\perl}{{Perl\index{Perl}}}
\newcommand{\java}{{Java\index{Java}}}
\newcommand{\python}{{Python\index{Python}}}
\newcommand{\fortran}{{Fortran\index{Fortran}}}

\newcommand{\lisp}{{LISP\index{LISP}}}
\newcommand{\scheme}{{Scheme\index{Scheme}}}
\newcommand{\haskell}{{Haskell\index{Haskell}}}
\newcommand{\mllessequal}{{ML$_{\le}$\index{ML$_{\le}$}}}
\newcommand{\fcpp}{{FC++\index{FC++}}}

\newcommand{\olucidop}[1]{{\bf \texttt{\textmd{\textsc{#1}}}}}
\newcommand{\lucidop}[1]{{\bf \texttt{#1}}}

\newcommand{\tab}[1]{\hspace{#1pt}}

\newcommand{\shrule}[0]{\vspace{3pt}\hrule\vspace{6pt}}
\newcommand{\ehrule}[0]{\vspace{6pt}\hrule\vspace{3pt}}

\newcommand{\nonterminal}[1]{$\mathtt{<\!\!#1\!\!>}$}

\newcommand{\source}[1]
{
	{\shrule}
	\scriptsize
	#1
	\normalsize
	\hrule
}

\newcommand{\sourcefloat}[3]
{
	\begin{figure}[!hp]
	\begin{centering}
	\begin{minipage}{0.5\textwidth}
	\source{#1}
	\end{minipage}
	\caption{\small{#3}}
	\label{#2}
	\end{centering}
	\end{figure}
}

\newcommand{\file}[1]{\texttt{#1}\index{Files!#1}}
\newcommand{\tool}[1]{\texttt{#1}\index{Tools!#1}}
\newcommand{\option}[1]{\texttt{#1}\index{Options!#1}}
\newcommand{\api}[1]{\texttt{#1}\index{API!#1}}

\newcommand{\codesegment}[1]{\texttt{\##1}\index{Segments!\##1}}

\newcommand{\marf}[0]{MARF\index{Tools!MARF}\index{Libraries!MARF}}
\newcommand{\javacc}[0]{JavaCC\index{Tools!JavaCC}}
\newcommand{\junit}[0]{JUnit\index{Tools!JUnit}}

\newcommand{\statement}[2]
{
	\vspace{7pt}
	\shrule
	{\bf #1}

	#2
	\ehrule
	\vspace{7pt}
}

\newcommand{\proposition}[2]
{
	\statement{Proposition #1}{#2}
}

\newcommand{\definition}[2]
{
	\statement{Definition #1}{#2}
}

\newcommand{\unix}{\index{Unix@{\sc{Unix}}}{\sc{Unix}}}
\newcommand{\macos}[1]{\index{Mac OS #1@{\sc{Mac OS #1}}}{\sc{Mac OS #1}}}
\newcommand{\linux}{\index{Linux@{\sc{Linux}}}{\sc{Linux}}}
\newcommand{\rhl}[1]{\index{Red Hat Linux #1@{\sc{Red Hat Linux #1}}}{\sc{Red Hat Linux #1}}}
\newcommand{\fcore}[1]{\index{Fedora Core #1@{\sc{Fedora Core #1}}}{\sc{Fedora Core #1}}}

\newcommand{\solaris}[1]{\index{Solaris #1@{\sc{Solaris #1}}}{\sc{Solaris #1}}}
\newcommand{\win}[1]{\index{Windows #1@{\sc{Windows #1}}}{\sc{Windows #1}}}

\newcommand{\lucidL}[1]{{$\mathit{Lucid}$}($L$) }

		{}

\def\myvert{\raise 2.27pt \hbox{\vrule depth 0pt height 8pt width 0.2mm}}
\def\myarrow{\hspace*{0.43mm}%
             \raise 2.29pt\hbox{\vrule depth 0pt height 8pt width 0.16mm}%
             \hspace*{-0.32mm}%
             $\longrightarrow$
             \ %
             }

\newcommand{\johndef}{\mathcal{D}}
\newcommand{\myid}{\textit{id}}

\newcommand{\mydagger}{\!\dagger\!}
\newcommand{\context}[2]{\mathcal{D},\mathcal{P} \vdash #1 : #2}

\newcommand{\qcontext}[2]{\mathcal{D},\mathcal{P} \vdash #1 \::\: #2}

\newcommand{\myifthenelse}{\mathtt{if}\;E\;\mathtt{then}\;E'\;\mathtt{else}\;E''}

\makeindex

\author{Serguei A. Mokhov}

\title
{
	Towards Hybrid Intensional Programming
	with JLucid, Objective Lucid, and
	General Imperative Compiler Framework
	in the GIPSY
}

\degree{Master of Computer Science}
\dept{Computer Science and Software Engineering}
\submitdate{October 2005}
\copyrightyear{2005}

\begin{document}


\begin{abstract}
Pure Lucid programs are concurrent with very fine granularity.
Sequential Threads (STs) are functions introduced to enlarge the grain
size; they are passed from server to workers by Communication Procedures
(CPs) in the General Intensional Programming System (GIPSY).
A JLucid program combines Java code for the STs with Lucid code
for parallel control.  Thus first, in this thesis, we describe the way in which the new
JLucid compiler generates STs and CPs. JLucid also introduces array support.

Further exploration goes through the additional transformations that the Lucid
family of languages has undergone to enable the use of Java objects and their members,
in the Generic Intensional Programming Language (GIPL),
and Indexical Lucid: first, in the form of JLucid
allowing the use of {\em pseudo-objects}, and then through the specifically-designed
the Objective Lucid language.
The syntax and semantic definitions of Objective Lucid and the meaning of
Java objects within an intensional program are provided with discussions and examples.

Finally, there are many useful scientific and utility routines written in many
imperative programming languages other than Java, for example in C,
C++, Fortran, Perl, etc. Therefore, it is wise to provide a framework to
facilitate inclusion of these languages into the GIPSY and their use by Lucid
programs. A General Imperative Compiler Framework and its concrete implementation
is proposed to address this issue.
\end{abstract}


\begin{acknowledgments}
I would like to thank my supervisor Dr. Joey Paquet and Dr. Peter Grogono
for ever lasting patience and caring guidance throughout the variety of learning experience and their advices
and insightful comments to make these contributions possible.
I would also like to thank my friendly team members with whom we together were
lifting the complex GIPSY system off the ground. Specifically, I would like
to mention Chun Lei Ren, Paula Bo Lu, Ai Hua Wu, Yimin Ding, Lei Tao, Emil Vassev,
and Kai Yu Wan for outstanding team work. Thanks to Dr. Patrice Chalin
for an in-depth introduction to semantics of programming languages.
Thanks to Dr. Sabine Bergler and Dr. Leila Kosseim for the journey through
the internals of natural language processing side related to this work.
Thanks to my beloved Irina for helping me to carry through.

This work has been sponsored by NSERC and the Faculty of Engineering and
Computer Science of Concordia University, Montr\'eal, Qu\'ebec, Canada.
This document was produced in {\LaTeX} with the guidance of Dr. Grogono's manual
in \cite{grogono2001} and Concordia University {\LaTeX} thesis styling
maintained by Steve Malowany, Stan Swiercz, and Patrice Chalin.
\end{acknowledgments}



\chapter{Introduction}
\label{chatp:introduction}
\index{Introduction}

%
%

\section{Thesis Statement}
\label{sect:problem-statement}
\index{Thesis!Statement}
\index{Introduction!Thesis Statement}

In the previous prototype of the General Intensional Programming System
({\gipsy}) there existed limitations to its potential
in distributed computing -- lack of sequential threads and
communication procedures.
Additionally, the capabilities of {\ilucid} and {\gipl}, the primary
GIPSY's languages, were limited
to only computing aspects without input/output, arrays, and some other essential features
(e.g. math, non-determinism, dynamic loading) that exist in imperative (e.g. {\java}) languages.
We discuss an extension to Generic Intensional Programming Language ({\gipl}) and
{\ilucid} with embedded {\java} -- {\jlucid}.
A few problems are solved as an example using the enhanced language.

{\jlucid} brings embedded {\java} and most of its powers into {\ilucid}
in the {\gipsy} by allowing intensional languages to manipulate Java methods
as first class values\footnote{The Java methods are not referred to
as ``functions'' as in functional programming -- the Java methods can
be passed around as values inside the Lucid part, but not to or from {\java}
part of a GIPSY program.}.
However, it is very natural to have objects with {\java}
and manipulate their members in scientific intensional computation, yet
{\jlucid} fails to support that {\java}'s capability. Hence, we design
{\olucid} to address this deficiency.
We define the operational semantics of {\olucid}, and give some examples of its application.

Existence of {\jlucid}, {\olucid}, and {\glu}
as well as many useful libraries written in other imperative
languages, such as {\C}/{\cpp}, {\perl}, {\python}, {\fortran} etc. demanded
ability to use code written in those languages by
intensional programs, naturally. Thus, we design a first version of the
General Imperative Compiler Framework ({\gicf}) as a
part of the {\gipsy} to allow GIPSY programs to use virtually
any combination of intensional and imperative languages at the meta level.
This is a very
ambitious goal; therefore, the proposal is the first iteration of
the framework open for later refinements as it matures along with
the corresponding changes to the run-time system.

%
%

\section{Contributions}
\index{Thesis!Contributions}
\index{Introduction!Contributions}

Primary contributions of this thesis are outlined below:

\begin{itemize}
\item
{\jlucid}
	\begin{itemize}
	\item
	Semantics of {\em pseudo-free} Java methods in Lucid programs
	\item
	Design and implementation of {\jlucid} and its compiler in the {\gipsy}
	\end{itemize}
\item
{\olucid}
	\begin{itemize}
	\item
	Semantics of the integration of Java objects in {\lucid} programs
	\item
	Design and implementation of the Objective Lucid compiler
	\end{itemize}
\item
General Imperative Compiler Framework
	\begin{itemize}
	\item
	Design and Implementation of the {\gicf}
	\item
	Embedding of a {\java} compiler in the {\gicf}
	\end{itemize}
\item
WebEditor to edit, compile, and run GIPSY programs online
\item
System Architecture Issues
	\begin{itemize}
	\item
	Rework and refactoring of most existing system design, both at the
	architectural and detailed design levels
	\item
	Major rework of the architecture and detailed design of {\gipc}
	\item
	Java sequential threads generation
	\item
	Threaded and {\rmi} communication procedures generation
	\item
	GIPSY Type System\footnote{Though the type system may seem not to be related
	to the architecture, but it impacted the design most of the main modules in it, so it
	was classified as architectural.}
	\item
	GIPSY Exceptions Framework
	\item
	Regression Testing Infrastructure
	\item
	Unit Testing Automation with \tool{JUnit}
	\end{itemize}
\end{itemize}

The last contributed items touch the rest of the {\gipsy},
the components and modules done by other team members. The integration
performed (outside of the main scope of this thesis) demanded
extensive testing. Without the integration and testing
work, these other contributions wouldn't be possible.
This also includes developing and enforcing
Coding Conventions and setting up project's CVS\index{CVS}
repository \cite{mokhovcoding, mokhovcvs, mokhovssh} for the entire project
as this work is to become a manual for the current and future GIPSY
developers and researchers.

%
%

\section{Scope of the Thesis}
\index{Thesis!Scope}
\index{Introduction!Scope of the Thesis}

%

While the Contributions section outlines the major work done,
the below explains what was not done or exhibits some limitations
at the time of this writing:

\begin{itemize}
\item
Integrated imperative compilers aren't native to the {\gipsy},
instead we call external compilers, such as \tool{javac}, \tool{gcc}, \tool{g++},
\tool{nmake.exe}, \tool{bc.exe}, \tool{perl}, etc. depending on a platform.

\item
Even though the mechanism was designed and implemented to generate CPs and STs,
only two of the concrete implementations of the actual CPs were done:
for local execution and distributed execution by extending the {\rmi} implementation done by Bo Lu.
The other implementations of CPs for {\jini}, {\complus},
{\corba} and others are being worked on by other team members at the time of this
writing.

\item
Semantic rules to have Java objects in {\olucid}
have been developed, but have not been formally proven to be correct.

\item
When presenting {\gicf} and the Preprocessor syntax, no semantic rules are given
for any of parts of the hybrid programs, except for {\jlucid} and {\olucid}, i.e.
the semantics of integrated {\java} itself or {\C} constructs, etc.

\item
{\jlucid} and {\olucid} are still in their experimental stage of development
and it will take some time before they mature.

\end{itemize}

%
%

\section{Structure of the Thesis}
\index{Thesis!Structure}
\index{Introduction!Structure of the Thesis}

The next chapter provides the necessary background on the Lucid family
of languages, its history, operational semantics, compiler frameworks, and hybrid programming.
Then, it gives the context of this thesis, the {\gipsy} system, and the tools and techniques
employed to make the contributions possible.
The core of this thesis is based on three publications, namely
\cite{mokhovjlucid2005, mokhovolucid2005, mokhovgicf2005}.
\xc{chapt:methodology} describes the approach and methodology
used to overcome and provide a solution to the problems stated
in \xs{sect:problem-statement}. Then, the design implementation
details are presented in \xc{chapt:implementation}.
\xc{chapt:testing} introduces the Regression Testing Suite
for {\gipsy} and what kinds of tests were performed and their limitations.
Finally,
\xc{chapt:conclusion} and \xc{chapt:future-work} conclude on the work done,
discuss the results and limitations
of the implementation,
and lay down
some paths towards enhancing the {\gipsy} in various
areas further. At the end, there is a list of references, Bibliography, and
an Appendix with most common abbreviations found in this work, CP and ST interfaces,
{\jlucid} and {\olucid} grammar generation scripts, etc.,
followed by an overall index.



\chapter{Background}
\label{chapt:background}
\index{Background}

%
%


While there is a complete and comprehensive set of references
in the Bibliography chapter that was a great deal of help to the
creation of this work, there are some keynotes that require special mention.
The following are some of the related readings that were sources
of inspiration and invaluable informational food for thought.
These include Joey Paquet's PhD thesis ``Scientific Intensional Programming''
\cite{paquetThesis}, related hybrid intensional-imperative programming
in various GLU-related work, such as \cite{glu1, glu2},
other recent hybrid programming papers, such as \cite{glu3, fcpp1, fcpp2},
the PhD thesis of Paula Bo Lu \cite{bolu04} and other theses of the GIPSY group,
such as \cite{chunleiren02, yimin04, leitao04, aihuawu02}, and
semantics of programming languages in \cite{semgrogono, javahoare, moellerhoare}.
Additionally, since this work also deals with compiler frameworks, a general
overview of existing frameworks is presented.
An on-line encyclopedia, Wikipedia \cite{wikipedia}, was a valuable resource for
the background and literature review, some of which is summarized
in the sections that follow.

%
%

\label{sect:ip}
\section{Intensional Programming}
\index{Intensional!Programming}
\index{Intensional Programming}

{\it Intensional programming} is a generalization of unidimensional contextual
(also known as {\it modal logic} \cite{carnap, kripke59, kripke63}) programming such as temporal programming,
but where the context is multidimensional and implicit rather than
unidimensional and explicit. Intensional programming is also called
{\it multidimensional} programming because the expressions involved are allowed
to vary in an arbitrary number of dimensions, the context of evaluation
is thus a {\it multidimensional context}. For example, in intensional programming,
one can very naturally represent complex physical phenomena such as
plasma physics (e.g. in {\tlucid} in \cite{paquetThesis}),
which are in fact a set of charged particles placed in
a space-time continuum that behaves according to a limited set of laws
of intensional nature. This space-time continuum becomes the different
dimensions of the context of evaluation, and the laws are expressed
naturally using intensional definitions \cite{paquetThesis}.
Joey Paquet's PhD thesis discusses the syntax and semantics of
the {\lucid} language, designs {\gipl} and {\tlucid}.
While we omit the {\tlucid} part, the reader
is reminded about the basic properties of the {\ilucid} and {\gipl} languages
in the follow up sections in greater detail to provide the necessary context
for the follow up work in \xc{chapt:methodology} and \xc{chapt:implementation}.

\subsection*{Intensional Logic}
\index{logic!intensional}

Intensional programming (IP) is based on intensional (or multidimensional or modal)
logic\index{intensional!logic}\index{logic!intensional} (where semantics was applied
first by \cite{carnap, kripke59, kripke63}), which, in turn, are based on Natural Language Understanding
(aspects, such as, {\it time}, {\it belief}, {\it situation}, and {\it direction} are considered).
IP brings in {\bf dimensions}\index{dimensions} and {\bf context}\index{context} to programs (e.g. space and time
in physics or chemistry). Intensional logic adds dimensions to logical
expressions; thus, a non-intensional logic\index{logic!non-intensional} can be seen as a constant or a
snapshot in all possible dimensions. {\it Intensions are dimensions} at which a
certain statement is true or false (or has some other than a Boolean value).
{\it Intensional operators}\index{intensional!operators} are operators that allow us to navigate within these
dimensions.

\subsection*{Temporal Intensional Logic}\index{logic!temporal}

{\it Temporal intensional logic} is an extension of temporal logic that allows
to specify the time in the future or in the past.

(1)\tab{5} $E_1$ := it is raining {\bf here} {\bf today}

\tab{20} Context: \{\texttt{place:}{\bf here}, \texttt{time:}{\bf today}\}

(2)\tab{5} $E_2$ := it was raining {\bf here} {\it before}({\bf today}) = {\it yesterday}

(3)\tab{5} $E_3$ := it is going to rain {\it at}(altitude {\bf here} + 500 m) {\it after}({\bf today}) = {\it tomorrow}

Let's take $E_1$ from (1) above. Then
let us fix {\bf here} to {\bf Montreal} and assume it is a {\it constant}.
In the month of March, 2004, with granularity of day, for every day, we can
evaluate $E_1$ to either {\it true} or {\it false}:

\begin{verbatim}
Tags:   1 2 3 4 5 6 7 8 9 ...
Values: F F T T T F F F T ...
\end{verbatim}

If you start varying the {\bf here} dimension (which could even be broken down
into $X$, $Y$, $Z$), you get a two-dimensional evaluation of $E_1$:

\begin{verbatim}
City / Day  1 2 3 4 5 6 7 8 9 ...
Montreal    F F T T T F F F T ...
Quebec      F F F F T T T F F ...
Ottawa      F T T T T T F F F ...
\end{verbatim}

The purpose of this example is to remind the reader the basic ideas
behind intensions and intensional programming and what dimensionality
is by using natural language. What follows is formalization of the
above in terms of the {\lucid} programming language.

%
%

\section{The Lucid Programming Language}
\index{Lucid!Introduction}

Let us begin by introducing the {\lucid} language history
and which features of it came at different stages of its
evolution to its present form. This is the necessary step
to further illustrate the purpose of this thesis.

\subsection{Brief History and The Family}
\index{Lucid!History}
\index{Lucid!Family}

From 1974 to {\lucid} Today:

\begin{enumerate}
\item
Lucid as a Pipelined Dataflow Language\index{Lucid!Pipelined Dataflows} through 1974-1977.
{\lucid} was introduced by Anchroft and Wadge in \cite{lucid76, lucid77}. Features:
	\begin{itemize}
	\item
	A purely declarative language for natural expression of iterative algorithms.

	\item
	Goals: semantics and verification of correctness of programming languages
	(for details see \cite{lucid76, lucid77}).

	\item
	Operators as pipelined streams: one for initial element, and then all for the
	successor ones.
	\end{itemize}

\item
Intensions, {\ilucid}, GRanular Lucid ({\glu}, \cite{glu1, glu2}), circa 1996.
More details on these two dialects are provided further in the chapter as
they directly relate to the theme of this thesis. Features:
	\begin{itemize}
	\item
	Random access to streams in {\ilucid}.

	\item
	First working hybrid intensional-imperative	paradigm
	({\C}/{\fortran} and {\ilucid}) in the form of {\glu}.

	\item
	Eduction or demand-driven execution (in {\glu}).
	\end{itemize}

\item
{\plucid}, {\tlucid}, 1999 \cite{paquetThesis}.
	\begin{itemize}
	\item
	{\plucid}
	is an intermediate experimental language used for
	demonstrative purposes in presenting the semantics of
	{\lucid} in \cite{paquetThesis}.

	\item
	{\tlucid} dialect was developed by Joey Paquet for plasma
	physics computations to illustrate advantages and expressiveness
	of {\lucid} over an equivalent solution written in {\fortran}.
	\end{itemize}

\item
{\gipl}, 1999 \cite{paquetThesis}.
	\begin{itemize}
	\item
		All Lucid dialects can be translated into
		this basic form of {\lucid}, {\gipl} through
		a set of translation rules. ({\gipl} is in the
		foundation of the execution semantics
		of {\gipsy} and its {\gipc} and {\gee} because its
		{\AST} is the only type of {\AST} {\gee} understands
		when executing a GIPSY program).
	\end{itemize}

\item RLucid, 1999, \cite{rlucid99}
	\begin{itemize}
	\item
		A Lucid dialect for reactive real-time
		intensional programming.
	\end{itemize}

\item
{\jlucid}, {\olucid}, 2003 - 2005
	\begin{itemize}
	\item
		These dialects introduce a notion of hybrid and object-oriented programming
		in the {\gipsy} with {\java} and {\ilucid} and {\gipl},
		and are discussed great detail in the follow up chapters
		of this thesis.
	\end{itemize}

\clearpage

\item
{\lucx} \cite{kaiyulucx}, 2003 - 2005
	\begin{itemize}
	\item
		Kaiyu Wan introduces a notion of contexts as first-class values
		in {\lucid}, thereby making {\lucx} the true intensional language.
	\end{itemize}

\item
{\onyx} \cite{grogonoonyx2004}, April 2004.
	\begin{itemize}
	\item
		Peter Grogono makes an experimental derivative of {\lucid} -- {\onyx}
		to investigate on lazy evaluation of arrays.
	\end{itemize}

\item
{\glusharp} \cite{glu3}, 2004
	\begin{itemize}
	\item
		{\glusharp} is an evolution of {\glu} where {\lucid}
		is embedded into {\cpp}.
	\end{itemize}
\end{enumerate}

%
%

\subsection{{\ilucid}}
\index{Lucid!Indexical}

When {\ilucid} came into existence,
it allowed accessing context properties in multiple dimensions.
Prior {\ilucid}, the only implied dimension was a set of natural numbers. With
{\ilucid}, we can have more than one dimension, and we can query for a part of
the context (any dimensions of it). Thus, the syntactic definition has been
amended to include an ability to specify which dimensions exactly
we are working on.

\subsubsection{Streams}
\index{Lucid!Streams}

Lucid variables and expressions are said to be {\it streams}
of values, through which one can navigate using some sort
of navigational operators. In the natural language example given
earlier the operators were {\it before}(), {\it after}(),
and {\it at}(); here we begin by introducing {\it first}()
and {\it next}() (very much like in {\lisp}).

If the following equations hold\footnote
{Note, these are initial conditions of a definition to illustrate the ideas
behind the streams and not an actual
declaration of constructs in the language one would normally write.}:

\begin{itemize}
\item
\olucidop{first} $X = 0$
\item
\olucidop{next} $X = X + 1$ (like \texttt{succ} in {\lisp})
\end{itemize}

\noindent
where $0$ is a stream of 0's: $(0, 0, 0, ..., 0, ...)$.
Likewise, $1$ is a stream of 1's, and the `$+$' operator performs
pair-wise addition of the elements in the streams according to
the implied current dimension index.
Thus, $X$ is defined as a stream, such that:

\begin{itemize}
\item
$x_0 = 0,  x_{i+1} = x_i + 1$, or
\item
$X = (x_0, x_1, ..., x_i, ...) = (0, 1, ..., i, ...)$
\end{itemize}

\noindent
Similarly, if:

\begin{itemize}
\item
\olucidop{first} $X = X$
\item
\olucidop{next} $Y = Y +$ \olucidop{next} $X$
\end{itemize}

\noindent
$Y$ here becomes a running sum of $X$:

\begin{itemize}
\item
$y_0 = x_0; y_{i+1} = y_i + x_{i+1}$
\item
$Y = (y_0, y_1, ..., y_i, ...) = (0, 1, ..., i(i + 1)/2, ...)$
\end{itemize}

%
%

\subsubsection{Basic Operators}
\index{Lucid!Basic Operators}

This section defines properties of basic Lucid operators,
which were proven by Paquet in \cite{paquetThesis}.

\paragraph{Operator \olucidop{fby}.}
\index{Indexical Lucid!fby}

Operator \olucidop{fby} stands for ``followed by''.
\olucidop{fby} allows simply to suppress dimension index
and switch to another stream.
As an example the previously shown
streams $X$ and $Y$ can be defined as follows using \olucidop{fby}:

\begin{itemize}
\item
$X = 0$ \olucidop{fby} $X + 1 = (0, 1, 2, ..., i, ...)$
\item
$Y = X$ \olucidop{fby} $Y +$ \olucidop{next} $X = (0, 1, ..., i(i + 1)/2, ...)$
\end{itemize}

\noindent
To provide an analogy to lists, we can say that
that the following operators
are equivalent:

\begin{itemize}
\item
\olucidop{first} and \texttt{hd}
\item
\olucidop{next} and \texttt{tl}
\item
\olucidop{fby} and \texttt{cons}
\end{itemize}

\paragraph{Informal Definition of \olucidop{first}, \olucidop{next}, \olucidop{fby}.}
\index{Indexical Lucid!first}
\index{Indexical Lucid!next}
\index{Indexical Lucid!fby}

\begin{itemize}
\item
Definitions:
	\begin{itemize}
	\item
		\olucidop{first} $X = (x_0, x_0, ..., x_0, ...)$
	\item
		\olucidop{next} $X = (x_1, x_2, ..., x_{i+1}, ...)$
	\item
		$X$ \olucidop{fby} $Y = (x_0, y_0, y_1, ..., y_{i-1}, ...)$
	\end{itemize}

\item
These are the three operators of the original {\lucid}.
\item
{\ilucid} has come into existence with the ability to access an arbitrary
element by some index $i$ in the stream.
\end{itemize}

\paragraph{Operators \olucidop{wvr}, \olucidop{asa}, and \olucidop{upon}.}
\index{Indexical Lucid!wvr}
\index{Indexical Lucid!asa}
\index{Indexical Lucid!upon}

The other three operators that are slightly more
complex informally defined below:

\begin{itemize}

\item
	$X$ \olucidop{wvr} $Y =$

	\tab{20} {\bf if} \olucidop{first} $Y \neq 0$

	\tab{20} {\bf then} $X$ \olucidop{fby} (\olucidop{next} $X$ \olucidop{wvr} \olucidop{next} $Y)$

	\tab{20} {\bf else} (\olucidop{next} $X$ \olucidop{wvr} \olucidop{next} $Y)$

\item
	$X$ \olucidop{asa} $Y =$ \olucidop{first} $(X$ \olucidop{wvr} $Y)$

\item
	$X$ \olucidop{upon} $Y =$

	\tab{20} $X$ \olucidop{fby}

	\tab{40} ({\bf if} \olucidop{first} $Y \neq 0$ {\bf then} $($\olucidop{next} $X$
			\olucidop{upon} \olucidop{next} $Y)$ {\bf else} $(X$ \olucidop{upon} \olucidop{next} $Y))$

\end{itemize}

\noindent
where \olucidop{wvr} stands for {\it whenever}, \olucidop{asa} stands for {\it as soon as}
and \olucidop{upon} stands for {\it advances upon}. \olucidop{wvr} chooses
from its left-hand-side operand only values in the current dimension where
the right-hand-side evaluates to {\em true}. \olucidop{asa} returns
the value of its left-hand-side as a first point in that stream as soon as the
right-hand-side evaluates to {\em true}. Unlike \olucidop{asa}, \olucidop{upon}
switches context of its left-hand-side operand uf the right-hand side
is {\em true}.

\subsubsection{Sequentiality Problem}
\index{Sequentiality Problem}

With tagged-token dataflows of the original Lucid operators one could
only define an algorithm with pipelined, or sequential, data flow:

\begin{itemize}
\item
It is wasteful use of computing resources (e.g. to compute an element $i$ we
need $i-1$, but $i-1$ may never be used/needed otherwise).
\item
Sequential access to the stream of values.
\end{itemize}

\subsubsection{Random Access to Streams}
\index{Stream!Random access to}
\index{Lucid!Operators @ and \#}

New intensional operators are introduced to remedy the sequentiality problem: @ and \#.
The operators are used as an index \# corresponding to the current position that
allows querying the current context, and @
is intensional navigation to switch the context.
With @ and \#:

\begin{itemize}
\item
the computation is defined according to a context (here a single integer),

\item
{\lucid} is no longer a data-flow language and is on the road to intensional programming, and

\item
the previously introduced intensional operators can be redefined in terms of
the operators \# and @.
\end{itemize}

\clearpage

\noindent
In terms of the three original operators of \olucidop{first}, \olucidop{next},
and \olucidop{fby} the operators @ and \# are defined as follows:

\definition{1}
{
	$\# = 0$ \olucidop{fby} $(\# + 1)$

	$X @ Y =$ {\bf if} $Y = 0$ {\bf then} \olucidop{first} $X$ {\bf else} (\olucidop{next} $X) @ (Y - 1)$
}

\noindent
Both $X$ and $Y$ in the above definition are variable streams, and their
current values are determined by their current context at the time of evaluation.
To redefine the meaning of @ and \# Paquet uses the denotational form, with
the following proposition:

\proposition{1}
{
	(1) $[\#]_i = i$

	(2) $[X @ Y]_i = [X]_{[Y]_i}$
}

\noindent
where (1) means the value of \# at the current context $i$ is $i$ itself
(i.e. we query the value of our current dimension), and (2) says that
evaluate $Y$ at the current context $i$ and then use $Y$ as a new context
for $X$.

\subsubsection{Definition of Lucid Operators By Means of @ and \#}
\index{Lucid!Operators @ and \#}

First we present the definition of the operators via @ and \# denoted in
\lucidop{monospaced font}, and then we will provide their equivalence to the original Lucid
operators, denoted as \olucidop{small caps}.

\definition{2}
{
	(1) \lucidop{first} $X$ = $X @ 0$

	(2) \lucidop{next} $X = X @ (\# + 1)$

	(3) $X$ \lucidop{fby} $Y =$ {\bf if} \# $= 0$ {\bf then} $X$ {\bf else} $Y @ (\# - 1)$

	(4) $X$ \lucidop{wvr} $Y = X @ T$ \lucidop{where}

		\tab{40} $T = U$ \lucidop{fby} $U @ (T + 1)$

		\tab{40} $U =$ {\bf if} $Y$ {\bf then} \# {\bf else} \lucidop{next} $U$

		\tab{15} \lucidop{end}

	(5) $X$ \lucidop{asa} $Y =$ \lucidop{first} $(X$ \lucidop{wvr} $Y)$

	(6) $X$ \lucidop{upon} $Y = X @ W$ 

		\tab{40} \lucidop{where} $W = 0$ \lucidop{fby} ({\bf if} $Y$ {\bf then} $(W + 1)$ {\bf else} $W)$ \lucidop{end}
}


%
%

\subsubsection{Abstract Syntax of {\lucid}}
\index{Lucid!Abstract Syntax}

Abstract and concrete syntaxes of {\lucid} for expressions, definitions, and operators are
presented in \xf{fig:expressions},
\xf{fig:gipl-definitions}, and \xf{fig:indexical-syntax} for
both {\ilucid} and {\gipl}.

\begin{figure}
\begin{minipage}[b]{\textwidth}
\begin{center}
\begin{tabular}{| r c l |} \hline
{\it op}             & $::=$ & {\it intensional-op} \\
                     & $|$   & {\it data-op} \\
& &\\
{\it intensional-op} & $::=$ & {\it i-unary-op} \\
                     & $|$   & {\it i-binary-op} \\
& &\\
{\it i-unary-op}     & $::=$ & \lucidop{first} $|$ \lucidop{next} $|$ \lucidop{prev} \\
{\it i-binary-op}    & $::=$ & \lucidop{fby} $|$ \lucidop{wvr} $|$ \lucidop{asa} $|$ \lucidop{upon} \\
& &\\
{\it data-op}        & $::=$ & {\it unary-op} \\
                     & $|$   & {\it binary-op} \\
& &\\
{\it unary-op}       & $::=$ & ! $|$ $-$ $|$ \lucidop{iseod} \\
{\it binary-op}      & $::=$ & {\it arith-op} \\
                     & $|$   & {\it rel-op} \\
                     & $|$   & {\it log-op} \\
{\it arith-op}       & $::=$ & $+$ $|$ $-$ $|$ $*$ $|$ $/$ $|$ \% \\
{\it rel-op}         & $::=$ & $<$ $|$ $>$ $|$ $<=$ $|$ $>=$ $|$ $==$ $|$ $!=$ \\
{\it log-op}         & $::=$ & \texttt{\&\&} $|$ \texttt{||} \\ \hline
\end{tabular}
\end{center}
\end{minipage}
\caption{Concrete {\ilucid} Syntax}
\label{fig:indexical-syntax}
\end{figure}

\begin{figure}
\begin{minipage}[b]{\textwidth}
\begin{center}
\begin{tabular}{c c l}
$E$ & $::=$ & $id$ \\
    & $|$   & $E(E_1,...,E_n)$ \\
    & $|$   & {\bf if} $E$ {\bf then} $E'$ {\bf else} $E''$ \\
    & $|$   & $\#E$ \\
    & $|$   & $E @ E' E''$ \\
    & $|$   & $E$ \lucidop{where} $Q$ \\
\end{tabular}
\end{center}
\end{minipage}
\caption{GIPL Expressions}
\label{fig:expressions}
\end{figure}

\begin{figure}
\begin{minipage}[b]{\textwidth}
\begin{center}
\begin{tabular}{c c l}
$Q$ & $::=$ & \lucidop{dimension} $id$ \\
    & $|$   & $id = E$ \\
    & $|$   & $id(id_1,id_2,...,id_n) = E$ \\
    & $|$   & $QQ$ \\
\end{tabular}
\end{center}
\end{minipage}
\caption{GIPL \lucidop{where} Definitions}
\label{fig:gipl-definitions}
\end{figure}

%
%

\subsubsection{Concrete GIPL Syntax}
\index{Syntax!GIPL}
\index{GIPL!Syntax}

The {\gipl} is the generic programming language of
all intensional languages, defined by the means
of only two intensional operators -- \texttt{@}
and \texttt{\#}. It has been proven that other intensional
programming languages of the Lucid family can be translated into the {\gipl} \cite{paquetThesis}.
The concrete syntax of the {\gipl} is presented in \xf{fig:gipl-syntax}.
It has been amended to support the \lucidop{isoed} operator of {\ilucid} for completeness
and influenced by the productions from {\lucx} \cite{kaiyulucx}
to allow contexts as first-class values while maintaining backward
compatibility to the {\gipl} language designed by Paquet in \cite{paquetThesis}.

\begin{figure}
\begin{center}
\begin{verbatim}
           E  ::=   id
               |    E(E,...,E)                      #LUCX
               |    E[E,...,E](E,...,E)             #GIPL
               |    if E then E else E fi
               |    # E
               |    E @ [E:E]                       #GIPL
               |    E @ E                           #LUCX
               |    E where Q end;
               |    [E:E,...,E:E]                   #LUCX
               |    iseod E;                        #INDEXICAL
           Q  ::=   dimension id,...,id;
               |    id = E;
               |    id(id,....,id) = E;             #LUCX
               |    id[id,...,id](id,....,id) = E;  #GIPL
               |    QQ
\end{verbatim}
\end{center}
\normalsize
\caption{Concrete {\gipl} Syntax}
\label{fig:gipl-syntax}
\end{figure}

\subsubsection{Semantic Rules}
\label{sect:lucid-semantics}
\index{Lucid!Semantics}

Paquet's PhD thesis \cite{paquetThesis} presents details of the operational semantics of {\gipl}
recited here for the unaware reader with a brief description.
\xf{fig:lucid-semantics} provides initial operational semantic rules for
{\ilucid} in Hoare Logic\index{logic!Hoare} \cite{moellerhoare, javahoare}.
Later on, these rules are extended to support free Java methods
and Java objects in {\jlucid} and {\olucid} respectively in \xc{chapt:methodology}.

\clearpage

\paragraph*{Notation}

\begin{itemize}
\item
$\johndef$ represents the definition environment where all symbols are defined
(a dictionary of identifiers).

\item
$\context{E}{a}$ represents current context of evaluation (a set of dimensions $\mathcal{P}$)
and the dictionary that yields a specified result $a$ under that context given
expression $E$.

\item
\texttt{const}, \texttt{op}, \texttt{dim}, \texttt{func}, 
and \texttt{var} represent what kind of construct types are
put into $\johndef$ as constants, operators, dimensions,
functions, and variables respectively.

\item
the $\mathbf{E_{Xid}}$ type of rules place different identifier
types listed above into the definition environment $\mathcal{D}$.

\item
the remaining $\mathbf{E_{xyz}}$-style rules correspond to the
execution (or rather application of) of the operators, functions,
and conditionals to their argument expressions given the
definition of them in $\mathcal{D}$ and the current context.
Thus, $\mathbf{E_{op}}$ specifies application of a defined 
operator function $f$ in the current context to its arguments
(usually one for unary operators and two for binary); $\mathbf{E_{fct}}$
applies the named function to its arguments translating the formal
arguments to actual; $\mathbf{E_{c_{T}}}$ and $\mathbf{E_{c_{F}}}$
correspond to conditional evaluation of the \texttt{then} and \texttt{else}
branching clauses; $\mathbf{E_{at}}$ and $\mathbf{E_{tag}}$ correspond
to the universal intensional operators @ and \# for switching of and querying for
the current context; and $\mathbf{E_{w}}$ corresponds to the scope definition
marked by the \texttt{where} clause.

\item
the $\mathbf{Q}$-style rules allow definitions within the scope of
the dimension $\mathbf{Q_{dim}}$ and variable identifier $\mathbf{Q_{id}}$ types
and their composition.

\end{itemize}

\begin{figure*}
\scriptsize
\begin{eqnarray*}
{\mathbf{E_{cid}}} &:& \frac
   {\johndef(\myid)=(\texttt{const},c)} {\context{\myid}{c}}\\\\
{\mathbf{E_{opid}}} &:& \frac
   {\johndef(\myid)=(\texttt{op},f)}
   {\context{\myid}{\myid}}\\\\
{\mathbf{E_{did}}} &:& \frac
   {\johndef(\myid)=(\texttt{dim})}
   {\context{\myid}{\myid}}\\\\
{\mathbf{E_{fid}}} &:& \frac
   {\johndef(\myid)=(\texttt{func},\myid_i,E)}
   {\context{\myid}{\myid}}\\\\
{\mathbf{E_{vid}}} &:& \frac
   {\johndef(\myid)=(\texttt{var},E)\qquad
    \context{E}{v}}
   {\context{\myid}{v}}\\\\
{\mathbf{E_{op}}} &:& \frac
   {\context{E}{\myid}\qquad
    \johndef(\myid)=(\texttt{op},f)\qquad
    \context{E_i}{v_i}
   }
   {\context{E(E_1,\ldots,E_n)}{f(v_1,\ldots,v_n)}}\\\\
{\mathbf{E_{fct}}} &:& \frac
   {\context{E}{\myid}\qquad
    \johndef(\myid)=(\texttt{func},\myid_i,E')\qquad
    \context{E'[\myid_i\leftarrow E_i]}{v}
   }
   {\context{E(E_1,\ldots,E_n)}{v}}\\\\
{\mathbf{E_{c_T}}} &:& \frac
  {\context{E}{\textit{true}}\qquad
    \context{E'}{v'}
   }
   {\context{\myifthenelse}{v'}}\\\\
{\mathbf{E_{c_F}}} &:& \frac
   {\context{E}{\textit{false}}\qquad
    \context{E''}{v''}
   }
   {\context{\myifthenelse}{v''}}\\\\
{\mathbf{E_{tag}}} &:& \frac
   {\context{E}{\myid}\qquad
    \johndef(\myid)=(\texttt{dim})
   }
   {\context{\#E}{\mathcal{P}(\myid)}}\\\\
{\mathbf{E_{at}}} &:& \frac
   {\context{E'}{\myid}\qquad
    \johndef(\myid)=(\texttt{dim})\qquad
    \context{E''}{v''}\qquad
    \mathcal{D},\mathcal{P}\mydagger[\myid\mapsto v''] \vdash E : v
   }
   {\context{E\;@E'\;E''}{v}}\\\\
{\mathbf{E_{w}}} &:& \frac
   {\qcontext{Q}{\mathcal{D}',\mathcal{P}'}\qquad
    \mathcal{D}',\mathcal{P}' \vdash E : v
   }
   {\context{E\;\mathtt{where}\;Q}{v}}\\\\
{\mathbf{Q_{dim}}} &:& \frac
   {}
   {\qcontext{\texttt{dimension}\;\myid}
    {\mathcal{D}\mydagger[\myid\mapsto(\texttt{dim})],
    \mathcal{P}\mydagger[\myid\mapsto 0]}
   }\\\\
{\mathbf{Q_{id}}} &:& \frac
   {}
   {\qcontext{\myid=E}
   {\mathcal{D}\mydagger[\myid\mapsto(\texttt{var},E)],
     \mathcal{P}}
   }\\\\
{\mathbf{QQ}} &:& \frac
   {\qcontext{Q}{\mathcal{D}',\mathcal{P}'}\qquad
    \mathcal{D}',\mathcal{P}' \vdash Q' : \mathcal{D}'',\mathcal{P}''
   }
   {\qcontext{Q\;Q'}{\mathcal{D}'',\mathcal{P}''}}
\end{eqnarray*}
\normalsize
\caption{Operational Semantics of GIPL}
\label{fig:lucid-semantics}
\end{figure*}

\clearpage

%
%

\subsubsection{Examples of Lucid Programs}
\index{Examples!Lucid}
\index{Lucid!Examples}

Two simple examples of Lucid programs are presented. The
examples demonstrate absence of iterative/sequential
operation as opposed to the traditional imperative programming
languages.

\paragraph*{Natural Numbers Problem}
\index{Examples!Natural Numbers Problem}

An example program in {\ilucid} that yields 44 as
the result is in \xf{fig:nat1}.
The way the program is expanded using the re-definitions of the Lucid operators,
such as \texttt{fby}, employing @ and \# in {\gipl} is shown in \xf{fig:nat2-intro}.

\sourcefloat
	{\begin{verbatim}
N @.d 2
where
    dimension d;
    N = 42 fby.d (N + 1);
end;
\end{verbatim}
}
	{fig:nat1}
	{Natural numbers problem in {\ilucid}.}

\sourcefloat
	{\begin{verbatim}
N @.d 2
where
    dimension d;
    N = if (#.d <= 0) then 42 else (N + 1) @.d (#.d - 1) fi;
end;
\end{verbatim}
}
	{fig:nat2-intro}
	{Natural numbers problem in {\gipl}.}

\paragraph*{The Hamming Problem}
\index{Examples!The Hamming Problem}

This example (see \xf{fig:merge-intro}) illustrates the simple use of
functions in {\lucid}.

\sourcefloat
	{\begin{verbatim}
H
where
    H = 1 fby merge(merge(2 * H, 3 * H), 5 * H);

    merge(x, y) = if(xx <= yy) then xx else yy
    where
        xx = x upon(xx <= yy);
        yy = y upon(yy <= xx);
    end;
end;
\end{verbatim}
}
	{fig:merge-intro}
	{Indexical Lucid program implementing the \texttt{merge()} function.}

\subsection{Lucid Now}
\index{Lucid!State of the Art}

To summarize, {\lucid} is a functional programming language where
a variable (stream), a function, a dimension, or even entire context
can be a {\it first class value} (i.e. can viewed and manipulated as data).
{\lucid} provides operators, such as @ and \#, to navigate within dimensions
and switch contexts. The language also exhibits the
eductive execution model (demand-driven distributed computation) that augments
the semantics with a warehouse (intensional value
cache) and its consistency\footnote{Paquet defines the augmented operational
semantics in \cite{paquetThesis} and Tao implements its first
incarnation in {\gipsy} \cite{leitao04}. This work has an impact on this
aspect by introducing the side effects with the imperative languages, which will
be discussed later.}.

%
%

\section{Hybrid Programming}
\index{Hybrid Programming}

There have been previous approaches to couple intensional or functional
and imperative and object-oriented paradigms prior to this work.
Some recent related work on the same issue is presented in
\cite{mllessequal, glu3, fcpp1, fcpp2} with the \cite{glu3} being the most
relevant. The two major approaches of addressing the OO
issue are -- either (1) to extend {\lucid} to become object-oriented or objects-aware
or (2) make a host imperative language be extended to embed {\lucid}.
The authors of \cite{glu3} chose the latter by extending
{\glu}-with-{\C} to GLU\#-with-{\cpp}, whereas this work
approaches the problem from {\lucid} to {\java}. This means a Lucid
program is the main one driving the computation. We will briefly
consider the following approaches to the hybrid programming:

\begin{itemize}
\item {\mllessequal}
\item {\fcpp}
\item {\glu}
\item {\glusharp}
\end{itemize}

\clearpage

\subsection{{\mllessequal}}

{\mllessequal} \cite{mllessequal} is a system introduced in 1996
that proposed to marry OOP and functional paradigms using their
own language and providing the details of the predicative and decidable typing rules
and operational semantics of such a system. Their main goal
is to be able to induce implicit polymorphism of functional languages
in objects. They do not extend an existing functional language with
the OO capabilities, instead they reinterpret all data types
as either abstract or concrete classes and use the dynamic
dispatch, a typical OO feature, on run-time types.

\subsection{{\fcpp}}

{\fcpp} \cite{fcpp1, fcpp2} tries to promote the functional paradigm
in {\cpp}. {\fcpp} is a library add-on to enable
higher-order polymorphic functions in a novel use
of {\cpp} type inference that is not very complex and
is still expressive. {\fcpp} adds support for both parametric and
subtype polymorphism policies for functions in order
to be able to fit FC++ functions within the
C++ object model and pass higher-order functions as parameters.
The FC++ functions are kept as
objects called {\em functoids} and use a reference counter machinery
for allocation and de-allocation. Closures in {\fcpp} (operation on a some state
and the state itself) can automatically be created
during functoid object creation, but their ``closing'' of that
state is not automatic and the state values have to be passed
explicitly during the creation process.
The library also
adds a set of functional operators from the {\haskell}
Standard Prelude. {\fcpp} comes more from the OOP-to-functional
point of view and conforms with standard software engineering
design patterns and is suitable for the common OO tasks.

\subsection{{\glu}}

{\glu} was the most general intensional programming tool recently available \cite{glu1}.
However, experience has shown that, while being very efficient, the {\glu}
system suffers from a lack of flexibility and adaptability \cite{paquetThesis}. Given that
{\lucid} is evolving continually, there is an important need for the successor
to {\glu} to be able to stand the heat of evolution \cite{paquetThesis}. The two major successors
of {\glu} are the {\gipsy} and {\glusharp} systems.

\subsubsection*{Eduction}
\index{eduction}
\index{eduction!GLU}
\index{GLU!eduction}

The earlier mentioned notion of eduction was
first introduced by the GLU compiler. {\glu}
supports so-called {\it tagged-token demand-driven dataflow}
where data elements (tokens) are computed on demand following a
dataflow network defined in {\lucid}.
Data elements flow in the normal flow direction (from producer to consumer)
and demands flow in the reverse order, both being
tagged with their current context of evaluation.
This form of lazy computation is inherited by {\gipsy} from {\glu}.

\subsection{{\glusharp}}

{\glusharp} \cite{glu3} is a successor of {\glu}, which enables
{\lucid} within {\cpp}. The authors argue for the embedding
small functional/intensional-language pieces of {\lucid}
into C++ programs allowing lazy (demand-driven) evaluation
of arrays and functions thereby making {\lucid} easily accessible
within a popular imperative programming language, such as {\cpp}.
Because {\glusharp} appeared quite recently (2004) to when this work was written,
its success compared to {\glu} is yet to be evaluated; however, it seems to suffer
from the same inflexibility {\glu} did and targets only {\cpp}
as a host language.

\section{Compiler Frameworks}
\index{Compiler Frameworks}
\index{Frameworks!Compiler}

A significant number of compiler frameworks emerged for
the past decade. All try to enable compilation of more
than one language, either hybrid or not, in an uniform
manner. Some frameworks or libraries became ``frozen'' (i.e. non-extendable)
and fixed to a specific set of languages, some other ones were build with the extension in mind,
so it is relatively easy to ``plug-in'' yet another compiler
into the system (a collection of compilers and the necessary tools)
with minimum integration work required. A brief
overview of different compiler frameworks is given next:

\begin{itemize}
\item
	{\glu} tried to accommodate {\fortran}, {\C}, and {\lucid}
	in one system, but was made so inflexible \cite{paquetThesis}
	that it would take a significant effort to extend it and
	add other languages to the system.

\item
	{\glusharp} merges {\lucid} and {\cpp}; however, makes
	no provisions for extension to other languages on either
	intensional or imperative side.

\item
	Microsoft .NET can also be thought of a commercial heterogeneous compiler
	framework (it is more than a compiler framework, but our focus
	is on compilers) that allows easy cooperation and application development
	between different language models, such as C\#, {\cpp}, Visual Basic,
	and Assembly in a homogeneous environment. However, none of these
	languages have natively any of the intensional or functional capabilities,
	so no native debugging support or other tools exist, even if one
	starts using {\fcpp} or {\glusharp} in this environment. Despite the fact
	that all programs can be compiled into the common bytecode, the debugging
	tools have to be aware of the functional paradigms on a higher level and
	they are not (at least at this writing).

\item
	The GNU Compiler Collection (GCC) can also be said as a compiler
	framework from the free software \cite{gcc}. It supports
	{\C}, {\cpp}, Objective-C, Objective-C++, {\java}, {\fortran}, and Ada.
	Again, these languages are more of an imperative nature, but it is
	far easier to add new language into GCC than to Microsoft .NET due
	to its openness.

\item
	Finally, the {\gipsy} presents the {\gipc} framework
	that is designed for expansion and integration of the intensional
	and imperative (and later functional) languages. This is presented
	through the rest of this thesis.

\end{itemize}

\clearpage

%
%

\section{General Intensional Programming System}
\index{General Intensional Programming System}
\index{GIPSY}

\subsection{Introduction}
\index{GIPSY!Introduction}
\index{Introduction!GIPSY}

\begin{figure}
	\begin{centering}
	\includegraphics[width=80pt]{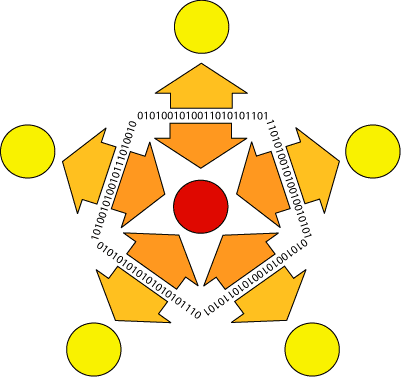}
	\caption{The {\gipsy} Logo representing the distributed nature of GIPSY.}
	\label{fig:gipsystar}
	\end{centering}
\end{figure}

{\gipsy} is broadly presented in \cite{wu03, bolu04, gipsy2005},
and others. Please refer to the online resources \cite{gipsy, gipsy2005, gipsywiki}
to obtain the most current status of the project.
{\gipsy} is primarily implemented in
Java. General GIPSY architecture is presented in \xf{fig:gipsy-general}.
The essence behind {\gipsy} is
demand-driven computation support for the intensional
programming languages, e.g.,
{\ilucid}\index{Lucid!Indexical},
{\tlucid}\index{Lucid!Tensor} \cite{paquetThesis},
etc.

\begin{figure}
	\begin{centering}
	\includegraphics[width=\textwidth]{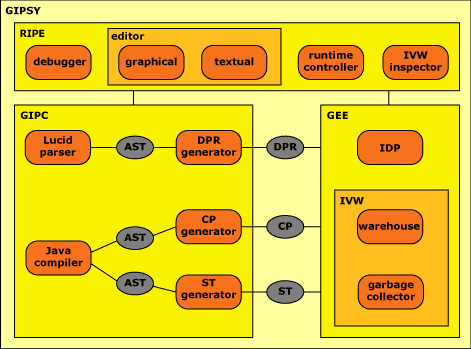}
	\caption{Structure of the GIPSY}\index{GIPSY!Structure}
	\label{fig:gipsy-general}
	\end{centering}
\end{figure}

The {\gipsy} consists in three modular
sub-systems: the General Intensional Programming Language Compiler ({\gipc});
the General Eduction Engine ({\gee}), and the Intensional Run-time Interactive
Programming Environment ({\ripe}). The sub-systems have to be modular so that
one implementation of parts of them or the whole can be replaced by another
without having major if any impact on the other modules.
Although the theoretical basis of the language has
been settled, the implementation of an efficient, general and adaptable
programming system for this language raises many questions. The following
sections outline the theoretical basis and architecture of the different
components of the system. All these components are designed in a modular
manner to permit the eventual replacement of each of its
components -- at compile-time or even at run-time -- to improve the
overall efficiency and productivity of the system \cite{paquetThesis}.

A GIPSY instance sends
out little bits of work to others to compute and then
gathers the results in distributed fashion. Of course,
synchronization, latency tolerance, and maximum utilization
of resources are primary goals for the system to be
productive. Unlike in most programming language models (see \cite{llModels})
considered for parallel computation, in {\gipsy}
several key concepts are considered:

\begin{itemize}
\item
Thread-Level Parallelism ({\tlp})
\item
Stream-Level Parallelism ({\slp})
\item
Cluster-Level Parallelism ({\clp})
\end{itemize}

{\gipsy}'s parallelism granularity takes into account the amount
of {\tlp}, {\slp}, and {\clp} available. {\tlp} determines the maximum number of
threads that should or can be created when a Lucid program
is being executed. In other words, {\tlp} defines on how many pieces of
terminal computational work we can chop a big
job into. The goal, as far as programming is concerned, is to program
for infinite {\tlp}, and later adjust (load-balance) at run-time
to the actual amount of {\slp}. {\slp} determines the maximum number of
{\it streams}\index{Stream}\index{Stream!hardware}
available
to execute the threads. Here, by ``streams'' we mean processors
but, with the invention of multithreaded CPUs for a
single processor, there may be several thread streams
available in parallel, and hence a more general notion of
{\slp}. The amount of {\slp} is machine-dependent and has to be discovered
at run-time on remote machines. If a job is to be run
on a single machine, GIPSY tries to maximize {\slp} utilization,
providing just enough {\tlp} for the machine in question with
the design goal of always assuming infinite {\tlp}.
Then load-balancing comes into play. {\clp} takes {\gipsy}
to another level --- distributed computing, involving
utilization of {\slp} of the machines across the network
nearby or across the globe over the Internet.

NOTE: the Lucid family of languages has also a notion of streams\index{Stream!Lucid variable} that refers
to Lucid variables that evaluate in multiple contexts.
Every Lucid stream (e.g. a variable) can potentially be evaluated on
any hardware stream available, but it is important not
to confuse the two kinds of streams. The reason for the existence of the two
notions is that both terms were used independently in each field. Now that
parallel architectures and language models such as {\lucid} came into
proximity, the terms clash.

\subsection{Goals}
\label{sect:gipsy-goals}
\index{GIPSY!Goals}

The system has to withstand the evolution of the tools, languages,
and underlying platforms, thus be flexible and adaptable to the changes.
That is one of the most
important and stringent requirements put on the development of {\gipsy}
\cite{paquetThesis}. Other subordinate requirements
in compiler design, run-time system, communication, and user
interfaces are presented in detail throughout the follow up sections.

\subsection{General Intensional Programming Compiler}
\index{GIPC}
\index{GIPC!Initial Conceptual Design}
\index{GIPC!Introduction}

\begin{figure}
	\begin{centering}
	\includegraphics[width=\textwidth]{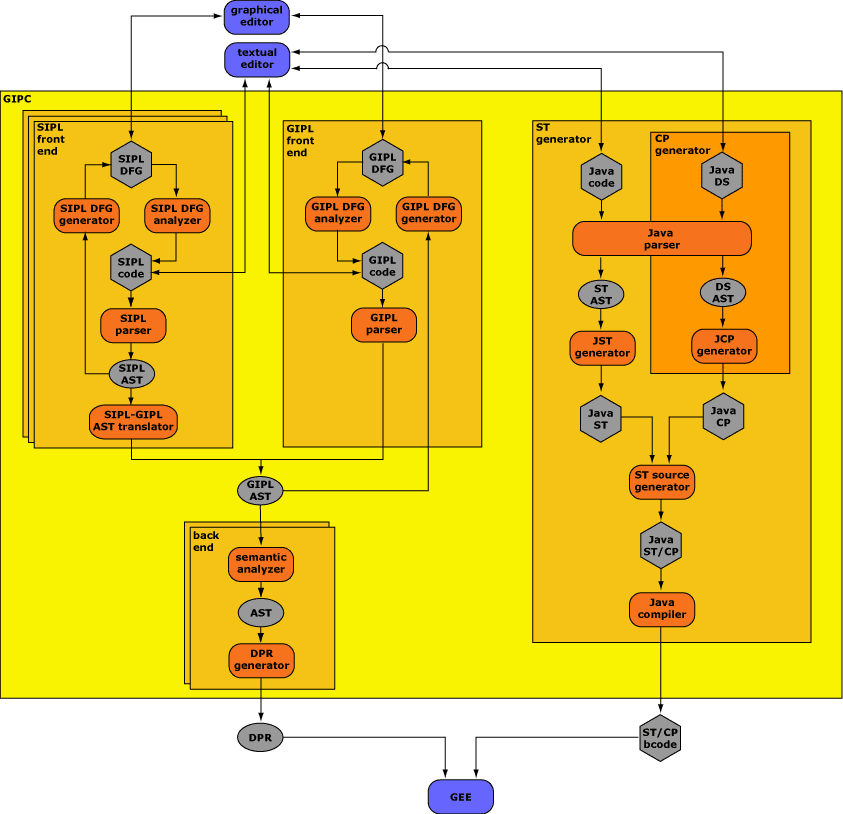}
	\caption{Initial Conceptual Design of the {\gipc}}
	\label{fig:gipc-initial-conceptual}
	\end{centering}
\end{figure}

GIPSY programs are compiled in a two-stage process (see \xf{fig:gipsy-program-compilation}, \xp{fig:gipsy-program-compilation}).
First, the intensional part of the GIPSY program is translated in Java,
then the resulting Java program is compiled in the standard way. 

The source code consists of two parts: the Lucid part that defines the
intensional data dependencies between variables and the sequential part
that defines the granular sequential computation units (usually written in any
imperative language, e.g. {\C} or {\java}).
The Lucid part is compiled into an intensional data dependency structure (IDS)
describing the dependencies between each variable involved in the Lucid part.
This structure is interpreted at run-time by the {\gee} following the demand
propagation mechanism. Data communication procedures used in a distributed
evaluation of the program are also generated by the {\gipc} according to the
data structures definitions written in the Lucid part, yielding a set of
communication procedures (CP). These are generated following
a given communication layer definition such as provided by
{\rpc} (or rather {\rmi} since GIPSY is implemented in {\java}), {\corba}, {\jini}, or
the WOS \cite{wos98}. The sequential functions defined in the second part of the GIPSY
program are translated into imperative code using the second stage imperative compiler syntax,
yielding imperative sequential threads (ST). Intensional function definitions,
including higher order functions, will be flattened using
a well-known efficient technique \cite{yotis94, paquetThesis}.
The closures in the higher order functions case are still applicable because
the function state and the operation on it are correctly passed to the functions
by expanding and using function definitions inline. The insignificant limitation here
is that self-referential closures for such functions cannot be made.
The function elimination in {\gipsy} pertinent to some of these aspects
was implemented by Wu in \cite{aihuawu02}.

The \xf{fig:gipc-initial-conceptual} presents the initial
conceptual design of the {\gipc}. Based on this design,
the GIPSY module integration and the development of the
STs and CPs support has begun. Later on the design was
refined in \cite{wugipc04, mokhovgicf2005} and its latest
reincarnation is shown in \xf{fig:gipc-preprocessor}
in \xc{chapt:implementation}, \xp{fig:gipc-preprocessor}; thus, the evolution description
is delayed until then.

Prior this work, {\gipc} supported only two Lucid dialects:
{\gipl} and {\ilucid}. The initial {\gipc} compiler was
implemented by Chun Lei Ren in \cite{chunleiren02}, and the
translation of the {\ilucid} into {\gipl} and the semantic
analysis was implemented by Aihua Wu in \cite{aihuawu02}.
A large integration and re-engineering effort went into
{\gipc} to approach it to the goals of the {\gipsy} (see \xs{sect:gipsy-goals})
and add more compilers for investigation of the underlying language models.
The results of this effort are presented in the Design and Implementation
chapter (\xc{chapt:implementation}).

\subsection{General Eduction Engine}
\index{GEE}
\index{GEE!Conceptual Design}
\index{GEE!Introduction}

\begin{figure}
	\begin{centering}
	\includegraphics[width=\textwidth]{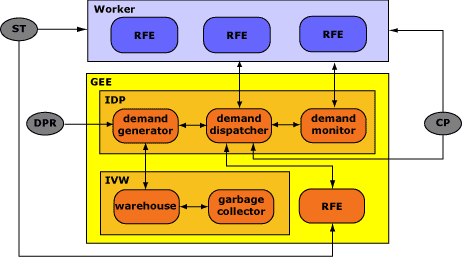}
	\caption{Conceptual Design of the {\gee}}
	\label{fig:gee-conceptual}
	\end{centering}
\end{figure}

The {\gipsy} uses a demand-driven model of computation, which is based on the
principle is that certain computation takes effect {\em only} if there is an explicit demand for it.
The {\gipsy} uses eduction, which is demand-driven computation in conjunction
with an intelligent value cache called a warehouse. Every demand can
potentially generate a procedure call, which is either computed locally
or remotely, thus eventually in parallel with other procedure calls. Every
computed value is placed in the warehouse, and every demand for an
already-computed value is extracted from the warehouse rather than computed
again and again (demands that may have side effects, e.g. if we cache
results of STs, shall not be cached).
Eduction, thus, reduces the overhead induced by the procedure calls needed
for the computation of demands sequentially. \xf{fig:gee-conceptual} describes the internal
conceptual structure and functioning of the {\gee}.

The {\gee} itself is composed of three main modules: the executor, the intensional demand propagator
(IDP), and the intensional value warehouse (IVW). First, the intensional
data dependency structure (IDS, which represents {\geer}) is fed to the demand generator (DG) by the
compiler ({\gipc}). This data structure represents the data dependencies between
all the variables in the Lucid part of the GIPSY program. This tells us
in what order all demands are to be generated to compute values from this program.
The demand generator receives the initial demand, that in turn raises the need for
other demands to be generated and computed as the execution progresses.
For all non-functional demands
(i.e. demands not associated with the execution of sequential threads (ST)),
the DG makes a request to the warehouse to see if this demand has already been
computed. If so, the previously computed value is extracted from the warehouse.
If not, the demand is propagated further, until the original demand resolves to
a value and is put in the warehouse for further use. This type of warehousing
was introduced by {\glu} due to its distributed nature to cut down on communication
costs, but it can certainly be applicable to any functional language, such as
{\lisp}, {\scheme}, {\haskell}, ML\index{ML} and others to improve efficiency even
on a single machine provided there are no any side effects whatsoever. The
garbage collector can run on the background to clean up old function-parameters-values
tuples periodically, and given that the large amounts of memory are cheap these days
functional languages may gain much more popularity with the increased performance.

For functional demands (i.e. demands associated with the execution of a
sequential thread), the demands are sent to the demand dispatcher (DD)
that takes care of sending the demand to one of the workers or to resolve it
locally (which normally means that a worker instance is running on the
processor running the generator process). If the demands are sent to a remote
worker, the communication procedures (CP) generated by the compiler are used
to communicate the demand to the worker. The demand dispatcher (DD) receives some
information about the liveness and efficiency of all workers from the demand
monitor (DM), to help it make better decisions in dispatching the demands.

The demand monitor, after some functional demands are sent to workers, starts to
gather various types of information about each worker, including, but not limited to:

\begin{itemize}
\item liveness status (is it still alive, not responding, or dead)
\item network link performance
\item response time statistics for all demands sent to it
\end{itemize}

These data points are accessed by the DD to make better decisions about the
load balancing of the workers, and thus achieving better
overall run-time efficiency.

Bo Lu was the first one to do the original design of the {\gee} framework \cite{bolu04}
and investigate its performance under threaded and {\rmi} environments.
She also introduced the notion of the Identifier Context (IC) classes -- demands
converted into {\java} code and using Java Reflection\index{Java!Reflection} \cite{java-reflection}
to compile, load, and execute them them at run-time. She also contributed
the first version of the interpreter-based execution engine.
Next, Lei Tao
contributed the first incarnation of the intensional value warehouse
and garbage collection mechanisms in \cite{leitao04} based on the
popular scientific library called NetCDF\index{NetCDF}. The author of this thesis
put an effort to modularize these all and make them easier to extend
and customize. He also provided the initial \api{GEE} application to
start available network services.
The {\gee} was also made aware of the STs and CPs as well as
the new type system, described in \xs{sect:gipsy-types}.
Further, Emil Vassev \cite{vas05} produced a very general and
functional framework for demand migration and its implementation,
Demand Migration System ({\dms}) that supports among other things
{\jini}, {\corba}, and {\dotnet} for fault-tolerant
demand transportation system, a part of the Demand Dispatcher.
The {\dms} is still pending integration as of this writing.

%
%

\subsubsection{Demand Propagation Resources for the GEE}
\label{sect:dpr}
\index{DPR}
\index{GEER}

The IDP generates and propagates demands according to the data dependence
structure ({\dpr}, now renamed to {\geer} in \cite{wu03}) generated by the {\gipc}.
If a demand requires some
computation, the result can be calculated either locally or on a remote
computing unit. In the latter case, the communication procedures (CP)
generated by the {\gipc} are used by the {\gee} to send the demand to the
worker. When a demand is made, it is placed in a demand queue, to be
removed only when the demand has been successfully computed. This way of
working provides a highly fault-tolerant system. One of the weaknesses of
{\glu} is its inability to optimize the overhead induced by
demand-propagation.
The IDP will remedy to this weakness by implementing
various optimization techniques:

\begin{itemize}
\item
Data blocking techniques used to aggregate similar demands at run time,
which will also be used at compile-time in the {\gipc} for automatic
granularization of data and functions for data-parallel applications

\item
The performance-critical parts (IDP and IVW) are designed as replaceable
modules to enable run-time replacements by more efficient versions adapted
to specific computation-intensive applications

\item
Certain demand paths identified (at compile-time or run-time) as critical
will be compiled to reduce their demand propagation overhead

\item
Extensive compile-time and run-time rank analysis (analysis of the
dimensionality of variables) \cite{dodd96}.
\end{itemize}

%
%

\subsubsection{Synchronization}
\label{sect:synchronization}
\index{Synchronization}

\subsubsection*{Distributed vs. Parallel}
\index{Synchronization!Distributed vs. Parallel}

It is important to make a distinction between parallel
and distributed computing. In parallel computing, SLP matters
and latency tolerance for memory references with mostly
UMA (uniform memory access) characteristics, whereas
in distributed computing communication is much more expensive (and perhaps even prohibitive)
and CLP matters as well. This setup largely exhibits
NUMA (non-UMA) characteristics (see \cite{probstBandwidth}) and latency
tolerance (and so also fault tolerance)
has a higher significance. This greatly impacts the
way we synchronize in parallel and distributed worlds.

\paragraph*{Synchronization in Distributed Environment}
\index{Synchronization!in Distributed Environment}

A distributed environment is a very popular domain these
days, so we'll start with it first.
Typically, the network is the scarce resource and is the bottleneck
for a distributed application because it implies communication (e.g., MPI\index{MPI}),
which is often unacceptable.
Therefore, many distributed applications choose not to communicate at all
or communicate very little through message passing.
This implies blocking on waiting for the network requests to propagate,
i.e. network latency.

\paragraph*{Synchronization in Parallel Environment}
\index{Synchronization!in Parallel Environment}

Synchronization in a parallel environment is more fine-grained, often
at the hardware level (e.g., a full/empty bit in memory cells).
Java does not give us control over such synchronization, so we have
to rely on the JVM built for an architecture that has such synchronization. The JVM
has to be developed to make use of the full/empty bits that
are usually represented as {\em future variables} \cite{probstBandwidth, jordan03}
in the languages specifically designed for parallel computing.

\subsubsection*{Secure Synchronization}
\index{Synchronization!Secure}

Secure synchronization is especially pertinent in a distributed environment.
Like any act of communication within worker-generator
architecture (see \xs{sect:worker}) and a warehouse (\xf{fig:gipsy-program-compilation}; \xs{sect:dpr}),
synchronization has to be secure to avoid (a) {\it over-demanding},
(b) incorrect results sent back, (c) loss of results and demands, and (d)
poisoning the warehouse with wrong data.
Secure synchronization implies fault tolerance. In GIPSY, we will rely on {\java}'s
{\rmi} and {\jini} over JSSE\index{JSSE} for secure communication in a distributed environment,
using {\java}'s synchronization primitives (see \xs{sect:implicit-synchronization})
to achieve the goal of secure synchronization. Thus, the reliability and accountability
of the results of a GIPSY program are dependent on these properties
of underlying Java Runtime Environment (JRE\index{JRE}) and the communication
protocols used.

\subsubsection*{Implicit vs. Explicit Synchronization}
\label{sect:implicit-synchronization}
\index{Synchronization!Implicit vs. Explicit}

One of the productivity metrics of a software completing
its task on time, is the efficiency of development of (see \cite{probstProgrammability}) such
a software, i.e., the amount of programmer's effort required to
create and debug the software. This is essentially a metric,
called time-to-solution (TTS\index{TTS}) \cite{probstProgrammability}; from creation until
the end result (e.g. completion of some scientific computation).
The goal is to minimize TTS\index{TTS}. One way to achieve this is ease of programming.
As the proportion of the work done by the compiler increases, so does the
reliability of the code, but we target scientific researchers, not just programmers.
Scientific researchers from math and physics should not care about these issues
and, thus, just be concerned mastering the basics of {\lucid}.
Therefore, the programmer
has to be freed from taking care of synchronization
explicitly, which a source of bugs and inefficiency
of programming (e.g., using {\java}'s synchronization primitives, such as \api{synchronized},
\api{Object.wait()}, \api{Object.notify()}, and \api{Object.notifyAll()},
\cite{javanuttshell}). The programmer should rather focus
on the problem being solved
and let the compiler/run-time system deal with
the synchronization pain. The {\gipsy} system, built around the Lucid family,
advocates implicit synchronization either by wrapping around the {\java}'s
synchronization primitives
or through the communication synchronization and data dependencies
(although a complete discussion is beyond the scope of this
thesis, see \cite{bolu04, vas05}).

\subsection{Run-time Interactive Programming Environment}
\index{RIPE}
\index{RIPE!Introduction}
\index{RIPE!Conceptual Design}

\begin{figure}
	\begin{centering}
	\includegraphics[width=\textwidth]{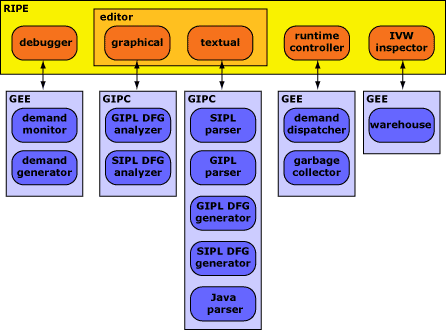}
	\caption{Conceptual Design of the {\ripe}}
	\label{fig:ripe-conceptual}
	\end{centering}
\end{figure}

The {\ripe} is a visual programming aid to the run-time environment ({\gee})
enabling the visualization
of a dataflow diagram corresponding to the Lucid part of the GIPSY program,
source code editing, launching the compilation and execution of GIPSY programs.
The original conceptual design of {\ripe} \cite{paquetThesis} is illustrated
in \xf{fig:ripe-conceptual}.
The user's points of interaction with the {\ripe} at run-time vary
in the following ways:

\begin{itemize}
\item
	Enable interactive editing of GIPSY programs
	via a variety of editors (textual, graphical, web).
\item
    Dynamic inspection of the IVW.
\item
    Modification of the input/output channels of the program.
\item
    Recompilation of the GIPSY programs.
\item
    Modification of the communication protocols.
\item
    Swapping of the parts of the {\gipsy} itself (e.g. garbage collection,
    optimization, warehouse caching etc. strategies).
\end{itemize}

Because of the interactive nature of the {\ripe}, the {\gipc} is modularly designed to
allow the individual on-the-fly compilation of either the IDS (by changing the
Lucid code), CP (by changing the communication protocol), or ST (by changing
the sequential code). Such a modular design even allows sequential threads to
be programs written in different languages (for now, we are concentrating on
Java sequential threads, but a provision is made for easy inclusion
of other languages with the {\gicf}, \xs{sect:gicf}).

The {\ripe} even enables the graphic
development of Lucid programs, translating the graphic version of the
program into a textual version that can then be compiled into an operational
version through a DFG generator of Yimin Ding \cite{yimin04}.
However, the development of this facility for graphical programming
posed many problems whose solution is not yet settled, for example representation
of the STs and CPs in the DFG nodes. An extensive and
general requirements analysis will be undertaken, as this interface will have
to be suited to many different types of applications. There is also the
possibility to have a kernel run-time interface on top of which we can
plug-in different types of interfaces adapted to different applications,
such as stand-alone, web-, or server-based.

\clearpage

%
%

\section{Tools}
\label{sect:tools}
\index{Tools}

This section presents a brief description of a variety of tools that
helped most with the implementation aspects of this work.

\subsection{Java as a Programming Language}
\index{Tools!Java}

The primary implementation language of {\gipsy} is {\java}.
This includes using Java's Reflection, JNI, and JUnit
frameworks and packages.
We have chosen to implement our project using the {\java} programming language
mainly because of
the binary portability of the Java applications as well as
its facilities, for e.g. memory management and communication tasks, so we can concentrate more on
the algorithms instead. {\java} also provides
built-in types and data-structures to manage
collections (build, sort, store/retrieve) efficiently \cite{javanuttshell, javasun}.
There is also source code written in other languages in the main GIPSY repository.
This includes LEFTY code for {\dfg} generation and the code of the test intensional
programs in various Lucid dialects.
The Java versions supported by {\gipsy} are 1.4 and 1.5. The {\gipsy}
will no longer build on 1.3 and earlier JDKs.

\subsubsection{Java Reflection}
\index{Tools!Java Reflection}
\index{Java!Reflection}

Java Reflection Framework \api{java.reflect.*} \cite{java-reflection}
allows us to load/query/discover a given class
for all of its API through enumeration of constructors, fields,
methods, etc. at run-time. This is incredibly useful for dynamic
loading and execution of our compilers, identifier context classes,
and sequential threads on local and remote machines.

The basic API from the reflection framework used in the implementation
of {\gipsy} is the \api{Class} class that allows getting arrays of
declared \api{Method} objects through the \api{getDeclaredMethods()} call that
will become the STs at the end, then for each \api{Method} the reflection
API allows getting parameter and return types via \api{getParameterTypes()}
and \api{getReturnType()} calls, which will become the CPs. The \api{Class.newInstance()}
method allows instantiating an object off the newly generated class. Likewise,
an enumeration of \api{Constructor} objects is acquired through the
\api{Class.getConstructors()} call. Constructors in {\java} are treated
differently from methods because they are not inherited and don't have
a return type (except that the type of the object they create). We still need
to enumerate them to allow Objective Lucid
programs to use the constructors, default or non-default, directly, so
we can get a handle on them similarly to STs.

\subsubsection{Java Native Interface (JNI)}
\label{sect:jni-intro}
\index{Tools!JNI}
\index{JNI}

The Java Native Interface (JNI) \cite{jni} is very useful for the
thread generation component of the {\gipc}.
We rely on JNI
to increase the number of popular imperative languages in which the
sequential threads could be written. Developers use the JNI to handle
some specific situations when an application cannot be written entirely
in {\java}, e.g. when the standard Java classes do not provide some
platform-dependent features an application may require, or use a library
written in another language be accessible to Java applications, or
for performance reasons a small portion of a time-critical code has
to be written say in {C} or assembly, but still be accessible from a
Java application \cite{jni}. In {\gipsy}, the second and third of the
listed cases are most applicable (e.g. to adopt {\glu} programs).
The JNI will allow us to avoid Lucid-to-C or Lucid-to-C++ type matching
as we can do it all through {\java} and maintain only Lucid-to-Java type
mapping table.

The JNI is made so that the native and Java sides of an application can
pass back and forth objects, strings, arrays and update their state on
either end \cite{jni}. The JNI is bi-directional, i.e., allows Java to use
the native libraries and applications and provide access to Java libraries
from the native applications.

The general methodology of creating a JNI application say that interacts
with a {\C} implementation is done in six steps \cite{jni}:

\begin{enumerate}
\item
Write a Java code with a \api{native} method to be implemented in {\C},
the \api{main()}, and the dynamic loading statement for a library
(to be compiled in the next steps).

\item
Compile the Java code with \tool{javac} and produce a \file{.class} file.

\item
Create a C header \file{.h} file from the compiled \file{.class} file by calling
\tool{javah}. This header file will provide the necessary \texttt{\#include}
directives along with the C-style prototype declaration of the \api{native}
method.

\item
Next, write the implementation of the function in regular
{\C} in a \file{.c} file.

\item
Then, create a shared library by compiling the \file{.h} and \file{.c} files
with a C compiler.

\item
Run the application regularly with the JVM (\tool{java}).
\end{enumerate}

\subsubsection{JUnit}
\label{sect:junit-intro}
\index{Tools!JUnit}
\index{JUnit}

JUnit is an open-source Java testing framework used to write and run automated
repeatable unit tests in a hassle-free manner \cite{junit}. The goal is to sustain application
correctness over time, especially when undergoing a lot of integration
efforts. JUnit is designed with software architecture patterns in mind
and follows best software engineering practices. It encourages developers
to write tests for their applications that withstand time and bit rot.

The main abstract class is \api{TestCase} that follows the Command design pattern
that implements the \api{Test} interface. This class maintains the name of
the tests (if it fails) and defines the \api{run()} method that has to be
overridden to do the actual testing work. The default Template Method
\api{run()} simply does three things: \api{setUp()}, \api{runTest()},
and \api{tearDown()}. Their default implementation is to do nothing,
so a developer can override them as necessary. Then, to collect the
test results they apply Collecting Parameter pattern. They use the
\api{TestResult} class for that.

JUnit makes a distinction between {\em errors} and {\em failures} in the following
way: errors to JUnit are mostly unexpected run-time or regular exceptions, whereas
failures are anticipated and are tested for using assertion checks.
The errors and failures are collected for further test failure reporting.

To run tests in a general manner from the point of view of the tester,
the test classes have with a generic interface using the Adapter pattern.
JUnit also offers a pluggable selector capability via the Java Reflection
API \cite{java-reflection}.
The \api{TestSuite} class represents a collection of tests to run.
In the {\gipsy}, the \api{Regression} application (see \xs{sect:regression})
comprises concrete implementation of such a test suite that tests
most of the feasible functionality of the {\gipc} and {\gee} modules.
See more details of application of JUnit to the {\gipsy} in \xc{chapt:testing}.

\subsection{\texttt{javacc} -- Java Compiler Compiler}
\index{Tools!JavaCC}

JavaCC \cite{javacc}, accompanied by JJTree, is the tool the GIPSY project is
relying on since the first implementation \cite{chunleiren02} to
create Java-language parsers and {\AST}s off
a source grammar files. The Java Compiler Compiler tool implements
the same idea for {\java}, as do \tool{lex}/\tool{yacc} \cite{louden97} (or \tool{flex}/\tool{bison}) for {\C} -- reading
a source grammar they produce a parser that complies with this grammar
and gives you a handle on the root of the abstract syntax tree.
The {\gipl}, {\ilucid}, {\jlucid}, {\olucid}, \api{PreprocessorParser}, and
\api{DFGGenerator} parsers are generated with the JavaCC/JJTree parser generation tools.
{\javacc} is a LL(K) \cite{louden97} parser generator, so the original {\gipl} and {\ilucid} grammars and the new
grammars had to be modified to eliminate or avoid the left recursion.

\subsection{MARF}
\label{sect:marf-in-tools}
\index{Tools!MARF}
\index{Frameworks!MARF}

Modular Audio Recognition Framework (MARF) library \cite{marf} provides a few
useful utility and storage classes {\gipsy} is using to manipulate threads, arrays,
option processing, and byte operations.
Despite {\marf}'s belonging to a voice/speech/natural language recognition and
processing library, it contains a variety of useful utility modules for
threading and options processing.

\subsection{CVS}
\index{Tools!CVS}
\index{CVS}

For managing the source code repository the Concurrent Versions System (CVS) \cite{cvs}
is used. The CVS allows multiple developers work on the {\em up-to-date} source tree
in parallel that keeps tracks of the revision history and works in an transactional
manner. The author produced a mini-tutorial on the CVS \cite{mokhovcvs} for
the GIPSY Research and Development team, which contains the necessary summary
for the team to work with the project repository.

While CVS has a comprehensive set of commands, the basic set includes:

\begin{itemize}

\item
\texttt{init} to initialize the repository
\item
\texttt{checkout} or \texttt{co} to checkout the source code
tree from the repository to a local directory
\item
\texttt{update} or \texttt{up} to make the local tree up-to-date
with the one on the server
\item
\texttt{add} to schedule a new file inside the existing local checkout
for addition to the repository
\item
\texttt{remove} to schedule a new file inside the existing local checkout
for removal from the repository
\item
\texttt{commit} to upload the changes done locally to the server
\item
\texttt{diff} to show the differences between the local and
the server versions of the tree
\end{itemize}

\subsection{Tomcat}
\index{Tools!Tomcat}

Apache Jakarta Tomcat \cite{tomcat} is an open-source Java application servlet and server pages container
project from Apache Foundation to
run web Java-based applications written in accordance with the Java Servlet and JavaServer Pages
\cite{servlets, jsp} specifications developed by Sun Microsystems.
Tomcat powers up the web front end to {\gipsy} to test intensional programs online.
The web frontend is represented by the \api{WebEditor} servlet as of this writing
a part of {\ripe} which is discussed later in \xc{chapt:implementation}. Tomcat
has an easy interface to deploy Java-based applications and their libraries, e.g.
through a manager presented in \xf{fig:tomcat}.

\begin{figure}
	\begin{centering}
	\includegraphics[width=\textwidth]{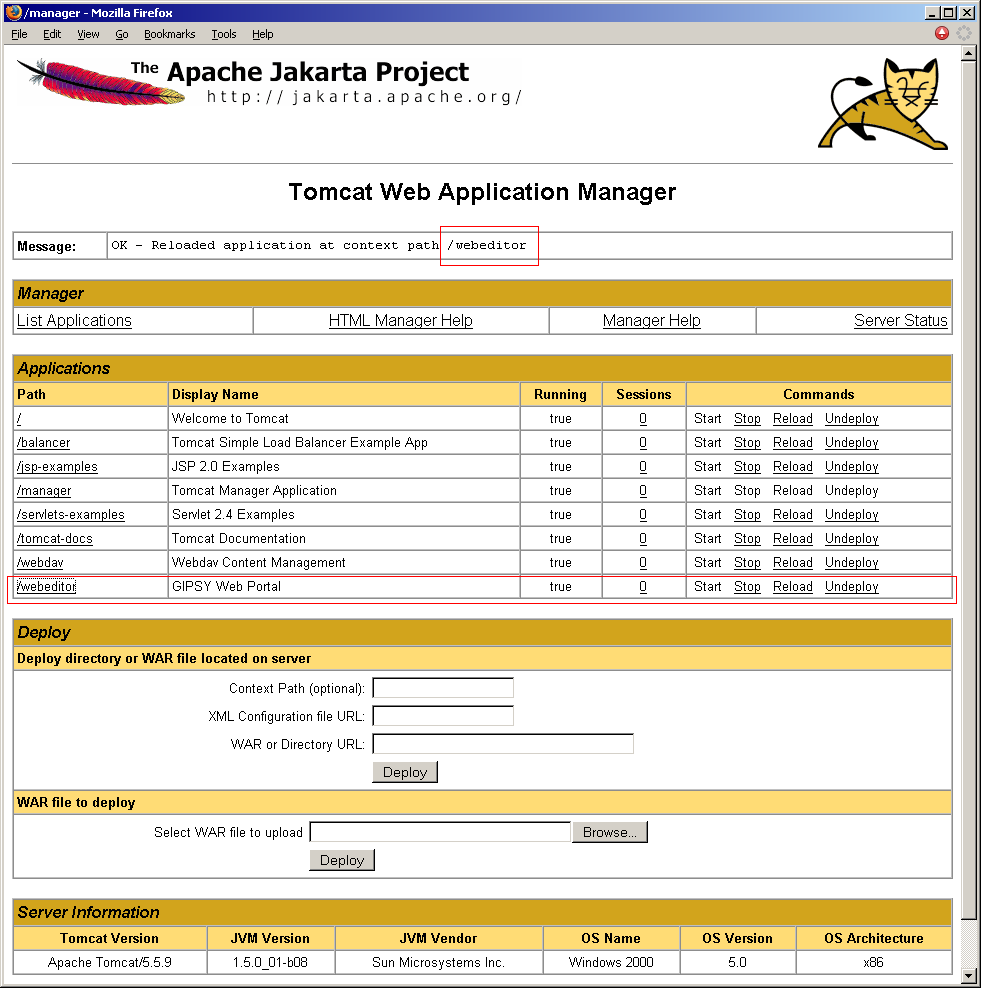}
	\caption{Tomcat Web Applications Manager}
	\label{fig:tomcat}
	\end{centering}
\end{figure}

Tomcat itself consists from a variety of modules that includes implementation
of the JSP (Jasper engine) and Servlet APIs, a webserver called Coyote, the application
server called Catalina, and many other things for logging, security, administration, etc.

\subsection{Build System}
\index{Build System}

The {\gipsy}'s sources can be built using a variety of
ways, using different compilers and IDEs on different platforms. This includes
Linux Makefiles, IBM's Eclipse, Borland's JBuilder, Apache's Ant, and Sun's
NetBeans.

\subsubsection{Makefiles}
\index{Build System!Makefiles}
\index{Tools!Makefiles}

{\unix}/{\linux} Makefiles are targeting all {\unix} systems that support {\gnu} \tool{make}
(a.k.a \tool{gmake}) \cite{gmake,mokhovmakefile}. Often, to compile all of the {\gipsy} is just enough to
type in \tool{make} and the system will be built. All {\unix} versions support
\tool{make}, and our system has been tested to build on \rhl{9}, \fcore{2},
\macos{X}, and \solaris{9}. There is a test script \tool{make-test.sh} that tests whether we are dealing
with the {\gnu} \tool{make} on {\unix} systems, as this is the only \tool{make}
supported.

\subsubsection{Eclipse}
\index{Build System!Eclipse}
\index{Tools!Eclipse}

There are project files \texttt{.project} and \texttt{.classpath}
that belong to this IDE from IBM \cite{eclipse}. The {\gipsy} build with this IDE
properly and has its library CLASSPATH set. Eclipse is another
open source tool available free of charge and provides extended
tools for Java projects development, refactoring, and deployment.

\subsubsection{JBuilder}
\index{Build System!JBuilder}
\index{Tools!JBuilder}

There is a project file \file{GIPSY.jpx}
that belongs to this IDE from Borland \cite{jbuilder}.
The {\gipsy} build with this IDE
properly and has its library CLASSPATH set.

\subsubsection{Ant}
\index{Build System!Ant}
\index{Tools!Ant}

There is a project file \file{build.xml}
that belongs to this build tool from the Apache Foundation \cite{ant}.
The {\gipsy} build with this tool
properly and has its library CLASSPATH set. In this case
\file{build.xml} is a portable way to write a Makefile
in XML.

\subsubsection{NetBeans}
\index{Build System!NetBeans}
\index{Tools!NetBeans}

There is a project file \file{nbproject.xml}
that belongs to this IDE from Sun \cite{netbeans}.
The {\gipsy} build with this IDE
properly and has its library CLASSPATH set.

\subsection{\tool{readmedir}}

This script generates a human-readable description
of a directory structure starting from some directory
with file listing and possibly descriptions (for this
there should be specially formatted file \file{README.dir} in
every directory traversed. The contents of this file will
be a part of the output and is a responsibility of the
directory creator/maintainer. The output formats of the
script are {\LaTeX}, HTML, and plain text.

\section{Summary}

In this chapter the reader was introduced to the necessary background
on the {\gipsy} project and how it is being managed starting from
the {\lucid} language origins to its implementations in the {\gipsy}
and the summary of the tools used to aid the advancement of the
project. In the {\gipsy} section the three main modules were introduced,
such as {\gipc}, {\gee}, and {\ripe}. While most of the remaining
work has gone into the {\gipc} in this thesis, the author had to
perform the necessary integration and adjustments to the {\gee}
and {\ripe}.



\chapter{Methodology}
\label{chapt:methodology}
\index{Methodology}

This chapter focuses on the methods and techniques proposed
to the solve the stated problems (see \xs{sect:problem-statement}).
The approaches described are based on three publications, namely
\cite{mokhovjlucid2005, mokhovolucid2005, mokhovgicf2005}.
\xs{sect:jlucid} introduces the {\jlucid} language and all
related considerations including the syntax and semantics.
Next, {\olucid} is introduced along with its syntax and semantics.
Further, the {\gicf} is introduced by providing the necessary
requirements for it to exist and the way to satisfy them. Lastly,
the summary is presented outlining the benefits and limitations
of the proposed solutions.

%
%

\section{JLucid: Lucid with Embedded Java Methods}
\label{sect:jlucid}
\index{JLucid}
\index{JLucid!Introduction}
\index{Introduction!JLucid}
\index{Lucid!JLucid}

%
%


%
%

\subsection{Rationale}
\index{JLucid!Purpose}
\index{JLucid!Rationale}

The name {\jlucid} comes from the {\gipc} component known as {\em Java Compiler}
within the Sequential Thread (ST) Generator of the {\gipsy}.
It subsumes all of {\ilucid} and General Intensional Programming Language
({\gipl}) \cite{paquetThesis} and syntactically allows embedded Java code.
In fact, a {\jlucid} program looks like a partial fusion of the intensional and Java code segments.
{\jlucid} gives a great deal of flexibility to Lucid programs by allowing
to use existing implementations of certain functions in {\java}, providing
I/O facilities and math routines (that {\lucid} entirely lacks), and other Java features
accessible to {\lucid}, arrays, and permits to increase the granularity of computations
at the operator level by allowing the user to define Java operators, i.e., functions manipulating
objects, thus allowing {\em streams of objects}\footnote{A more precise meaning
of Java objects within {\lucid} is explored further in the {\olucid} language,
including the meaning of an object stream and how object members are manipulated
(see for example \xs{sect:objective-lucid} and \xs{sect:objective-arrays}).
Additionally, since
the actual Java objects are flattened into primitive types, it would be possible
to access object members in parallel manner.}
in {\lucid}. {\jlucid} more or less
achieves the same goals and mechanisms as provided by {\glu}. What we are proposing
is a flexible compiler and run-time system that permits the evolution of
languages through a framework approach \cite{mokhovgicf2005, gipsy2005}.

\subsubsection{Modeling Non-Determinism}
\index{Non-Determinism}
\index{Lucid!Non-Determinism}
\index{JLucid!Non-Determinism}

{\lucid}, by its nature, is deterministic,
so introduction of imperative languages, such as {\java}, may allow us
to model non-determinism in Lucid programs for example by providing access to
random number generators available to the imperative languages. Non-determinism
can also be introduced as a result of side effects from for example reading
a different file each time an ST is invoked, or making a database query against
a table where data regularly changes, or say by reading the current time of day
value. Of course, a special care should be taken {\em not} to cache the results
of such STs in the warehouse.

%
%

\subsubsection{Loading Existing Java Code with \api{embed()}}
\label{sect:embed}
\index{embed()}

In a nutshell, we want to make the following possible for the Indexical Lucid program
in \xf{fig:merge} (replicated here from \xc{chapt:background} for convenience)
to become something as in \xf{fig:mergejavainline}
or, alternatively as in \xf{fig:mergejavaembed}.
The latter form would allow us to include objects from any types of URLs, local, HTTP\index{HTTP}, FTP\index{FTP}, etc.
The idea behind \api{embed()} is to include or to import the code written
already by someone and not to rewrite it in {\lucid} (which may not be a trivial task).
It is not meant to adjust to URL's existence at run-time as all embed-referenced
resources are resolved at compile time.
We ``include'' the pointed-to resource and attempt to compile it
where the original program-initiator resides. If the URL is invalid at compile time,
then there will be a compile error and no computation will be started.
\api{embed()} by itself does not necessarily provoke a remote function call.

\sourcefloat
	{}
	{fig:merge}
	{{\ilucid} program implementing the \texttt{merge()} function.}

\sourcefloat
	{\begin{verbatim}
#JAVA

void merge(int x, int y)
{
    // java code here
}

#JLUCID

H
where
    H = 1 fby merge(merge(2 * H, 3 * H), 5 * H);
end;
\end{verbatim}
}
	{fig:mergejavainline}
	{{\ilucid} program implementing the \texttt{merge()} function as inline Java method.}

\sourcefloat
	{\begin{verbatim}
H
where
    H = 1 fby merge(merge(2 * H, 3 * H), 5 * H);
    merge(x, y) =
        embed("file://path/to/class/Merge.class", "merge", x, y);
end;
\end{verbatim}
}
	{fig:mergejavaembed}
	{{\ilucid} program implementing the \texttt{merge()} function as \api{embed()}.}

\sourcefloat
	{\begin{verbatim}
F
where
    dimension d;
    F = foo(#d);
    where
        foo(i) = embed("file://my/classes/Foo.class", "foo", i);
    end;
end;
\end{verbatim}
}
	{fig:origEmbed}
	{Illustration of the \api{embed()} syntax.}


Existing Java code, in either \file{.class} or \file{.java} form, can be loaded with
\api{embed()}.
Intuitively, we would prefer the approach presented in
\xf{fig:origEmbed}. That added flexibility requires
syntactical extension of Lucid and is not portable.
For the program in \xf{fig:origEmbed} to work, \texttt{foo()}
has to
return a Java type of \texttt{int}, \texttt{byte}, \texttt{long}, \texttt{char}, \texttt{String},
or \texttt{boolean}, as per \xt{tab:datatypes}, \xp{tab:datatypes}.
A wrapper class will be created to extend from the \texttt{Foo}
and implement the \texttt{ISequentialThread} interface
(see \xa{sect:st-iface}).
General \texttt{embed()} syntax would be defined as follows:

\noindent
\fbox
{
	\begin{minipage}[t]{0.97\textwidth}
	\texttt{id(id, id, ...) ::= embed(URI, METHOD, id, id, ...);}
	\end{minipage}
}

\noindent
where \texttt{id} is the Lucid function name being defined that is mapped
to a Java's method named \texttt{METHOD} (which may or may not be of
the same name as the first \texttt{id}). The \texttt{URI}
is pointing to either \file{.class} or \file{.java} file.
Example URI's would be:

\noindent
\fbox
{
	\begin{minipage}[t]{0.97\textwidth}
	\texttt{foo(a,b) = embed("file://files/Foo.java","bar",a,b);}

	\texttt{bar(a,b) = embed("http://www.java.com/Foo.class","foo",a,b);}

	\texttt{baz(a,b) = embed("ftp://ftp.file.com/pub/Foo.java","zee",a,b);}
	\end{minipage}
}

These declarations associate Lucid functions with Java implementations.
Name clashes may be avoided, if necessary, by using different function
names.  Above, for example, Lucid \texttt{baz()} is implemented by Java \texttt{zee()}.

\sourcefloat
	{\begin{verbatim}
public class <filename>_<machine_name>_<timestamp>
extends my.classes.Foo
implements ISequentialThread
{
    // The definition is provided later in the text
}
\end{verbatim}
}
	{fig:embedST}
	{Generated corresponding ST to that of \xf{fig:origEmbed}.}

There are several ways of making this work. We could
extract either a textual or a bytecode definition of \texttt{foo()},
wrap it in our own class and, (re)compile it. However,
there is an issue here. What about other functions it may use, like
shown in \xf{fig:nat2java} with two methods calling each other? That
would mean extracting those dependencies as well along with the
method of interest. This won't scale very efficiently. Thus, alternate
approaches include: to either inherit from the desired class
as in \xf{fig:embedST}, encapsulate this class instance, or
attempt to wrap the entire class as done for the \texttt{JAVA} segment
in \xs{sect:java-segment} below. The former approach would imply having a class variable
instance of the type of that class encapsulated into the wrapper. The latter approach was
chosen as more feasible to implement,
although it doesn't deal with user-defined classes and subclass and packages
the \file{.class} or \file{.java} file may require at the moment.
Thus, the \api{embed()}
acts in a way similar to \texttt{\#include} in C/C++ or \texttt{import} in Java
of a set of Java definitions to be used in a JLucid program. Therefore,
\api{embed()} has to be resolved at compile time. Similar
technique may be taken towards other languages than {\java} at a later time.
{\lucid}'s syntax has to be extended to support \api{embed()}.

%
%

\subsubsection{The \texttt{\#JAVA} and \texttt{\#JLUCID} Code Segments}
\label{sect:java-segment}

This section explores ways of mixing Java and Lucid source code
segments in a single text file and ways of dealing with such a merge.

\sourcefloat
	{\begin{verbatim}
F
where
    dimension d;
    F = foo(#d);
    where
        foo(i) = int foo(int i) { return i + 1; }
    end;
end;
\end{verbatim}
}
	{fig:javafooinline}
	{Inline Java function declaration.}

An attempt to use Java's methods inline, such as in \xf{fig:javafooinline}
would be intuitive, but does not justify the effort spent on syntax analysis.
Therefore, we take the inline definition out of the Lucid part, and make
it a separate outer definition of the same method. Additionally, we explicitly
mark the \texttt{JLUCID} and \texttt{JAVA} code segments to simplify pre-processing of the JLucid code
as presented in \xf{fig:javafooouter}.

\sourcefloat
	{\begin{verbatim}
#JAVA

int foo(int i)
{
    // Some i + PI
    return (int)(java.lang.Math.PI + i);
}

#JLUCID

F
where
    dimension d;
    F = foo(#d);
end;
\end{verbatim}
}
	{fig:javafooouter}
	{Java method declaration split out from the Lucid part.}

Given the Natural Numbers Problem (see \cite{paquetThesis}) in \xf{fig:nat2} (replicated
here for convenience), one could imagine
the function definition for $N$ to be implemented in {\java} in two functions. To illustrate the point
when two separate functions can call each other in the \texttt{JAVA} segment or several
\texttt{JAVA} segments. This modified
JLucid code along with line numbers is shown in \xf{fig:nat2java}.
Since we allow one Java method to call another within, we have to
wrap them both into the same class.

\sourcefloat
	{}
	{fig:nat2}
	{Natural numbers problem in plain {\gipl}.}

The JLucid code segments after ``\texttt{\#JAVA}'' constructs
will be grouped together by the compiler. For all definitions (functions, classes, variables)
in these segments, their original location in the JLucid source
recorded and statically put in the wrapper class. These definitions
will end up in that wrapper
class as well.

It would be possible to have a class defined within a wrapper class or any other valid Java
declaration; even a data member can be included. To summarize, the Java
segments in the JLucid code are a body of a generated class that implements the \texttt{ISequentialThread}
interface.

\sourcefloat
	{\begin{verbatim}
     1    #JAVA
     2    
     3    int getN(int piDimension)
     4    {
     5        if(piDimension <= 0)
     6            return get42();
     7        else
     8            return getN(piDimension - 1) + 1;
     9    }
    10    
    11    int get42()
    12    {
    13        return 42;
    14    }
    15    
    16    #JLUCID
    17    
    18    N @d 2
    19    where
    20        dimension d;
    21        N = getN(#d);
    22    end;
\end{verbatim}
}
	{fig:nat2java}
	{Natural numbers problem with two Java methods calling each other.}

For the example in \xf{fig:nat2java} the parser would proceed as follows:

\begin{itemize}
\item
In the preprocessing step the source code is split into
two parts: the Java part and the Lucid part. For both
parts original source's line numbers and length of the definitions
are recorded.

\item
Then they both are fed to the respective parsers. Java's part
requires extra handling: the Java methods (one or more) defined in the code,
have to be wrapped into a class and then
\api{JavaCompiler} class that takes the Java portion of the source and feeds it to
\tool{javac}
for syntactic
and semantic analyses and byte code generation. They will become
parts of a Sequential Thread, ST (see \xs{sect:st}) definition fed to Workers
(see \xs{sect:worker}).

\item
The Lucid part is processed by the modified Lucid compiler (to include the
syntactical modifications for arrays and \api{embed()}) and comes up
with the main {\AST} from that.

\item
The Java STs are then linked into the main {\AST} in place of nodes where
the identifiers of these appear in the Lucid part of the program prior
semantic analysis.

\end{itemize}

Any method or other definition in the \texttt{JAVA} segment
is
wrapped into a class.
The generated wrapper class will contain a \api{Hashtable}
that maps method signature strings to their starting line in the original
JLucid code plus the length of the definitions in lines of text they occupy
statically generated and initialized. This is needed for the error reporting
subsystem in case of syntax/semantic errors, report back correctly the line
in the original JLucid program and not in the generated class. The class name
is created automatically from the original program name, the machine name
it's being compiled on, and a timestamp to guarantee enough uniqueness
to the generated class' name to minimize conflict for multiple such generated classes.
Thus, the \texttt{JAVA} segment in \xf{fig:nat2java} will transform into the generated
class as in \xf{fig:gen-st-short}.
This is a short version; for more detailed one please refer to the \xs{sect:genst-long}.
In fact, after generating this class (and possibly compiling it)
this situation can be viewed as a special case for \api{embed()}, \xs{sect:embed}
or vice versa. Note, since we have no guarantee the Java methods are
side-effects free in {\jlucid}, their results are not cached in the warehouse.

\sourcefloat
	{\begin{verbatim}
public class <filename>_<machine_name>_<timestamp>
implements gipsy.interfaces.ISequentialThread {
    private OriginalSourceCodeInfo oOriginalSourceCodeInfo;

    // Inner class with original source code information 
    public class OriginalSourceCodeInfo {
        // For debugging / monitoring; generated statically 
        private String strOriginalSource = ... 

        // Mapping to original source code position for error reporting 
        private Hashtable oLineNumbers = new Hashtable();

        // Body is filled in by the preprocessor statically 
        public OriginalSourceCodeInfo() {
            Vector int_getN_int_piDimension = new Vector();
            // Start line and Length in lines
            int_getN_int_piDimension.add(new Integer(3));
            int_getN_int_piDimension.add(new Integer(7));
            oLineNumbers.put("int getN(int piDimension)", 
                             int_getN_int_piDimension); 
            Vector int_get42 = new Vector();
            int_get42.add(new Integer(11));
            int_get42.add(new Integer(4));
            oLineNumbers.put("int get42()", int_get42); 
        }
    }

    // Constructor 
    public <filename>_<machine_name>_<timestamp>() {
        oOriginalSourceCodeInfo = new OriginalSourceCodeInfo();
    }
    /*
     * Implementation of the SequentialThread interface
     */
    // Body generated by the compiler
    public void run() {
        Payload oPayload = new Payload();
        oPayload.add("d", new Integer(42));

        work(oPayload);
    }

    // Body generated by the compiler statically
    public WorkResult work(Payload poPayload) {
        WorkResult oWorkresult = new WorkResult();
        oWorkresult.add(getN(poPayload.getVaueOf("d")));
        return oWorkResult;
    }

    /*
     * The below are generated off the source file nat2java.ipl
     */
    public static int getN(int piDimension) {
        if(piDimension <= 0) return get42();
        else return getN(piDimension - 1) + 1;
    }

    public static int get42() {
        return 42;
    }
}
\end{verbatim}
}
	{fig:gen-st-short}
	{Generated Sequential Thread Class.}

In \cite{mokhovjlucid2005} we required \texttt{foo()} in the previous examples
to be \texttt{static}.
In fact, any method or other definition in the \texttt{JAVA} segment
were to be transformed to become \texttt{static} while being
wrapped into a class.
For example, ``\texttt{int foo() \{return 1;\}}'' would become
``\texttt{public static int foo() \{...\}}''. We insisted on static
declarations only because the sequential threads were not instantiated
by the workers when executed.
This restriction has been lifted during implementation as we instantiate and serialize
the sequential thread class as needed.

%
%

\subsubsection{Is JLucid an Intensional Language?}
\index{JLucid!SIPL}
\index{SIPL!JLucid}

We treat {\jlucid} as a separate specific intensional programming language
({\sipl}) rather
than a part of a GIPSY program within existing Indexical Lucid implementation.
Here are some pros and cons of this approach and {\jlucid} as a separate
SIPL approach is the winner.
Why extend it as a separate {\sipl}?

\begin{itemize}

\item
This would serve as an example on how to add other SIPLs.

\item
This would allow us to keep the original {\ilucid} clean and working.

\item
This would allow functions with Java syntax to be used within a Lucid
program as well as binary Java function calls of pre-compiled classes.

\item
It can be extended to other languages as it turns out
to be a successful approach.
\end{itemize}

Why {\em not} to treat is as a separate {\sipl}?

\begin{itemize}
\item
We might want to have embedded {\java} (or other language) in any intensional language, not
just {\ilucid}. How to make that possible?

\item
It is not truly an {\sipl}, but a hybrid\index{hybrid!JLucid}.
\end{itemize}

%
%

\subsection{Syntax}
\index{JLucid!Syntax}
\index{Syntax!JLucid}

In {\jlucid}, we extend the syntax of both {\gipl} and {\ilucid} to
support arrays.  For example, it is useful to be able to evaluate
several array elements under the same context.
This is included by the last $E$ rules of $E[E,...,E]$ and $[E,...,E]$ in both
syntaxes. Arrays are useful to manipulate a collection
Lucid streams under the same context. JLucid arrays are mapped
to Java arrays on the element-by-element basis with the appropriate
element type matching and may only correspond to arrays of primitive
types in {\java}.
The syntax also includes the \api{embed()} extension
to allow including external Java code.
The JLucid syntax extensions to {\gipl} and {\ilucid} are presented
in \xf{fig:jgipl} and \xf{fig:jindexical}.

\begin{figure}
\begin{minipage}[t]{\textwidth}
\scriptsize
\begin{center}
\begin{verbatim}
                    E  ::=   id
                        |    E(E,...,E)
                        |    if E then E else E fi
                        |    # E
                        |    E @ E
                        |    E where Q end;
                        |    [E:E,...,E:E]
                        |    embed(URI, METHOD, E, E, ...)
                        |    E[E,...,E]
                        |    [E,...,E]
                    Q  ::=   dimension id,...,id;
                        |    id = E;
                        |    id(id,...,id) = E;
                        |    QQ
\end{verbatim}
\end{center}
\normalsize
\end{minipage}
\caption{{\small JLucid Extension to GIPL Syntax}}
\label{fig:jgipl}
\end{figure}

\begin{figure}
\begin{minipage}[t]{\textwidth}
\scriptsize
\begin{verbatim}
                E      ::=   id
                        |    E(E,...,E)
                        |    if E then E else E fi
                        |    # E
                        |    E @ E E
                        |    E where Q end;
                        |    E bin-op E
                        |    un-op E
                        |    embed(URI, METHOD, E, E, ...)
                        |    E[E,...,E]
                        |    [E,...,E]
                Q      ::=   dimension id,...,id;
                        |    id = E;
                        |    id.id,...,id(id,...,id) = E;
                        |    QQ
                bin-op ::= fby | upon | asa | wvr
                un-op  ::= first | next | prev
\end{verbatim}
\normalsize
\end{minipage}
\caption{{\small JLucid Extension to Indexical Lucid Syntax}}
\label{fig:jindexical}
\end{figure}

%
%

\subsection{Semantics}
\label{sect:jlucid-semantics}
\index{JLucid!Semantics}

The {\jlucid} extension to the operational semantics of {\lucid}
(see \xs{sect:lucid-semantics} on \xp{sect:lucid-semantics})
is defined in \xf{fig:jlucid-semantics}.
As in the original {\lucid} semantics, each type of
identifier can only be used in the appropriate situations.
Notation:

\begin{itemize}
\item
{\texttt{freefun, ffid, {{\underline{ffdef}}}}} mean a type
of identifier is a hybrid free (i.e. object-free) function
\texttt{freefun}, where \texttt{ffid} is its identifier
and \texttt{\underline{ffdef}} is its definition (body).

\item
The ${\mathbf{E_{ffid}}}$ rule defines {\jlucid}'s free functions.

\item
The {\jlucid} $\mathbf{\#JAVA_{ffid}}$ rule add free function definition
to the definition environment.
\end{itemize}

\begin{figure*}
\scriptsize
\begin{eqnarray*}
{\mathbf{E_{ffid}}} &\!\!\!\!\!\!\!\!\!\!:\!\!\!\!\!\!\!\!\!\!& \frac
   {
    \begin{array}{c}
    \context{E}{id}\qquad
    \context{E_1,\ldots,E_n}{v_1,\ldots,v_n}\\
    \johndef(id)=({\texttt{freefun, ffid, {{\underline{ffdef}}}}})\\
    \context{<\!\!{\mathtt{ffid}}(v_1,\dots,v_n)\!\!>}{v}
    \end{array}
   }
   {
    \context{E(E_1,\ldots,E_n)}{v}
   }\\\\
{\mathbf{\#JAVA_{ffid}}} &\!\!\!\!\!\!\!\!\!\!:\!\!\!\!\!\!\!\!\!\!& \frac
   {
      {{\mathtt{\underline{ffdef}}}} = \mathit{frttype}\texttt{ ffid}(\mathit{fargtype_1 }\;farg_{id_1},\dots,\mathit{fargtype_n }\;farg_{id_n})
   }
   {
    \qcontext{
      {{\mathtt{\underline{ffdef}}}}
    }
    {
     \mathcal{D}\mydagger[\texttt{ffid}\mapsto(\texttt{freefun, ffid, {\underline{ffdef}}})], \mathcal{P}}
    }\\\\
\end{eqnarray*}
\normalsize
\caption{{\small Additional basic semantic rules to support JLucid}}
\label{fig:jlucid-semantics}
\end{figure*}


%
%

\section{Objective Lucid: JLucid with Java Objects}
\label{sect:objective-lucid}
\index{Lucid!Objective}
\index{Objective Lucid}

\subsection{Rationale}
\index{Objective Lucid!Introduction}

{\olucid} is a direct extension of {\jlucid}.
The original syntax of {\ilucid} (and also for {\jlucid} and {\gipl}) is
augmented to support a so-called {\em dot-notation}. This allows {\lucid} to
manipulate grouped data by using object's methods. In fact,
the idea is similar to manipulating arrays in {\jlucid}. The difference with the arrays
is that they are manipulated as a collection of ordered data of elements of
the same type, to be evaluated in the same context. However, an object that varies
in some dimension implies that all its members, possibly of different types, also
potentially vary along this dimension, but across objects, i.e. the objects themselves
are not intensional. An object can be thought of as a heterogeneous collection of different
types of members, which you can access individually using their name, whereas arrays
can be thought of as a homogeneous collection of members that can be accesses
individually using their index.

Just like {\jlucid} \cite{mokhovjlucid2005}, {\olucid} is being developed as a separate
specific intensional programming language ({\sipl}) within the {\gipsy} for the same
reasons: keeping the other implementations undisturbed and working while
experimenting on this particular implementation.

%
%

\subsubsection{Pseudo-Objectivism in JLucid}
\label{sect:jlucid-pseudo-oop}
\index{JLucid!Pseudo-Objectivism in}

A pseudo-object-oriented approach is already present in {\jlucid}. The
program presented in \xf{fig:ret-object1} gives an example of
a Java function returning an object of type \api{Integer}.
\sourcefloat
	{\begin{verbatim}
#JAVA

Integer f()
{
    return new Integer("1234");
}

int g()
{
    return f().intValue();
}

#JLUCID

A
where
    A = g();
end;
\end{verbatim}
}
	{fig:ret-object1}
	{Pseudo-objectivism in JLucid.}
In {\jlucid} we are not able to manipulate this object
directly in intensional programming as {\java} does, though we can provide methods, such as \texttt{g()}
to access properties of a particular Java object from within
{\jlucid}. However, that reduces legacy Java code reusability by forcing the
programmer to add such functions in his code to be able to use it in the {\gipsy}.
Another example in \xf{fig:ret-object2} shows how one can
make use of objects in {\jlucid} by providing pseudo-free Java
accessors similar to \texttt{getComputedBar()} in the example. They are pseudo-free because they
don't appear as a part of any Java class to a {\jlucid} programmer explicitly, but internally they get
wrapped into a class when the code is compiled.
\sourcefloat
	{\begin{verbatim}
#JAVA

class Foo
{
    private int bar;

    public Foo()
    {
        bar = (int)(Math.random() * Integer.MAX_VALUE);
    }

    public int getBar()
    {
        return bar;
    }

    public void computeMod(int piParam)
    {
        bar = bar % piParam;
    }
}

int getComputedBar(int piParam)
{
    Foo oFoo = new Foo();
    oFoo.computeMod(piParam);
    System.out.println("bar = " + bar);
    return oFoo.getBar();
}

#JLUCID

Bar
where
    Bar = getComputedBar(5);
end;
\end{verbatim}
}
	{fig:ret-object2}
	{Using pseudo-free Java functions to access object properties in JLucid.}
In {\olucid} such explicit workarounds are not necessary anymore, but
this gives us some ideas about how to actually implement some features of {\olucid}
in practice, i.e., the compiler  can {\em generate} a number of pseudo-free accessors to
object's members and use {\jlucid}'s implementation of Java functions internally.

%
%
\subsubsection{Stream of Objects}
\index{Stream!of Objects}

An interesting question could be to ask:
``What is an object stream?''
Is it that the members of this object vary in the same
dimension(s) or they can have ``substreams''? In {\olucid}
we answer this as decomposing public object's data members into
primitive types and varying them or in simplified manner
we employ object's effectors. Thus, when there is a demand
say for the object's state (data members) at some time $t$, there will have
to be generated demands for all of $t$ between $[0,t]$ where
at time $0$ an instance of the object is created. Therefore,
the object state changes in the $[0,t]$ interval represent
the object stream in the context of this thesis. There are
two possible outcomes of this evaluation: either a portion
of object's state is altered by an intensional program
or the entire object. In the former case, {\lucid} only
accesses {\em some} object's members via the dot-notation
in the intensional manner, whereas in the latter case {\em all} the
members of an object are altered in the intensional context
implicitly. The examples presented in \xf{fig:cl}, \xf{fig:car-oop}, \xp{fig:car-oop},
and \xf{fig:Nat42}, \xp{fig:Nat42} work on portions of an object, whereas
the examples in \xs{sect:objective-arrays}, \xp{sect:objective-arrays}
work on all the members of an object at the same time.

\clearpage

%
%

\subsubsection{Pure Intensional Object-Oriented Programming}

Objective Lucid has presented a way for Lucid programs to use Java objects.
This may seem rather restrictive and may look like a workaround (though practical!).
An interesting
concept would be to extend the Lucid language itself to create and manipulate
pure Lucid objects, not Java objects. This will allow addressing issues
like inheritance and polymorphism and other attributes of object-oriented programming
and will solve the problem of matching Lucid and Java data types.
This is
not addressed in this work, but attempted to be solved
in \cite{wu05}.

%
%

\subsection{Syntax}
\index{Objective Lucid!Syntax}
\index{Syntax!Objective Lucid}
\index{Objective Lucid!The Dot-Notation}

The parser is extended to support the \verb+<objectref>.<feature>+
{\em dot-notation}
for the Lucid part of reference data types. The semantic
analysis is augmented to accommodate objects and user-defined
data types. In doing so, {\lucid} is able to manipulate Java
objects as well as access public variables and methods of these
objects. An example is shown in \xf{fig:cl}. This example
manipulates a simple object \texttt{E} by evaluating its state at some time ``2''.
The program begins with the construction of the object with \texttt{f1()}
(or one could call the object constructor directly), and then the rest
of the expressions access public members \texttt{x} and \texttt{foo()}
of the object during expression evaluation.

\sourcefloat
	{\begin{verbatim}
#JAVA
class ClassXB
{
    public int x;
    public float b;

    public ClassXB()
    {
        x = 0; b = 1.2;
    }

    public int foo(int a, float c)
    {
        return x = (int)(x * a + b * c);
    }

    ClassXB addx(int b)
    {
        x += b;
        return this;
    }
}

ClassXB f1()
{
    return new ClassXB();
}

#OBJECTIVELUCID

/*
 * The result of this program should be the object E
 * to be evaluated at time dimension 2 with its 'x'
 * member modified accordingly.
 */

E @time 2
where
    dimension time;
    E = f1() fby.time A;
    A = E.addx(B);
    B = E.foo(A @time C, A) + 3;
    C = E.x * 2;
end;
\end{verbatim}
}
	{fig:cl}
	{Objective Lucid example.}

The Objective Lucid syntax is in \xf{fig:olucid-syntax}. It is a
direct extension of the JLucid syntax in \xf{fig:jindexical} to support
the dot-notation. Essentially, the extension is the \texttt{E.id} productions.
Any \texttt{E} on the left-hand-side can evaluate to an
object type, but the right-hand-side is always an identifier
(Java class' data member or method).

\begin{figure}
\begin{centering}
\begin{minipage}[b]{\textwidth}
\scriptsize
\begin{verbatim}
                E      ::=   id
                        |    E(E,...,E)
                        |    if E then E else E fi
                        |    # E
                        |    E @ E E
                        |    E where Q end;
                        |    E bin-op E
                        |    un-op E
                        |    embed(URI, METHOD, E, E, ...)
                        |    E[E,...,E]
                        |    [E,...,E]
                        |    E.id
                        |    E.id(E,...,E)
                Q      ::=   dimension id,...,id;
                        |    id = E;
                        |    E.id = E;
                        |    id.id,...,id(id,...,id) = E;
                        |    QQ
                bin-op ::= fby | upon | asa | wvr
                un-op  ::= first | next | prev
\end{verbatim}
\end{minipage}
\caption{{\small Objective Lucid Syntax}}
\label{fig:olucid-syntax}
\end{centering}
\end{figure}

%
%

\subsection{Semantics}
\index{Objective Lucid!Semantics of}

To support these extensions to {\jlucid}, the Semantic Analyzer of {\jlucid} requires more non-trivial changes
than the syntax analysis and the dot-notation implementation due to
arbitrary object data types. In order to perform type checks and
apply the semantic rules of {\lucid}, we place the object data types
into the definition environment $\mathcal{D}$, which is in fact a semantic
equivalent to the data dictionary part of the {\geer}. This is partly solved by using
the pseudo-free Java functions, which {\em de-objectify} the object members,
but in order to be able to do so, we need to have the object types in the
definition environment. The corresponding operational semantic rules from
\cite{paquetThesis} can be extended as follows.

The {\olucid} extension to the operational semantics of {\lucid} is defined in
\xf{fig:olucid-semantics}.  As in the original {\lucid} semantics, each type of
identifier can only be used in the appropriate situations.
Notation:

\begin{itemize}
\item
    {\texttt{class, cid, {{\underline{cdef}}}}} means
    it is a Class type of identifier with name \texttt{cid}
    and a definition \texttt{\underline{cdef}}.

\item
	{\texttt{classv, cid.cvid, {{\underline{vdef}}}}} means
	that the variable is a member variable of a class \texttt{classv}
	with identifier \texttt{cid.cvid} given the variable
	definition {\underline{vdef}} within the class.

\item
	$\mathtt{<\!\!cid.cvid\!\!>}$ means object-member
	reference within an intensional program.

\item
    {\texttt{classf, cid.cfid, {{\underline{fdef}}}}} means
    that the function is a member function of a class \texttt{classf}
    with identifier \texttt{cid.cfid} given the variable
    definition {\underline{fdef}} within the class.

\item
	$<\!\!{\mathtt{cid.cfid}}(v_1,\dots,v_n)\!\!>$ represents
	a object-function call within an intensional program with
	actual parameters.

\item
	{\texttt{freefun, ffid, {{\underline{ffdef}}}}} mean a type
	of identifier is a hybrid free (i.e. object-free) function
	\texttt{freefun}, where \texttt{ffid} is its identifier
	and \texttt{\underline{ffdef}} is its definition (body).

\item
	By ${{\mathtt{\underline{cdef}}}} = {\mathtt{Class\; cid\; \{\ldots\}}}$
	we declare a class definition. A class can contain member variable \texttt{\underline{vdef}}
	and member functions definitions \texttt{\underline{fdef}}.

\end{itemize}

\noindent
The rules:

\begin{itemize}
\item
The ${\mathbf{E_{c-vid}}}$ rule defines an object member variable for
an expression for the dot-notation. It is independent from the
language in which we define and express our objects. The rule
says that under some context given two expressions $E$ and $E'$ that evaluate to a
class-type identifier $id$ and a variable type identifier $id'$
respectively and if the two together via a dot-notation represent an object-data-member
reference, then the expression $E.E'$ evaluates to a value $v$.

\item
Member function calls are resolved by the
$\mathbf{E_{c-fct}}$ rule. Similarly to the ${\mathbf{E_{c-vid}}}$ rule,
it defines that given two expressions $E$ and $E'$ under some context that evaluate to a
class-type identifier $id$ and a member function type identifier $id'$
and a set of intensional expressions ${E_1,\ldots,E_n}$ evaluates
to some values ${v_1,\ldots,v_n}$ and the two identifiers via a dot-notation
represent a member function call with parameters ${v_1,\ldots,v_n}$, then
we say the expression $E.E'(E_1,\ldots,E_2)$ is a member function
call that under the same context evaluates to some value $v$, i.e.
the function {\em always} returns a value. Here we see why it is
necessary for {\lucid} to map a \api{void} data type to implicit
Boolean \api{true}. This choice may seem a bit arbitrary (for example,
one could pick an integer $1$), but aside from practicality aspect
the mere choice of \api{true} may signify a successful termination
of a method.

\item
The ${\mathbf{E_{ffid}}}$ rule defines {\jlucid}'s free functions.
The rule is a simpler version of $\mathbf{E_{c-fct}}$ with no
class type identifiers present.

\item
The ${\mathbf{\#JAVA_{objid}}}$ rule places class definition
into the definition environment.

\item
The $\mathbf{\#JAVA_{obvjid}}$ and $\mathbf{\#JAVA_{objfid}}$ rules add \texttt{public} Java object
member variable and function identifiers along with their definitions to the
definition environment.

\item
The {\jlucid} $\mathbf{\#JAVA_{ffid}}$ rule add free function definition
to the definition environment.
\end{itemize}

\begin{figure*}
\scriptsize
\begin{eqnarray*}
{\mathbf{E_{c-vid}}} &\!\!\!\!\!\!\!\!\!\!:\!\!\!\!\!\!\!\!\!\!& \frac
   {
    \begin{array}{c}
    \context{E}{id}\quad
    \context{E'}{id'}\quad \\
    \johndef(id)=({\texttt{class, cid, {{\underline{cdef}}}}})\quad
    \johndef(id')=({\texttt{classv, cid.cvid, {{\underline{vdef}}}}})\quad \\
    \context{<\!\!{\mathtt{cid.cvid}}\!\!>}{v}
    \end{array}
   }
   {
    \context{E.E'}{v}
   }\\\\
{\mathbf{E_{c-fct}}} &\!\!\!\!\!\!\!\!\!\!:\!\!\!\!\!\!\!\!\!\!& \frac
   {
    \begin{array}{c}
    \context{E}{id}\qquad
    \context{E'}{id'}\qquad
    \context{E_1,\ldots,E_n}{v_1,\ldots,v_n}\\
    \johndef(id)=({\texttt{class, cid, {{\underline{cdef}}}}})\qquad
    \johndef(id')=({\texttt{classf, cid.cfid, {{\underline{fdef}}}}})\\
    \context{<\!\!{\mathtt{cid.cfid}}(v_1,\dots,v_n)\!\!>}{v}
    \end{array}
   }
   {
    \context{E.E'(E_1,\ldots,E_n)}{v}
   }\\\\
{\mathbf{E_{ffid}}} &\!\!\!\!\!\!\!\!\!\!:\!\!\!\!\!\!\!\!\!\!& \frac
   {
    \begin{array}{c}
    \context{E}{id}\qquad
    \context{E_1,\ldots,E_n}{v_1,\ldots,v_n}\\
    \johndef(id)=({\texttt{freefun, ffid, {{\underline{ffdef}}}}})\\
    \context{<\!\!{\mathtt{ffid}}(v_1,\dots,v_n)\!\!>}{v}
    \end{array}
   }
   {
    \context{E(E_1,\ldots,E_n)}{v}
   }\\\\
{\mathbf{\#JAVA_{objid}}} &\!\!\!\!\!\!\!\!\!\!:\!\!\!\!\!\!\!\!\!\!& \frac
   {
    {{\mathtt{\underline{cdef}}}} = {\mathtt{Class\; cid\; \{\ldots\}}}
   }
   {
    \qcontext{
      {{\mathtt{\underline{cdef}}}}
   }
   {
    \mathcal{D}\mydagger[{\mathtt{cid}}\mapsto(\mathtt{class,\;cid,\;{{\underline{cdef}}}})],\;\mathcal{P}}
   }\\\\
{\mathbf{\#JAVA_{objvid}}} &\!\!\!\!\!\!\!\!\!\!:\!\!\!\!\!\!\!\!\!\!& \frac
   {
    {{\mathtt{\underline{cdef}}}} = {\mathtt{Class\; cid\; \{\ldots {\mathtt{\underline{vdef}}}\ldots \}}}\qquad
    {{\mathtt{\underline{vdef}}}} = {{\mathtt{public}\; type\; {\mathtt{vid};}}}
   }
   {
    \qcontext{
      {{\mathtt{\underline{cdef}}}}
   }
   {
    \mathcal{D}\mydagger[{\mathtt{cid.vid}}\mapsto(\mathtt{classv,\;cid.vid,\;{\underline{vdef}}})], \mathcal{P}}
   }\\\\
{\mathbf{\#JAVA_{objfid}}} &\!\!\!\!\!\!\!\!\!\!:\!\!\!\!\!\!\!\!\!\!& \frac
   {
      {{\mathtt{\underline{cdef}}}} = {\mathtt{Class\; cid}}\;\{\ldots{\mathtt{\underline{fdef}}}\ldots\} \qquad
      {{\mathtt{\underline{fdef}}}} = {\mathtt{public}}\;\mathit{frttype}\texttt{ fid}(\mathit{fargtype_1 }\;farg_{id_1},\dots,\mathit{fargtype_n }\;farg_{id_n})
   }
   {
    \qcontext{
      {{\mathtt{\underline{cdef}}}}
    }
    {
     \mathcal{D}\mydagger[\texttt{cid.fid}\mapsto(\texttt{classf, cid.fid, {\underline{fdef}}})], \mathcal{P}}
    }\\\\
{\mathbf{\#JAVA_{ffid}}} &\!\!\!\!\!\!\!\!\!\!:\!\!\!\!\!\!\!\!\!\!& \frac
   {
      {{\mathtt{\underline{ffdef}}}} = \mathit{frttype}\texttt{ ffid}(\mathit{fargtype_1 }\;farg_{id_1},\dots,\mathit{fargtype_n }\;farg_{id_n})
   }
   {
    \qcontext{
      {{\mathtt{\underline	{ffdef}}}}
    }
    {
     \mathcal{D}\mydagger[\texttt{ffid}\mapsto(\texttt{freefun, ffid, {\underline{ffdef}}})], \mathcal{P}}
    }\\\\
\end{eqnarray*}
\normalsize
\caption{{\small Additional basic semantic rules\index{Objective Lucid!Semantic Rules} to support {\olucid}}}
\label{fig:olucid-semantics}
\end{figure*}


%
%

\section{General Imperative Compiler Framework}
\index{GICF}

\subsection{Rationale}

Having to deal with {\jlucid}, {\olucid}, and {\java} and a future likely
possibility to include other than {\java} imperative languages into intensional ones
prompted invention of a general mechanism to handle that and
simplify addition of new languages into the {\gipsy} for research
and experiments. This generalization touches several critical aspects
exposed by the {\jlucid} and {\olucid} languages involving such
a hybrid programming model. Thus, a core redesign of the {\gipc}
was necessary to enable this feature. The General Imperative
Compiler Framework (\gicf) addresses the generalization issues (split
among this Methodology and Design and Implementation chapters)
for the imperative compilers and suggests later development of a
similar framework for the intensional languages.

The core areas in the hybrid compilation process affect
the way an intensional language program (which now syntactically allows
having any number of code segments written in one
or more imperative languages) is compiled. This kind of program has
to be preprocessed first to extract the code segments
to be compiled by the appropriate language compilers
and at the same time maintains syntactic and semantic
links between the parts of a hybrid program. This influences
the general intensional compiler instrumentation, such as
generation of sequential threads and communication procedures,
function elimination, {\gipl}-to-{\sipl} translation, semantic analysis,
and linking (and later interpreting/executing) of a GIPSY program.

Requirements for any such a framework like {\gicf} imply at least
the following considerations:

\begin{itemize}
\item
having a number of compiler interfaces
known to the system that any concrete compiler implements,

\item
ability
to pick such compilers at runtime based on a hybrid program being
compiled,

\item
have a generalized {\AST} that is capable of holding
intensional and imperative nodes,

\item
have the semantic analyzer
understand possible data types that any language may expose (which is a very challenging goal to do correctly),
and deal with function elimination for the imperative parts of
the {\AST},

\item
preprocess by breaking down a hybrid GIPSY program's source code
to be fed to the appropriate compilers gives us flexibility of
allowing to include any imperative language we want, but complicates
maintenance of semantic links between the intensional and imperative
parts for later linking and semantic analysis.
This necessitates development of the two other special segments
that can declare in a uniform manner for GIPSY providing some meta information about embedded
imperative sequential threads, like function and type identifiers,
parameter and return types for communication procedures, and
user data types. Thus, for the former we need a function prototype
declaration segment, that lists all free functions declared
within imperative segments to be used by {\lucid} and the type
declaration segment for the user-defined types possibly declared
in those same imperative segments. The purpose of this meta-information is two-fold: it
will help us maintaining the semantic links via a dictionary and create so-called ``imperative stubs''.
The former prompts the development of the GIPSY Type System\index{GIPSY!Type System}
(see \xs{sect:gipsy-types}, \xp{sect:gipsy-types}) as understood
by the {\lucid} language and its incarnation within the {\gipsy} to
handle types in a more general manner. The latter stubs have to be
produced in order for the intensional language compilers (that stay intact
with the introduced framework) not to choke on ``undefined'' symbols
that really were defined in the imperative parts, which an existing
intensional compiler running in isolation fails to see.

\item
After all
involved compilers are finished doing compilation of their code segments,
they all produce a partial {\AST}. For intensional compilers that means
the main {\AST} with the intensional and stub nodes. For imperative compilers
it is the appropriate imperative {\AST} for each sequential thread. The
imperative {\AST}, in fact, need not to be a real tree and may contain
a single imperative node that would hold a payload of STs (compiled object
or byte code), CPs, type information,
and some meta-information (e.g. what language the STs and CPs are in and
for which operating system and native compiler environment).

\item
Then,
the imperative stubs have to be replaced by the real imperative
nodes at the linking stage before the semantic analysis.

\item
Once
the main tree is formed, the semantic analyzer would use the
type system to verify type information of the intensional-imperative
calls within taking into consideration imperative nodes when
doing function elimination and producing the final ``executable''
tree, or Demand {\AST}, or DAST, a component of the {\geer}.
\end{itemize}

All this work is motivated by the desire to simplify the addition of new compilers into
the GIPSY environment with minimal integration hassle. The follow up
sections explore some of the issues about primary matching of the
Java and GIPSY data types, followed by the definition of sequential
threads and communication procedures in the {\gipsy}, and their \api{Worker}
aggregator. While the below
are sections that lay down a concrete example based on {\jlucid} and {\java},
the discussion addressing
the generalization of the design and implementation of these issues
are presented in the chapter that follows with the actual sequence
diagram showing implementation details of the above hybrid compilation
process.

%
%

\subsection{Matching Lucid and Java Data Types}
\label{sect:datatypes-matching}
\index{data types!matching Lucid and Java}

Allowing {\lucid} to call Java functions brings a new set of issues related to
data types. Additional work is required on the semantic analyzer,
especially when it comes to type checks between Lucid and Java
parts of a JLucid program. This is pertinent when Lucid
variables or expressions are used as parameters to Java functions and when
a Java function returns a result to be assigned to a Lucid
variable or used in an IP expression. The sets of types in both cases are not exactly
the same. The basic set of Lucid data types as defined by Grogono
\cite{gipcincrements} is \api{int}, \api{bool}, \api{double}, \api{string},
and \api{dimension}. {\lucid}'s \api{int} is of the same size
as Java's \api{int}, and so are \api{double}, \api{boolean}, and \api{String}.
Lucid \api{string} and Java \api{String} are simply mapped to each other since internally
we implement the former as the latter; thus, one can think of the Lucid \api{string}
as a reference when evaluated in the intensional program. Based on this fact, the lengths
of a Lucid \api{string} and Java \api{String} are the same. Java \api{String} is also
an object in {\java}; however, at this point, a Lucid
program has no direct access to any object properties.
We also distinguish the \api{float} data type for single-precision floating point operations.
The \api{dimension} index type is said to be an integer for the time being, but
might become
a float when higher precision of points in time, for example, will
be in demand, or it could even be an enumerated type of unordered values
(though \api{float} dimensions will introduce some very interesting problems).
Therefore, we perform data type matching as presented in \xt{tab:datatypes}.
The return and parameter types matching sets are not the same because
of the size of the types. Additionally, we allow \api{void} Java return type
which will always be matched to a Boolean expression \api{true} in {\lucid} as
an expression has to always evaluate to something.

\begin{table}
\caption{{\small Matching data types between Lucid and Java.}}
\begin{minipage}[b]{\textwidth}
\begin{center}
\begin{tabular}{|c|c|} \hline
{\scriptsize Return Types of Java Methods}  & {\scriptsize Types of Lucid Expressions}\\ \hline\hline
\texttt{int}, \texttt{byte}, \texttt{long}  & \texttt{int} \\
\texttt{float}                              & \texttt{float} \\
\texttt{double}                             & \texttt{double} \\
\texttt{boolean}                            & \texttt{bool} \\
\texttt{char}, \texttt{String}              & \texttt{string} \\
\texttt{void}                               & \texttt{bool::true} \\ \hline\hline

{\scriptsize Parameter Types Used in Lucid} & {\scriptsize Corresponding Java Types}\\ \hline\hline
\texttt{string}                             & \texttt{String}\\
\texttt{float}                              & \texttt{float}\\
\texttt{double}                             & \texttt{double}\\
\texttt{int}, \texttt{dimension}            & \texttt{int} \\
\texttt{bool}                               & \texttt{boolean}\\ \hline
\end{tabular}
\end{center}
\end{minipage}
\label{tab:datatypes}
\end{table}

The table does not reflect the fact that JLucid is able to manipulate arrays of values (streams), but these arrays are not Java arrays
({\java}'s arrays are objects). In {\olucid} (see \xs{sect:objective-lucid}), we also have Java object data types will also
be manipulated by a Lucid program with the Lucid part being able to access object's properties and methods and have them as
return types and arguments. As for now our types mapping and restrictions are as per \xt{tab:datatypes}.

%
%

\subsection{Sequential Thread and Communication Procedure Generation}
\label{sect:stcp}
\index{Communication Procedure}
\index{Sequential Thread}

%
%

\subsubsection{Java Sequential Threads}
\label{sect:st}
\index{Sequential Thread}

Sequential threads are imperative functions that can be called in the {\lucid} part of a
GIPSY program.
The data elements of a Lucid program are integers and the like.  Using
them as such would result in a very inefficient computation due to the
overhead in generation and propagation of demands.  STs overcome this
problem.
The notion of sequential thread and
granularization of data was introduced by the {\glu} (Granular LUcid\index{Lucid!GLU} system
\cite{glu1, glu2}.

\begin{figure}
	\begin{centering}
	\includegraphics{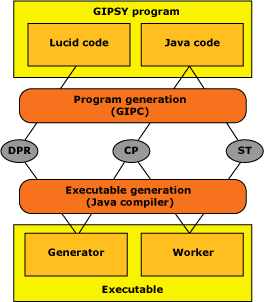}
	\caption{Hybrid GIPSY Program Compilation Process}\index{GIPSY!Compilation process}
	\label{fig:gipsy-program-compilation}
	\end{centering}
\end{figure}

Each GIPSY program potentially defines several Java methods that can be
called by the {\lucid} part of the program. Each of these functions are coded
in the Java part of the GIPSY program; thus, a sequential thread represents by itself a bit of work to compute
split into one or more Java methods.
They are compiled (see \xf{fig:gipsy-program-compilation}) to Java byte code
by the compiler
({\gipc}, \xf{fig:gipsy-general})
and packed into one executable, along with the
Communication Procedures (CP) (see \xs{sect:cp}) needed for the communication between the
generator and worker (\xs{sect:worker}, \xf{fig:gen-worker}).
The notion of worker is thus very close to the notion
of sequential threads, where a worker is basically the aggregation of the
(potentially) several sequential threads that can be executed by a worker,
along with the communications procedures needed for the generator-worker
communication.

Notice that the Generator-Worker Architecture may well be extended so that
the worker and the generator are fused into one; this is under review
and is discussed in \cite{bolu04} and in \cite{vas05}. This gives us distributed generators
as outlined in \cite{gipcincrements}, but as yet is only a topic for discussion.

%
%

\subsubsection{Java Communication Procedures}
\label{sect:cp}
\index{Communication Procedure}

The functional demands (i.e., demands that raise the need for a Java
function call) are potentially computed by remote workers, upon demand by
the generator. The demand is sent via the network by the generator to the
worker, along with the data representing the parameters of this Java
function call. Sending this data through the network requires the breaking
of the data structure into packets transmissible via a network. This
packing of the demand's input data is done by the Communication Procedures,
along with some kind of remote procedure call to the worker using, for
example, {\tcpip} {\rpc}. Once the function (the sequential thread) resolves,
the worker
(\xs{sect:worker})
is responsible for sending back the result to the generator that
called for this demand. That is also done by the CPs.

The CPs are generated by the compiler ({\gipc}) using the first part of the
GIPSY program: the definition of the data structures sent over the network
(i.e., the parameter and return types of the Java functions). The {\gipc} parses these
Java data structures and translates them into an abstract syntax tree. This
tree is then traversed by the CP generator, which generates byte code for
the communication procedures, following the communication protocol that was
selected.
{\it Serialization} summarizes much of this and Java helps us do it.

The CP generator has to be extremely flexible, as it has to be able to
generate code that uses various kinds of communication schemes.
In a nutshell, CPs determine the way a ST should be delivered to the computing host's worker
depending on the communication environment. For the localhost, it is plain
TLP (i.e., we create Java threads on a local machine) so
\texttt{NullCommunicationProcedure}
(\xs{sect:cp-iface})
is used. For distributed
environment CPs wrap transport functions over {\jini}, {\complus}, {\corba}, PVM, and {\rmi}
(see \cite{bolu04, vas05}) protocols.
Both CP and ST interfaces are presented in \xs{sect:interfaces}.

%
%

\subsubsection{C Sequential Threads and Communication Procedures with the JNI}
\label{sect:c-jni}
\index{JNI}

This is the methodology of how to extend the Java ST/CP generation concepts
to {\C} (and similarly can be done for {\cpp}) with the JNI \cite{jni} introduced
in \xs{sect:jni-intro}, \xp{sect:jni-intro}. This approach was designed, but
not implemented as of this writing; however, it may serve as a good head
start on the implementation of the \api{CCompiler} in {\gicf}.

Much of the ST wrapper class generation code for {\C} will be similar
to that of {\java}. The main difference is the bodies of the sequential
thread functions will not be present in the generated class as-is, but
they will be declared as \api{native} with no Java implementation.
The C code chunks will be saved to a \file{.c} file and the corresponding
\file{.h} fill will be generated declaring all the needed prototypes
with the \tool{javah} tool provided with the standard distribution of the JDK.
After that, we call an external C compiler to compile the C chunks into a shared
library. Thus, the other modification to the generated wrapper class the
\api{CCompiler} has to do, is to add a static initializer with the
\api{System.loadLibrary()} call for the newly compiled library with the C
implementation of our ST(s). The generated ST class and the compiled mini-library
can be stored together (e.g. the binary library file can be loaded into a byte
array of the class and deserialized back when about to be executed) in the
imperative node and later be communicated just like Java STs. A more sophisticated
alternative is to do the compilation and dynamic loading {\em after} communication
by the engine, but this can be a next step.

As far as type matching concerned, we still can use the same mapping
rules defined in \xs{sect:datatypes-matching} (and subsequently the \api{TypeMap}
class of the \api{JavaCompiler} presented later on) because with the JNI
with still work with {\java} and the JVM can do Java-to-native type translation
to {\C} or {\cpp} for us, not only for primitive types, but also for arrays,
objects, and strings.

%
%


\subsubsection{Worker Aggregator Definition in the Generator-Worker Architecture}
\label{sect:worker}
\index{Worker}
\index{Worker!Definition}

The {\gipsy} uses a generator-worker execution architecture as shown in \xf{fig:gen-worker}.
The {\geer} generated by the {\gipc} is interpreted (or executed) by the generator
following the eductive model of computation. The low-charge ripe
sequential threads are evaluated locally by the generator. The
higher-charge ripe sequential threads are evaluated on a remote worker.
The generator consists of two systems: the Intensional Demand Propagator
(IDP) and the Intensional Value Warehouse (IVW) \cite{leitao04}. The IDP implements the
demand generation and propagation mechanisms, and the IVW implements the
warehouse. A set of semantic rules that outlines the theoretical aspects
of the distributed demand propagation mechanism has been defined in \cite{paquetThesis}.
The worker simply consists of a ``Ripe Function Executor'' (RFE),
responsible for the computation of the ripe sequential threads as demanded
by the generator. The sequential threads are compiled and can be either
downloaded/uploaded dynamically by/to the remote workers. Better efficiency
can be achieved by using a shared network file system.

\begin{figure}
	\begin{centering}
	\includegraphics[width=\textwidth]{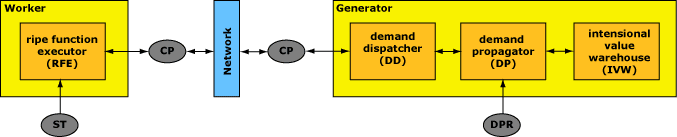}
	\caption{Generator-Worker Architecture}
	\label{fig:gen-worker}
	\end{centering}
\end{figure}

An example: a GIPSY screen saver would be a sample worker running when the an ordinary PC
is going into an idle mode and normally launches ordinary dancing bears
screensavers, it can actually run our downloaded worker instead
and contribute to computation. When such a worker starts, it has to
register it within a system somehow (see \cite{vas05}), so that the generators are aware of
its presence and can send demands to it. In the event of merging of
semantics of a worker and a generator, such a screensaver would also
be able to generate demands and maintain a local warehouse.

%
%

\section{Summary}

This chapter presented methodology behind concrete implementations
of the first two hybrid languages in the {\gipsy} -- {\jlucid} and {\olucid}.
Semantic rules were presented for free Java functions and Java objects to be included into the Lucid
programs and evaluated by the eduction engine in the hybrid environment.
Furthermore, operational semantics of {\olucid} is
clearly defined and is compatible with the semantics of {\lucid}.
The general requirements
for the {\gicf}, a tool simplifying imperative compiler management within {\gipc},
are introduced.
The follow up chapter details the architectural and detailed designs and concrete implementation of the languages
as well as General Intensional Compiler Framework and overall module integration and their interfaces.
Some immediate benefits and limitations are outlined below.

\clearpage

\subsection{Benefits}

\begin{itemize}
\item
{\jlucid} opens the door for STs and CPs and first hybrid programming
paradigm in the {\gipsy}.
\item
{\jlucid} provides ability to either write Java code alongside the Lucid code or embed
existing one via \api{embed()}.
\item
{\olucid} introduces Java objects and their semantics in the {\gipsy}.
\item
{\gicf} generalizes the \api{embed()} mechanism to all languages
in the {\gipsy}.
\item
{\gicf} promotes general type handling
in the {\gipsy}.
\item
{\gicf} promotes general compiler handling
in the {\gipsy}.
\item
{\gicf} generalizes the notion of the STs and CPs for all compilers.
\end{itemize}

\subsection{Limitations}

\begin{itemize}
\item
{\jlucid} is limited only to {\gipl}-{\java} and {\ilucid}-{\java} hybrids.
\item
{\jlucid} does not allow Java objects.
\item
{\jlucid} restricts the \api{embed()} mechanism only to itself and
its derivative -- {\olucid}.
\item
{\olucid} is primarily an experimental language to research on Java
objects in the intensional environment.
\item
{\gicf} addresses mostly the imperative compilers, but a similar approach
can be applied to the intensional and functional ones.
\end{itemize}



\chapter{Design and Implementation}
\label{chapt:implementation}
\index{Implementation}

%
%


This chapter combines the architectural\index{Design!Architectural} and detailed\index{Design!Detailed}
designs and integration of
the modules contributed not only by the author of this thesis but also by the other
GIPSY team members.
\xs{sect:internal-design} explores the GIPSY architecture
and implementation of the major components and frameworks.
Then, \xs{sect:external-design}
focuses on the user interface and external library interfaces.
User interfaces, class and sequence diagrams are provided mostly
following the top-down approach.
For GIPSY Java packages, directory structure with description of each package, and
\file{.jar} file packaging please refer to \xa{chapt:arch-layout}.

%
%

\section{Internal Design}
\label{sect:internal-design}
\index{Design!Internal}

The GIPC framework redesign along
with the realization of the two children frameworks of {\gicf} and {\iplcf}
are presented first followed by the design and implementation
of {\jlucid} and {\olucid} integrated into the new frameworks.

%
%

\subsection{General Intensional Programming Compiler Framework}
\index{Frameworks!GIPC}
\index{Internal Design!GIPC}

The GIPC Framework experienced several iterations of refinements
as a result of this research. Two new frameworks emerged,
namely General Imperative Compiler Framework ({\gicf}) to handle
all imperative languages within the {\gipsy} and, its counterpart
Intensional Programming Languages Compiler Framework ({\iplcf}).

\subsubsection{General Imperative Compiler Framework}
\label{sect:gicf}
\index{GICF}
\index{Frameworks!GICF}
\index{Internal Design!GICF}
\index{Introduction!GICF}
\index{GICF!Introduction}

{\glu} \cite{glu2, glu1}, {\jlucid} \cite{mokhovjlucid2005}, and later
{\olucid} \cite{mokhovolucid2005} prompted the
development of a General Imperative Compiler Framework ({\gicf}).
The framework targets integration (embedding of) different imperative
languages into GIPSY (see \cite{gipsy}) programs for portability and extensibility reasons.
{\glu} promoted {\C} and {\fortran} functions within; {\jlucid}/{\olucid} promote
embedded {\java}. Since {\gipsy} targets to unite all intensional
paradigms in one research system, we try to be as general as possible and as compatible
as possible and pragmatic at the same time.

For example, if we want to be able to run GLU programs with minimum (if at all)
modifications to the code base, {\gipsy} has to be extended somehow to support
{\C}- or {\fortran}-functions just like it does for Java. What if later on we would
need to add {\cpp}, {\perl}, {\python}, shell scripts, or some other language for example?
The need for a general ``pluggable'' framework arises to add imperative
code segments within a GIPSY program.
We could go even support multi-segment multi-language
(with multiplicity of 3 or more languages) GIPSY programs. Two examples are
presented in \xf{fig:multilang} and in \xf{fig:language-mix}.

\sourcefloat
	{\begin{verbatim}
#funcdecl

Integer f();
void gee();
void z();

#typedecl

Integer;

#JAVA

Integer f()
{
    return new Integer("123");
}

#CPP

#include <iostream>

void gee()
{
    cout << "gee" << endl;
}

#PERL

sub z()
{
    while(<STDIN>)
    {
        s/\n//;
        print;
    }
}

#OBJECTIVELUCID

A @.d 5
where
    dimension d;
    A = B fby.d (A - 1);
    B = C fby.d (B + f().intValue());
    C = z() && gee();
end;
\end{verbatim}
}
	{fig:multilang}
	{Example of a hybrid GIPSY program.}

\sourcefloat
	{\begin{verbatim}
/**
 * Language-mix GIPSY program.
 *
 * $Id: language-mix.ipl,v 1.5 2005/04/25 00:16:30 mokhov Exp $
 * $Revision: 1.5 $
 * $Date: 2005/04/25 00:16:30 $
 *
 * @author Serguei Mokhov
 */
#typedecl

myclass;

#funcdecl

myclass foo(int,double);
float bar(int,int):"ftp://newton.cs.concordia.ca/cool.class":baz;
int f1();

#JAVA
myclass foo(int a, double b)
{
     return new myclass(new Integer((int)(b + a)));
}

class myclass
{
    public myclass(Integer a)
    {
        System.out.println(a);
    }
}

#CPP
#include <iostream>

int f1(void)
{
    cout << "hello";
    return 0;
}

#OBJECTIVELUCID

A + bar(B, C)
where
    A = foo(B, C).intValue();
    B = f1();
    C = 2.0;
end;

/*
 * in theory we could write more than one intensional chunk,
 * then those chunks would evaluate as separate possibly
 * totally independent expressions in parallel that happened
 * to use the same set of imperative functions.
 */

// EOF
\end{verbatim}
}
	{fig:language-mix}
	{Another example of a hybrid GIPSY program.}

%
%

\subsubsection{Generalization of a Concrete Implementation}

Thus, the \api{JavaCompiler} component (see \xf{fig:gipsy-program-compilation}),
part of {\gipc}, has to be generalized,
and the \api{JavaCompiler} itself be a concrete implementation of this
generalization. The generalization would express itself by
having an abstract class \api{ImperativeCompiler}, the generic
\api{Preprocessor} (vs. \api{JLucidPreprocessor} in \xs{sect:jlucid-design}) should be
able to cope with all PLs and know what PLs are supported through enumerating
them.
Another thing the {\gicf} buys us is an ability to have any supported
imperative programming language embedded in any supported intensional
programming language. Though this may seem impractical at
the first glance, but the framework is designed such that a lot of syntax,
semantics, and type mapping work is performed by the individual concrete
compiler implementations and not by the generic machinery. The goal here is that
as long as any given
compiler within the framework conforms to the designed interface specification
and produces the required data structures,
there should be least possible effort to enable such a compiler in {\gipsy}.
Thus, the compilation process, semantic checks, linking, and execution at the
meta level of implementation of the {\gipc} and {\gee} can be reasonably
generalized without loss of practicality as we shall see.
With this great deal of flexibility, we have several issues:

\begin{itemize}
\item
Binary portability of compiled languages, such as {\C}/{\cpp} on a different host
(this problem theoretically does not exist for {\java}).

\item
Though some languages, such as {\perl}, {\python}, shell scripts, are interpreted,
a version mismatch may happen.

\item
A compiler for interpreted languages other than {\java} would be rather simple
because should we want to pass the ST code to a remote host, all we need is to pass the
source itself. Of course, in both compiled and interpreted variant there is
a large potential of security vulnerability exploits (e.g. with malicious code
injection), which will have to be dealt with as a part of the future work.
As of this writing, there are
no embedded checks in {\gipsy} for that; instead a guide of a sandboxed
installation of {\gipsy} will be provided when the system is released.

\item
Another important issue is having imperative PL nodes in the {\AST}.
The issue is in what such nodes should contain in order for them to be
linked back into the main {\AST}, how to perform semantic analysis
of the hybrid code based on the contents of such nodes, and {\gee}
should go about executing this code.

\item
Various languages define their own set of types and typing rules,
gluing them all together is a very difficult task for semantic
analysis and type inference.
\end{itemize}

The follow up sections clarify and address most of these issues.


\subsubsection{Resolving Generalization Issues and Binary Compatibility}
\index{GICF!Binary Compatibility}
\index{GICF!Generalization Issues}

In order to fully support {\gicf}\index{Frameworks!GICF}, the original GIPC framework\index{Frameworks!GIPC}
in \xf{fig:gipc-orig-wu} (discussed in detail by Wu and Paquet in
\cite{wugipc04}) has to be altered
in the following way: the Preprocessor has to be added on top of all the front-end modules, and new
links drawn between the Preprocessor and the other modules \xf{fig:gipc-preprocessor}.
This also changes the data structures flow between the components.
For the unaware reader, what follows
is the brief description of the layers, components, and abbreviations of the conceptual
design present in \xf{fig:gipc-preprocessor}:

The front-end and back-end layers are the two bottom ones represent the main
machinery of the {\gipc}. The front-end compilers and parsers are responsible
for parsing, producing initial syntax trees, STs, and CPs. At this layer,
the main abstract syntax tree {\AST} is always compliant to the one of
Generic Intensional Programming Language ({\gipl}). If the source code
program was written in some specific intensional programming language (SIPL, e.g. {\ilucid}
or {\tlucid}), its {\AST} has to be translated first into {\gipl}. Both,
{\gipl} and {\sipl} type components may translate a Lucid dialect source code
into a data flow ({\dfg}) graph language and back; hence, there is a variety
of the {\dfg} translators. Next, the other two types of conceptual components at the front-end layer
are the data type (DT) and the sequential thread (ST) front-ends. These
correspond to the imperative language compilers and their modules in the implementation.
The DT front-end is responsible for analyzing data-type definitions in the
ST code and producing {\em native} (i.e. compiled) representation of communication
procedures (NPCs). The ST front-end is responsible for compilation an ST code
and producing some equivalent of the native compiled code (NST) as the end result.

The GIPC back-end layer performs finalization of a GIPSY program compilation
by doing semantic analysis and eliminating Lucid functions and producing
the demand {\AST} (DAST) along with linking in the generated STs and CPs from
the imperative side. The GEER generator then produces the final linked version
of a GIPSY program as a resource usable by the {\gee} ({\geer}).

The first two layers are meta-level layers that prepare information for the
front-end and back-end layers. The second layer is the GIPC Preprocessor
layer discussed in depth through the rest of this chapter. The top level has
to do with some language specification processing and creating corresponding
parsers and data structures for the front-end layer. SIPL and GIPL front-end
generators have to do with the fact that our SIPL and GIPL parsers are generated
out of a source grammar specification by \tool{javacc}. Thus, a GIPL specification
corresponds to the GIPL grammar in the \file{GIPL.jjt} file and the GIPL
spec processor is the \tool{javacc} tool. The DT and ST front-end generators
exist for the same idea as the GIPL and SIPL ones do. However, in the current
implementation they are not present either because they are hand-written or
we rely on the external compiler tools (e.g. \tool{javac} to compile Java STs)
to do the processing for us. The design however implies that these components
may eventually be converted to the genuine imperative compilers within {\gipsy}
giving greater control and flexibility over the imperative parts than relying on
external tools. Therefore, we may acquire a \file{Java.jjt} one day, for example,
and generate a Java parser out of it.

\begin{figure*}
	\begin{centering}
	\includegraphics[width=6.5in,totalheight=4in]{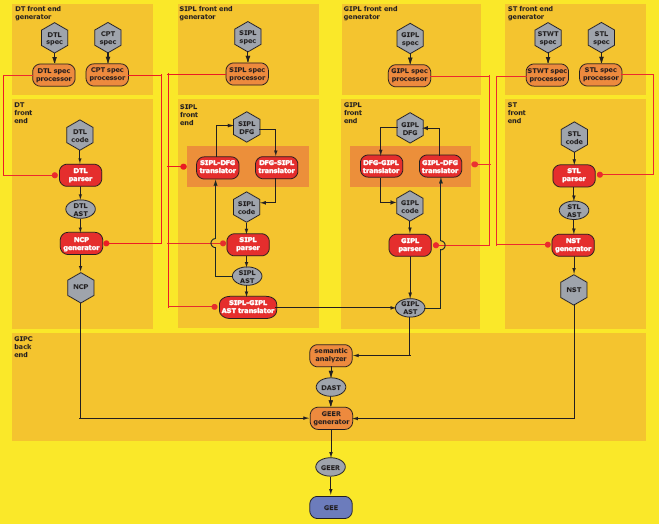}
	\caption{Original Framework for the General Intensional Programming Compiler in the {\gipsy}}
	\index{GIPSY!Original GIPC Framework}
	\label{fig:gipc-orig-wu}
	\end{centering}
\end{figure*}

\begin{figure*}
	\begin{centering}
	\includegraphics[totalheight=6in]{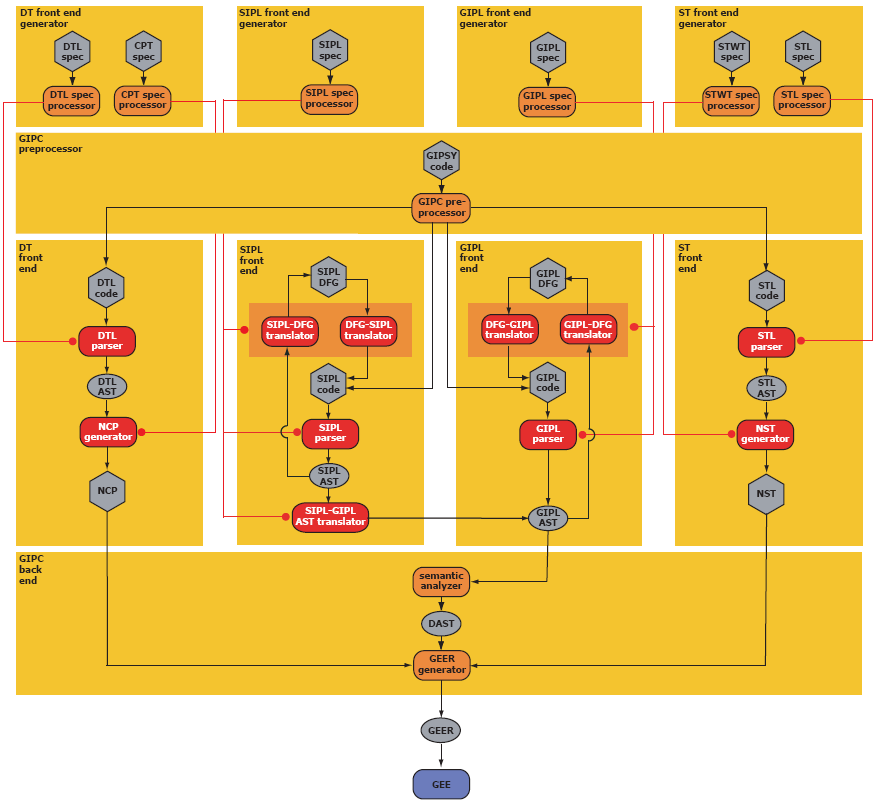}
	\caption{Modified Framework for the General Intensional Programming Compiler in the {\gipsy}}
	\index{GIPSY!GIPC Framework with Preprocessor}
	\label{fig:gipc-preprocessor}
	\end{centering}
\end{figure*}

%
%


\paragraph{Format Tag}
\index{Format Tag}
\index{GICF!Format Tag}

To address some binary compatibility issues we invent a notion
of a format tag attached to the STs and CPs. The format tag's purpose
is to include meta-information about STs and CPs such that
it includes the programming language, the object code format, the operating system, compiler,
and their versions. This is important if we are sending platform-dependent compiled
code, such as that of {\C} or {\cpp} from one host to another with different architectural
platforms.
The \api{FormatTag} API is in \xf{fig:formattag-cl}.

\begin{figure}
	\begin{centering}
	\includegraphics[width=\textwidth]{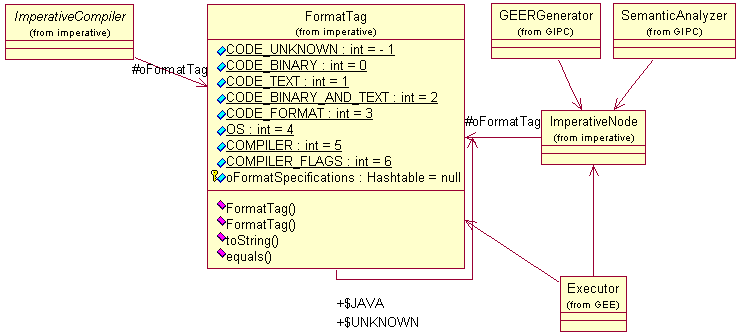}
	\caption{The \api{FormatTag} API.}
	\label{fig:formattag-cl}
	\end{centering}
\end{figure}

We implement format specifications as a hashtable.
We also predefine some common format tags, such as \api{JAVA},
for conveniences as most frequently used. The class overrides
\api{toString()} and \api{equals()} of \api{Object} to define
that the two format tags are only equal if the string representation
of all their specifications are identical.

\paragraph{Sending Source Code Text}
\index{GICF!Sending Source Code Text}

Not all non-intensional languages require compilation, e.g. {\perl}, {\python}, etc.
These can be sent over as plain source code text; thus, the format tag will
indicate the fact. We can go even further with this and send {\em any} language as plain
text and compile it on the target host instead prior invocation.
For the task of the source code inclusion we
reserved the \api{SequentialThreadSourceGenerator}.
Of course, this won't work
for embed-included binary code via a URI parameter because that code was already
compiled by someone else on some specific platform.
As far as current implementation concerned, the generated ST class does always contain
the source code of STs from the GIPSY program code segments, but it is unused
by the {\gee} except for debugging as of this writing.

\paragraph{Dictionary}
\index{GICF!Dictionary}

The Preprocessor's dictionary will initially be constructed based on
the \codesegment{funcdecl} and \codesegment{typedecl} program segments.
The dictionary will serve as an input to three other components: the NST
generator (for error reporting and pointers to the nodes in the {\AST} and
the compiled code), to the NCP generator (to analyze the data structures used
by STs and generate CPs accordingly), and to the semantic analyzer,
to perform data type matching between the intensional and imperative parts.
Both NCP and NST generators work under the command of some imperative
language compiler and are referred to as \api{SequentialThreadGenerator}
and \api{CommunicationProcedureGenerator} in their most general forms,
which are subclassed by a concrete language implementation.

\subsubsection{GIPC Preprocessor}
\label{sect:gipc-preprocessor}
\index{Preprocessor}
\index{GIPC!Preprocessor}
\index{Preprocessor!GIPC}

The \api{Preprocessor} is something that is invoked first by the \api{GIPC}
on incoming GIPSY program's source code stream. The \api{Preprocessor}'s
job is to do preliminary program analysis, processing, and splitting into
chunks. Since a GIPSY program is a hybrid program consisting
of different languages in one source file, there ought to be an interface
between all these chunks. Thus, the \api{Preprocessor} after initial
parsing and producing the initial parse tree, constructs
a preliminary dictionary of symbols used throughout the program. This is
important for type matching and semantic analysis later on.
The \api{Preprocessor} then splits the code segments of the GIPSY
program into chunks preparing them to be fed to the respective
concrete compilers for those chunks. The chunks are represented
through the \api{CodeSegment} class that the \api{GIPC} collects.
The corresponding class diagram of is in \xf{fig:preprocessor-cl}.

\begin{figure}
	\begin{centering}
	\includegraphics[width=\textwidth]{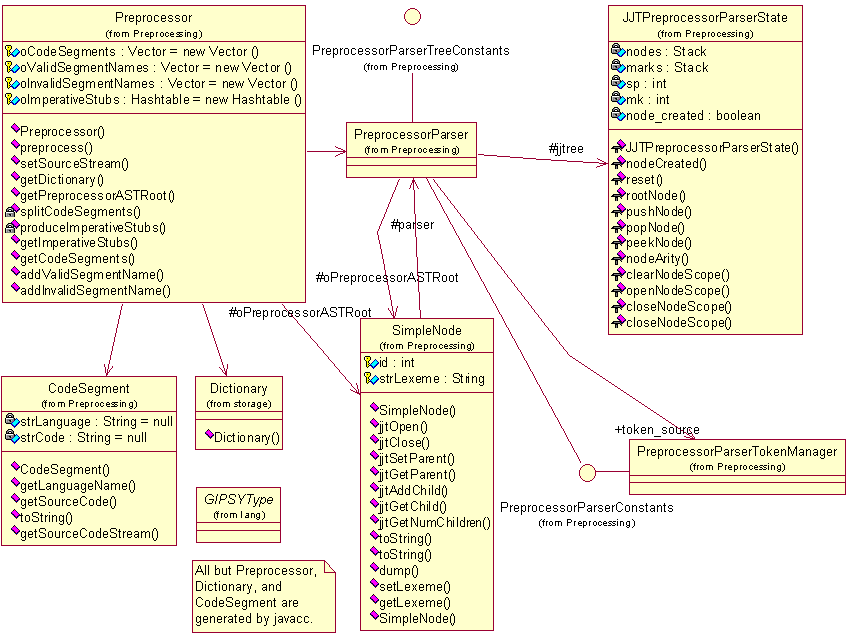}
	\caption{The GIPC \api{Preprocessor}.}
	\label{fig:preprocessor-cl}
	\end{centering}
\end{figure}

The \api{Preprocessor} can also be told to report certain
code segments are invalid at the preprocessing
stage rather delaying the error until the
compiler discovery stage through the
\api{addInvalidSegmentName()} and \api{addValidSegmentName()} methods and maintaining
internal vector of the strings with invalid segment
names. This feature is for example used in \api{Preprocessor}'s
extensions of \api{JLucidPreprocessor} and
\api{ObjectiveLucidPreprocessor} later on that filter out code
segments that do not belong to the languages. The filtering logic
works like this:

\begin{itemize}
\item
	if no valid and invalid segments are specified, all segments
	are accepted as valid at the preprocessing stage. This is the default
	for general \api{GIPC} work.

\item
	if some invalid and no valid segments are specified, the
	\api{Preprocessor} will error out on the invalid segments

\item
	if only valid segments are specified, everything else
	will be treated as invalid

\item
	if both valid and invalid segments are present; the invalid set
	segments are ignored and everything that it is not mentioned
	in the valid set is said to be invalid.
\end{itemize}

\paragraph{GIPSY Program Segments}
\index{GIPSY Program!Segments}

Here we define four basic types of segments
to be used in a GIPSY program. These are:

\begin{itemize}
\item \codesegment{funcdecl}
program segment declares function prototypes
of imperative-language functions defined later
or externally from this program to be used by
the intensional language part. These prototypes are
syntactically universal for all GIPSY programs and
need not resemble the actual function definitions
they describe in their particular programming language.

\item \codesegment{typedecl}
segment lists all user-defined data types
that can potentially be used by the intensional
part; usually objects. These are the types that
do not appear in the matching table in \xt{tab:datatypes}.

\item \codesegment{$<$IMPERATIVELANG$>$}
segment declares that this is a code segment
written in whatever IMPERATIVELANG may be, for example
\codesegment{JAVA} for {\java}, \codesegment{CPP} for {\cpp},
\codesegment{PERL} for {\perl}, \codesegment{PYTHON} for {\python}, etc.

\item \codesegment{$<$INTENSIONALLANG$>$}
segment declares that this is a code segment
written in whatever INTENSIONALLANG may be, for example
\codesegment{GIPL}, \codesegment{INDEXICALLUCID}, \codesegment{JLUCID}, \codesegment{OBJECTIVELUCID},
\codesegment{TENSORLUCID},
\codesegment{ONYX}\footnote{See \cite{grogonoonyx2004} for details on the {\onyx} language.},
etc. as understood by the {\gipsy}.

\end{itemize}

\paragraph{Preprocessor Grammar}
\index{Grammar!Preprocessor}
\index{Preprocessor!Grammar}

The initial grammar for the \api{Preprocessor} to be able to parse
a GIPSY program is shown in \xf{fig:prepgrammar}. After having parsed a program, we have
a Preprocessor AST (PAST) that will be used further by the compilation
process in the {\gipc} and its submodules. The grammar and the framework
were designed in such a way so all the previous neat features
of {\jlucid} \cite{mokhovolucid2005}/{\olucid} \cite{mokhovolucid2005} still be present,
such as \api{embed()} and are accessible to other dialects.
In the {\gicf}, we generalize our function prototype declaration
to be able to include external code of any imperative language.

\begin{figure}
\begin{minipage}[b]{\textwidth}
\scriptsize
\begin{center}
\begin{tabular}{|l c l|} \hline
\nonterminal{GIPSY}        & ::= & \nonterminal{DECLARATIONS} \nonterminal{CODESEGMENTS}\\
&&\\
\nonterminal{DECLARATIONS} & ::= & \nonterminal{FUNCDECLS} \nonterminal{DECLARATIONS} \\
                           & $|$ & \nonterminal{TYPEDECLS} \nonterminal{DECLARATIONS} \\
                           & $|$ & $\epsilon$ \\
&&\\
\nonterminal{FUNCDECLS}    & ::= & \texttt{\#funcdecl} \nonterminal{PROTOTYPES}\\
&&\\
\nonterminal{TYPEDECLS}    & ::= & \texttt{\#typedecl} \nonterminal{TYPES} \\
&&\\
\nonterminal{PROTOTYPES}   & ::= & \nonterminal{PROTOTYPE} \texttt{;} \nonterminal{PROTOTYPES}\\
                           & $|$ & $\epsilon$ \\
&&\\
\nonterminal{PROTOTYPE}    & ::= & \nonterminal{PSTART} \nonterminal{EMBED}\\
&&\\
\nonterminal{PSTART}       & ::= & [ \texttt{immutable} ] \nonterminal{TYPE} [ \texttt{[]} ] \nonterminal{ID} \texttt{(} \nonterminal{TYPELIST} \texttt{)}\\
&&\\
\nonterminal{EMBED}        & ::= & $\epsilon$ \\
                           & $|$ & \texttt{:} \nonterminal{LANGID} \texttt{:} \nonterminal{URI}\\
                           & $|$ & \texttt{:} \nonterminal{LANGID} \texttt{:} \nonterminal{URI} \texttt{:} \nonterminal{ID}\\
&&\\
\nonterminal{TYPES}        & ::= & \nonterminal{TYPE} \texttt{;} \nonterminal{TYPES} \\
                           & $|$ & $\epsilon$ \\
&&\\
\nonterminal{TYPELIST}     & ::= & \nonterminal{TYPE} [ \texttt{[]} ]\\
                           & $|$ & \nonterminal{TYPE} [ \texttt{[]} ] \texttt{,} \nonterminal{TYPELIST}\\
                           & $|$ & $\epsilon$ \\
&&\\
\nonterminal{CODESEGMENT}  & ::= & \nonterminal{LANGDATA} \nonterminal{LANGID}\\
                           & $|$ & \nonterminal{LANGDATA} \nonterminal{EOF} \\
&&\\
\nonterminal{CODESEGMENTS} & ::= & \nonterminal{CODESEGMENT} \nonterminal{CODESEGMENTS} \\
                           & $|$ & $\epsilon$ \\
&&\\
\nonterminal{URI}          & ::= & \nonterminal{CHARACTERLITERAL} \\
                           & $|$ & \nonterminal{STRINGLITERAL} \\
&&\\
\nonterminal{ID}           & ::= & \nonterminal{LETTER} (\nonterminal{LETTER} $|$ \nonterminal{DIGIT})*\\
&&\\
\nonterminal{LANGID}       & ::= & \texttt{\#}\nonterminal{CAPLETTER} (\nonterminal{CAPLETTER})*\\
&&\\
\nonterminal{TYPE}         & ::= & \nonterminal{ID} \\
                           & $|$ & \texttt{int} \\
                           & $|$ & \texttt{double} \\
                           & $|$ & \texttt{bool}\\
                           & $|$ & \texttt{float}\\
                           & $|$ & \texttt{char}\\
                           & $|$ & \texttt{string}\\
                           & $|$ & \texttt{void}\\ \hline
\end{tabular}
\end{center}
\normalsize
\end{minipage}
\caption{{\small Preprocessor Grammar\index{Grammar!Preprocessor} for a GIPSY program.}}
\label{fig:prepgrammar}
\end{figure}

The lexical elements, such as LETTER, LANGDATA, DIGIT, CAPLETTER, and *LITERALs
are not listed for brevity as they are merely standard and self-explanatory
lexical tokens except probably LANGDATA -- this is character data allowing any
character sequence within except LANGID that serves as a terminator of a
code segment chunk.

Notice, the grammar is not bound to our current set of supported intensional
and imperative languages. Rather, the {\gipc} attempts to look up appropriate
compiler for each code segment automagically using LANGID for mapping at run-time.
The {\javacc} version of the grammar can be found the \file{PreprocessorParser.jjt} file.

The grammar has been amended from what was published
in \cite{mokhovgicf2005} to include LANGID in the EMBED production, the
\texttt{immutable} keyword\index{immutable}\index{side effect!immutable}
and arrays subscript operator \texttt{[]} in the PSTART production.
LANGID in EMBED is needed to be able
to pick the appropriate compiler for the included code as it may be written
in any imperative language. The \texttt{immutable} keyword is needed to allow a programmer
to assert that certain STs are immutable meaning given the same parameters
they always return the same result, and, therefore, their result can be
safely cached in the warehouse as such functions are declared side-effects
free (e.g. as the \api{get42()} method in \xf{fig:nat2java}, \xp{fig:nat2java} can be marked
as immutable). This marking
of methods will allow more efficient caching of the ST results of STs {\em known} not to
have side effects and has to be explicitly set by the programmer. If the
programmer by mistake marks a method with side effects as \texttt{immutable},
then a program may exhibit erroneous execution at run-time by returning a
possibly incorrect value from the warehouse. There is no way
to automatically discover immutability of STs in {\gipsy} at this time
(it may only be possible when genuine imperative compilers are implemented).
The array subscript operator \texttt{[]} has been added to PSTART and TYPELIST
productions to allow GIPSY arrays (as a generalization of JLucid arrays) that
are composed of the elements of GIPSY types. The concrete imperative compilers
implementing the mapping (if possible) will have to do appropriate conversions
from the native arrays to GIPSY arrays.

%
%

\subsubsection{GIPSY Type System}
\label{sect:gipsy-types}
\index{GIPSY Type System}
\index{GIPSY!Types}
\index{Types}
\index{Frameworks!GIPSY Type System}

While the main language of {\gipsy}, {\lucid}, is polymorphic
and does not have explicit types, co-existing with other languages
necessitates definition of GIPSY types and their mapping to a particular
language being embedded. \xf{fig:gipsy-types} presents the design
aspects of the GIPSY Type System.

\begin{figure*}
	\begin{centering}
	\includegraphics[width=6.5in]{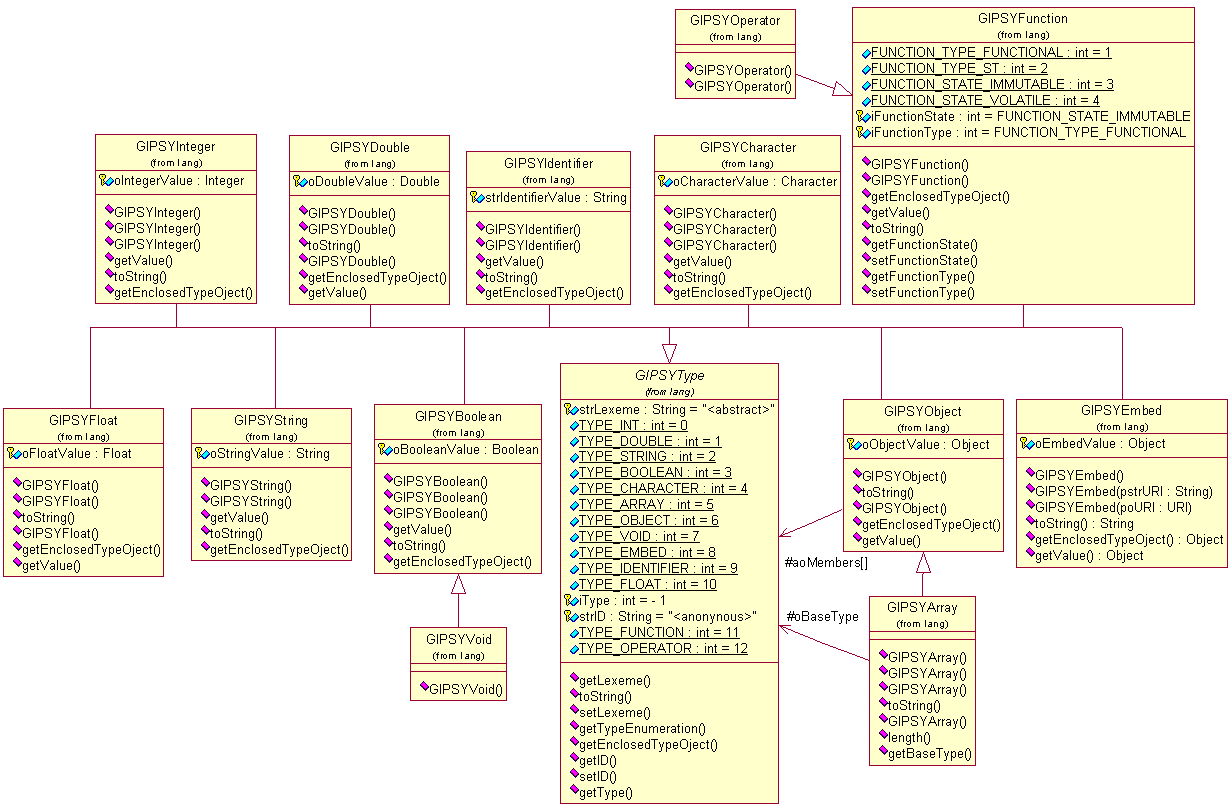}
	\caption{{\small GIPSY Type System.}}
	\label{fig:gipsy-types}
	\end{centering}
\end{figure*}

Each class is prefixed with \texttt{GIPSY} to avoid possible
confusion with similar definitions in the \api{java.lang} package.
The \api{GIPSYVoid} type always evaluates to the Boolean \texttt{true},
as described earlier in \xs{sect:datatypes-matching}.
The other types wrap around the corresponding
Java object wrapper classes for the primitive types, such as \api{Integer},
\api{Float}, etc.
Every class
keeps a lexeme (a lexical representation) of the corresponding type
in a GIPSY program and overrides \api{toString()} to show the lexeme
and the contained value. These types are extensively used by the
\api{Preprocessor}, imperative and intensional (for constants) compilers,
the \api{SequentialThreadGenerator},
\api{CommunicationProcedureGenerator}, \api{SemanticAnalyzer} for
the general type of GIPSY program processing, and by the GEE \api{Executor}.

The other special types that have been created are either experimental
or do not correspond to a wrapper of a primitive type. \api{GIPSYIdentifier}
type case corresponds to a declaration of some sort of an identifier in a GIPSY
program to be put into the dictionary, be it a variable or a function name
with the reference to their definition. This is an experimental type and may
be removed in the future. Constants and conditionals may be anonymous and
thereby not have a corresponding identifier. \api{GIPSYEmbed} is another
special transitional type that encapsulates embedded code via the URL parameter
and later is exploded into multiple types corresponding to STs and their CPs.
\api{GIPSYFunction} and its descendant \api{GIPSYOperator} correspond to the
function types for regular operators and user defined functions. A \api{GIPSYFunction}
can either encapsulate an ordinary Lucid function (as in functional programming
an which is immutable) or an ST function (e.g. a Java method), which may easily
be volatile (i.e. with side effects). These four types are not directly exposed
to a GIPSY programmer and at this point are managed internally.
The rest of the type system is exposed to the GIPSY programmer in the preamble
of a GIPSY program, i.e., the \texttt{\#funcdecl} and \texttt{\#typedecl}
segments, which result in the embryo of the dictionary for linking,  semantic analysis,
and execution. Once ST compilers return, the type data structures (return and parameter types) declared
in the preamble are matched against what was discovered by the compilers and
if the match is successful, the link is made.

%
%

%
%

\subsubsection{{\gicf} Design}
\index{Design!GICF}
\index{GICF!Design}
\index{Frameworks!GICF}

The {\gicf} is the first generalization framework of hybrid programming
in the {\gipsy}. Implementation-wise, only {\java} is
implemented as an imperative language with an external compiler. However,
provision was made for {\C}/{\cpp}, {\perl}, {\fortran} and {\python} with stub compilers.
The class diagram describing {\gicf} is shown in \xf{fig:gicf-cl}. On this diagram
the interaction between a given imperative compiler and the \api{SequentialThreadGenerator}
and \api{CommunicationProcedureGenerator} only shown for \api{JavaCompiler} to keep
the clearer picture, but the same kind of association will have to be
maintained for all imperative compilers as the \api{IImperativeCompiler}
interface mandates. The \api{EImperativeLanguages} is a Java interface
enumerating all available imperative language compilers. It is used
by the \api{GIPC} to discover a given compiler for a language
dynamically. As of this writing, the enumeration is maintained
by hand; however, it is planned to be generated in the near future
with a command-line-driven script or a RIPE GUI automagically to facilitate
addition of new languages.

\begin{figure}
	\begin{centering}
	\includegraphics[width=\textwidth]{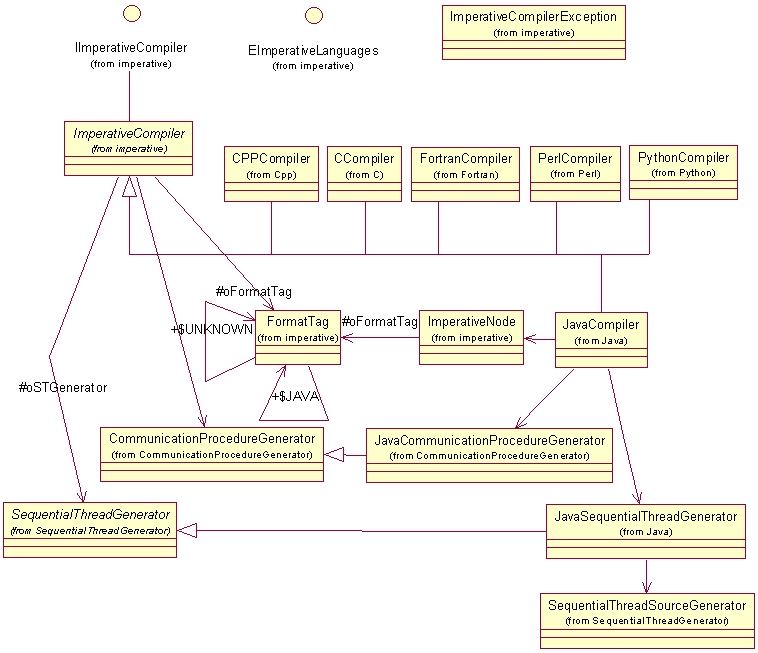}
	\caption{{\gicf} Design.}
	\label{fig:gicf-cl}
	\end{centering}
\end{figure}

%
%

\subsubsection{Intensional Programming Languages Compiler Framework}
\index{Frameworks!IPLCF}
\index{Internal Design!IPLCF}

As a consequence of {\gicf}, a similar approach was applied to the
intensional compilers in the form of {\iplcf}. See the corresponding
class diagram in \xf{fig:iplcf-cl}.
The \api{IIntensionalCompiler} was designed
and implemented by all the intensional compilers we have.
An enumeration \api{EIntensionalLanguages} of all supported intensional languages was
created, so the \api{GIPC} can pick needed compiler at run-time
as determined by the \api{Preprocessor}.

\begin{figure}
	\begin{centering}
	\includegraphics[width=\textwidth]{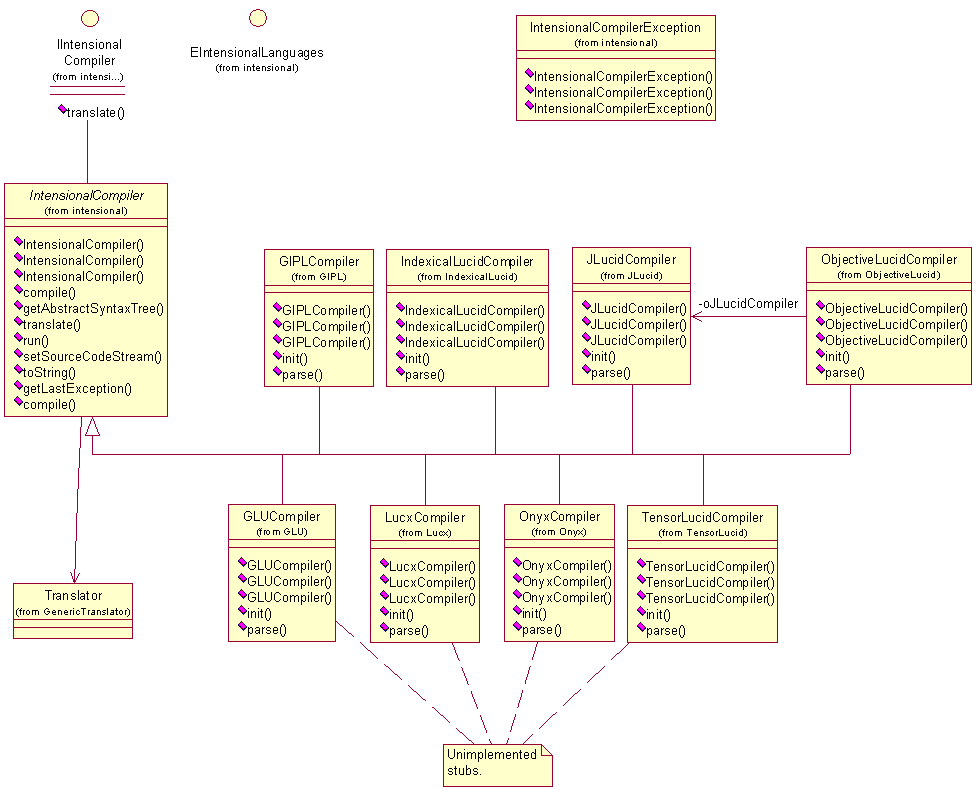}
	\caption{{\iplcf} Design.}
	\label{fig:iplcf-cl}
	\end{centering}
\end{figure}

\begin{figure}
	\begin{centering}
	\includegraphics[width=\textwidth]{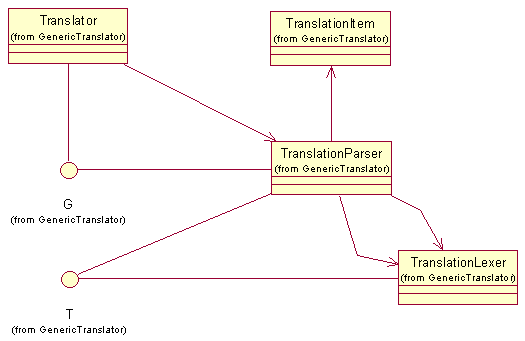}
	\caption{{\sipl} to {\gipl} Translator Integration.}
	\label{fig:translator-cl}
	\end{centering}
\end{figure}

Translation for all intensional compilers is done through the generic \api{Translator}
implemented by Aihua Wu in \cite{aihuawu02}. The \api{Translator} has been integrated into
the \api{GIPC.intensional.GenericTranslator} package and split and renamed as in \xf{fig:translator-cl}.
Thus, every SIPL compiler refers to this translator to acquire a GIPL AST at the end
via generic implementation of \api{IntensionalCompiler.translate()}. The \api{Translator}
was refactored and augmented to understand GIPSY Types (see \xs{sect:gipsy-types}) and \api{ImperativeNode}
for imperative languages. The \api{TranslationParser} and \api{TranslationLexer}
collaborate to compile intensional language translation rules (e.g. \file{IndexicalLucid.rul})
files provided by each SIPL author.

%
%

\subsubsection{Sequential Thread and Communication Procedure Interfaces}
\label{sect:interfaces}

This section details Sequential Thread and Communication Procedure
interfaces. The related class diagram is in \xf{fig:stcp-cl}.
The \api{ICommunicationProcedure} and \api{ISequentualThread} are
the core interfaces. Both extend \api{Serializable} in order for us to
be able to dump their concrete implementations to disk or distributed storage using Java's
object serialization machinery.
This is needed for the \api{GIPSYProgram} container to be saved to disk
or for an ST to be able to reside in JavaSpaces \cite{javaspaces}
implementation of the demand space \cite{vas05}.
The \api{ISequentialThread} also
extends \api{Runnable} to be true thread when materialized, especially
for the case of local execution.
The \api{Runnable} interface makes it possible for an implementing class to become
a thread in multithreaded environment in {\java}.
The \api{ICommunicationProceduresEnum}
is an enumeration of all known to the {\gipsy} communication procedure types.
The \api{NullCommunicationProcedure} and \api{RMICommunicationProcedure}
represent concrete implementations for local threaded processing as well
as {\rmi}. Therefore, the\\ \api{SequentialThreadGenerator} is an abstract factory
for all sequential threads that has to be overridden by a language-specific
sequential thread generator, e.g. such as \api{JavaSequentialThreadGenerator}.
Likewise, \api{CommunicationProcedureGenerator} is a factory for CPs.
The \api{WorkResult} class represents the result of (computation) work
done, which is also has to be \api{Serializable}. Upon various communication
needs the \api{CommunicationStats} is returned by the \api{ICommunicationProcedure}
API or the \api{CommunicationException} is thrown indicating an error. The
\api{Worker} class represents a collection of STs and CPs being executed.

\begin{figure}
	\begin{centering}
	\includegraphics[width=\textwidth]{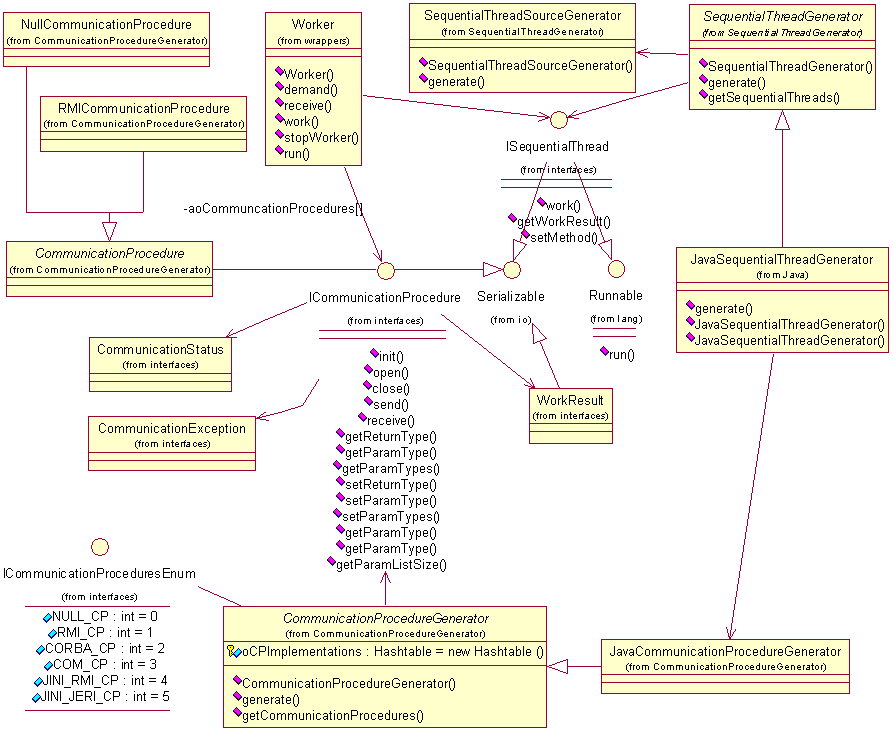}
	\caption{Sequential Thread and Communication Procedure Class Diagram.}
	\label{fig:stcp-cl}
	\end{centering}
\end{figure}

%
%

\subsubsection{{\gipc} Design}
\index{Design!GIPC}
\index{Frameworks!GIPC}


In \xf{fig:gipc-cl} there is a hierarchy that all imperative
and intensional compilers should adhere to.
The \api{IImperativeCompiler} interface
is something every imperative compiler
implements to ease up the job of {\gipc}. A similar
interface has been invented for intensional
languages -- \api{IIntensionalCompiler}
for consistency.


A set of interfaces has been designed for all the present
and future compilers to implement. There are three interfaces
so far:

\begin{enumerate}
\item
	\api{ICompiler} is a superinterface for all compiler
	interfaces. It is implemented by \api{GIPC} itself
	and by \api{DFGAnalyzer}, as shown in \xf{fig:gipc-compilers-cl}.
\item
	\api{IIntensionalCompiler} is a subinterface of \api{ICompiler}
	designated to differentiate intensional compilers. It is implemented
	in part by the \api{IntensionalCompiler} abstract class that
	most (for now all) intensional compilers implement.
\item
	\api{IImperativeCompiler} is a counterpart of \api{IIntensionalCompiler}.
	Its purpose is similar to that of \api{IIntensionalCompiler}
	for imperative languages.
\end{enumerate}

\begin{figure}
	\begin{centering}
	\includegraphics[width=\textwidth]{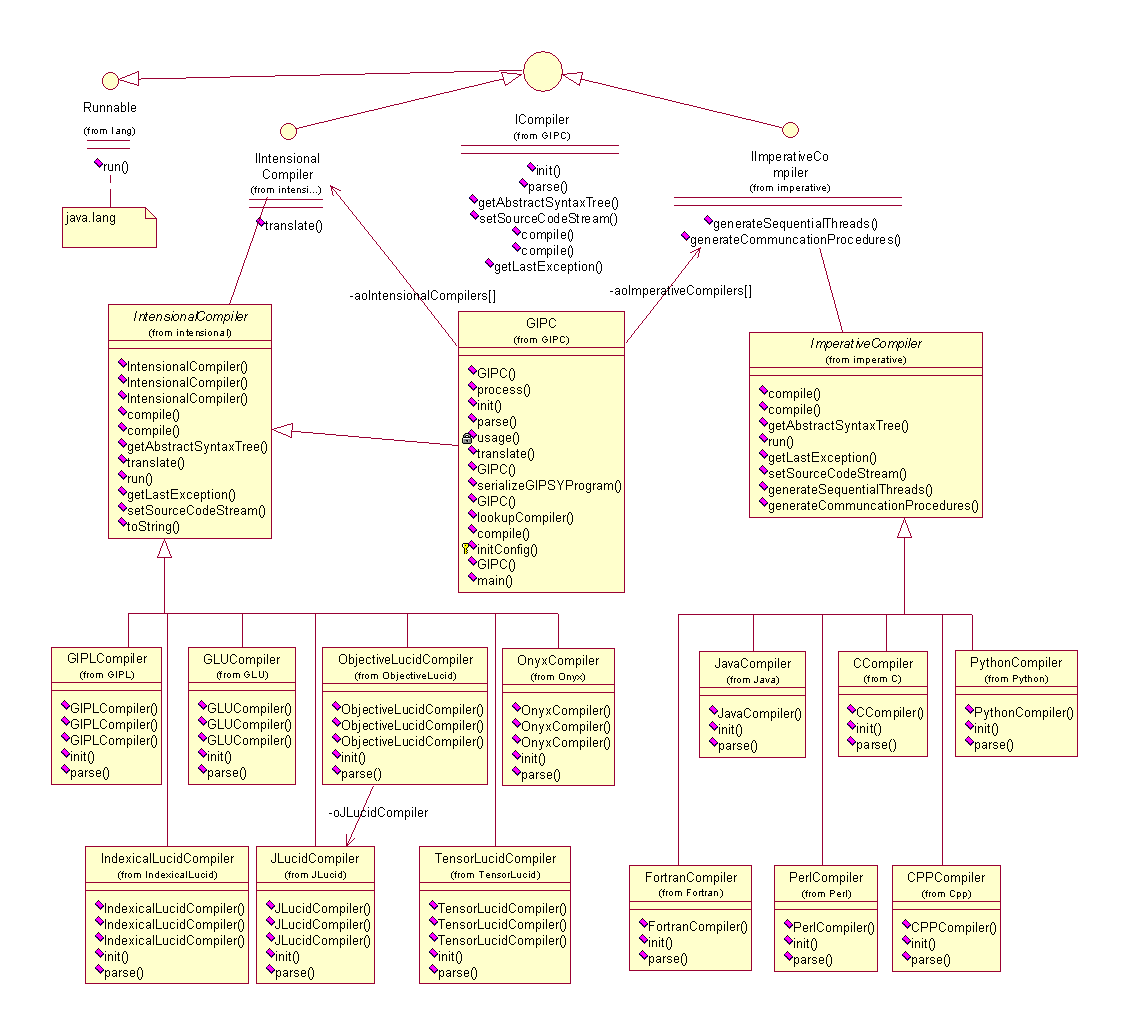}
	\caption{All {\gipc} Compilers.}
	\label{fig:gipc-compilers-cl}
	\end{centering}
\end{figure}

The core difference between \api{IIntensionalCompiler} and \api{IImperativeCompiler}
versus the general \api{ICompiler} is that most (except for {\gipl}) of the intensional compilers have
to perform SIPL-to-GIPL translation; hence, the \api{translate()} method, and
all imperative compilers must produce communication procedures and
sequential threads as the result of their work; hence, \api{generateSequentialThreads()}
and \api{generateCommunicationProcedures()} methods are provided.
The abstract classes \api{IntesionalCompiler} and \api{ImperativeCompiler}
provide the most common possible implementation for all intensional and imperative
compilers respectively, so the underlying concrete compilers only have to override
some parts specific to the language they are to compile. If extension of
these classes is not possible for some reason (e.g. when writing external
GIPSY plugins when a compiler class already inherits from some other class), they
must implement their corresponding interface. Out of the concrete classes on the
diagram the author of this thesis fully implemented \api{GIPC}, \api{GIPLCompiler}, \api{IndexicalLucidCompiler},
\api{JLucidCompiler}, \api{ObjectiveLucidCompiler}, and \api{JavaCompiler}. The \api{DFGAnalyzer} of Yimin Ding
was made to implement \api{ICompiler} as it in fact compiles the ``DFG code''
out of {\gipl} or {\ilucid}.

\begin{figure}
	\begin{centering}
	\includegraphics[width=6.5in]{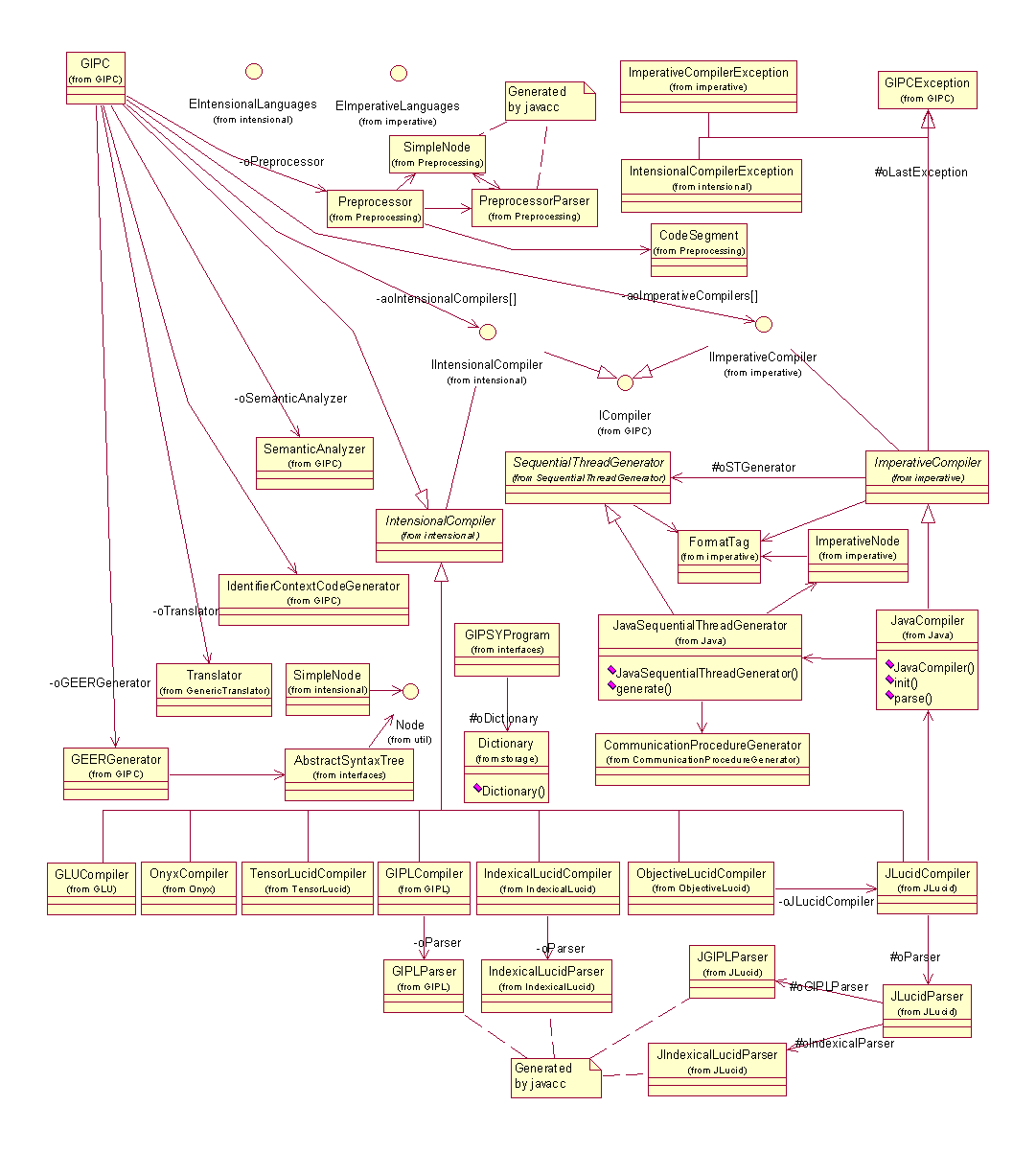}
	\caption{Overall {\gipc} Design.}
	\label{fig:gipc-cl}
	\end{centering}
\end{figure}

The overall design and integration of the GIPC participants is illustrated in \xf{fig:gipc-cl}.
The \api{GIPC} class is the main compiler application that drives
the compilation process, so in the general case in invokes the \api{Preprocessor},
intensional and imperative compilers required, the \api{SemanticAnalyzer},
\api{IdentifierContextCodeGenerator}, \api{Translator}, and the \api{GEERGenerator}
linker. It also acts like a facade to other GIPSY modules.
The major data structures, such as \api{AbstractSyntaxTree}, \api{Dictionary}, \api{CodeSegment},
\api{FormatTag}, \api{ImperativeNode}, and \api{SimpleNode} are created, accessed, or modified
throughout the modules during the compilation process. Out of imperative languages
only \api{JavaCompiler} is mentioned as it is the most advanced in this category.
The \api{JLucidCompiler}'s \api{JLucidParser} underneath invokes both \api{JGIPLParser}
and \api{JIndexicalLucidParser} as {\jlucid} \xs{sect:jlucid} provides extensions
to both of these languages. A number of association links have been removed from
the diagram to maintain clarity as these links are intuitive or present in detail diagrams.

\subsubsection{\api{GIPC} Class as a Meta Processor}
\index{GIPC!as a Meta Processor}

The \api{GIPC} (a concrete class) acts here as so-called ``meta processor''
that drives the entire compilation process and invokes appropriate
submodules in order to come up with a compiled version of a
GIPSY program. This involves calling the \api{Preprocessor}, then
feeding its output to whatever concrete compilers for the code
segments of the GIPSY program, collecting the output of them (various
ASTs, dictionaries), performing semantic analysis, and linking all the
parts back together in a binary form. This portable binary version
of the GIPSY program is to either be serialized as an executable
file for later execution by the {\gee} or optionally to be fed directly
to the {\gee}.

\subsubsection{Calling Sequence}
\index{GIPC!Sequence Diagram}

The sequence diagram in \xf{fig:sequence} illustrates
the entire compilation process and the
data structures passed between the modules.
This is the roundtrip description of the implementation efforts. The two followup diagrams detail the differences in the
compilation process between the imperative and intensional languages. The general compilation process begins by reading the
source GIPSY program and converting it into a meta token stream of types, declarations, and code segments by the
\api{Preprocessor}. The \api{Preprocessor} takes that input and with its own parser produces a preprocessor {\AST} and an embryo of
a dictionary with the identifiers and types declared in the imperative code segments for further semantic linking. The latter
is used to produce imperative stubs for cross-segment type checks. The former contains primarily code segments written in
various languages. The \api{GIPC} takes these code segments and creates appropriate compiler threads, one for each code
segment. Then, each compiler tries to compile its own chunk and produces a portion of a main {\AST}. Since we treat the IPL part
as a main program, its {\AST} is considered to be the main skeleton tree. The {\AST}s
produced by the imperative compilers (which really contain a single \api{ImperativeNode}) are secondary and should be merged into the main when appropriate. Once
all the compiler threads are successfully done, the \api{GIPC} collects all the {\AST}s and performs linking via the
\api{GEERGenerator}. The combined {\AST} is now a subject to the semantic analysis and the function elimination. Once semantic
analysis is complete, the final post-linking is performed where all the pieces of the \api{GIPSYProgram} are combined together and its
instance is serialized to disk. Optionally, right after compilation the \api{GEE} may be invoked to start the execution of the just
compiled program.

\begin{figure*}
	\begin{centering}
	\includegraphics[angle=90,width=6.5in]{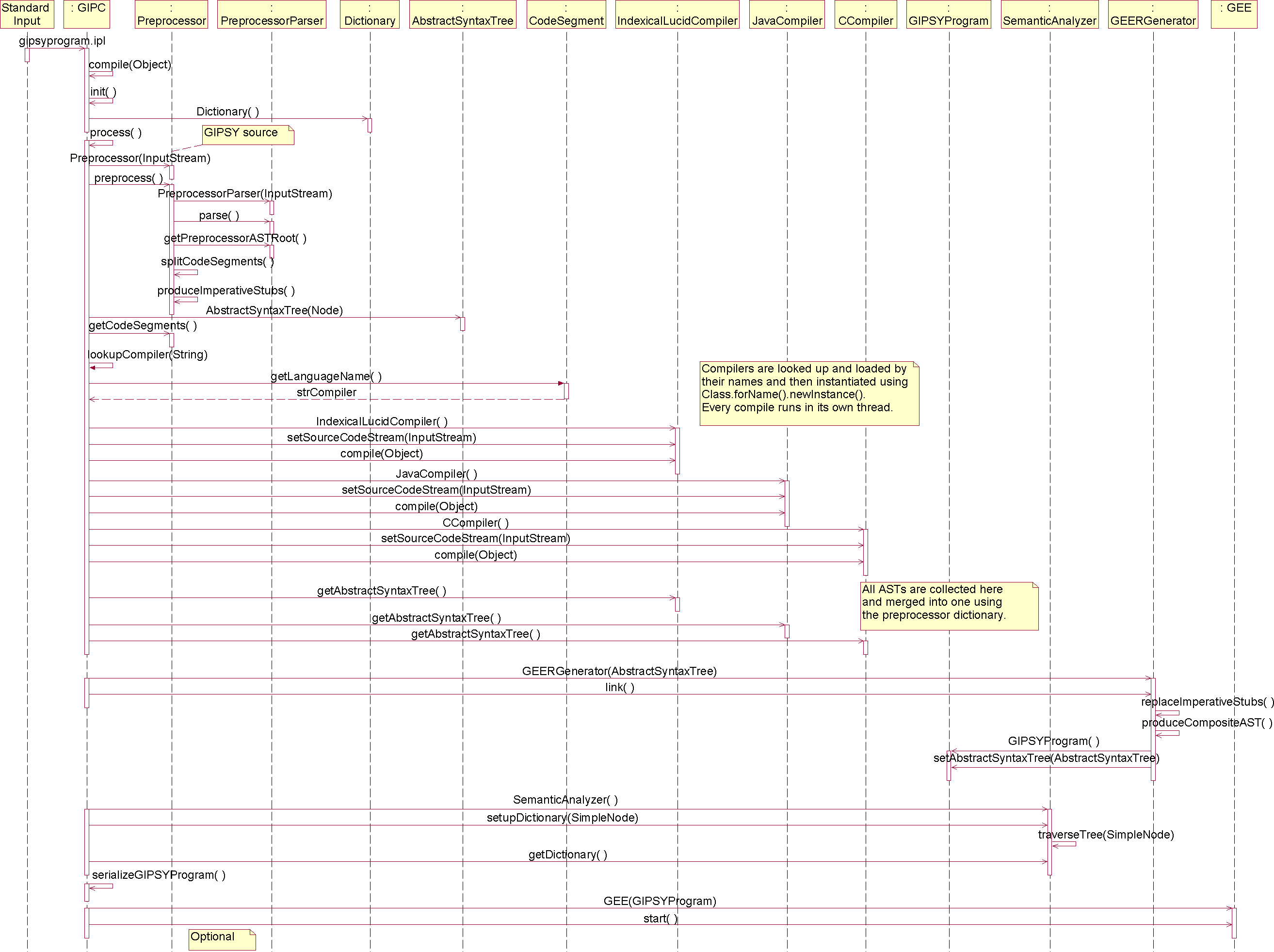}
	\caption{{\small Sequence Diagram of GIPSY Program Compilation Process.}}
	\label{fig:sequence}
	\end{centering}
\end{figure*}

There is no any preference made in \api{GIPC} on the number and the order of intensional and imperative compilers executed. This may
result in several main intensional programs (if the source code contained more than one intensional code segment) or unused
imperative nodes (an imperative segment is declared but the code from it is unused). For the former we maintain an array of
{\AST}s in the \api{GIPSYProgram}, so that when the actual program is executed, the same number of the GEE \api{Executor} threads are started and
all main {\AST}s are evaluated in parallel providing the result set of a computation instead of a single result.
Detailed sequence diagrams of the intensional and imperative compilation processes are
in \xf{fig:intensional-sequence} and \xf{fig:imperative-sequence} to illustrate the
differences in compiling intensional and imperative code segments.

\begin{figure*}
	\begin{centering}
	\includegraphics[width=\textwidth]{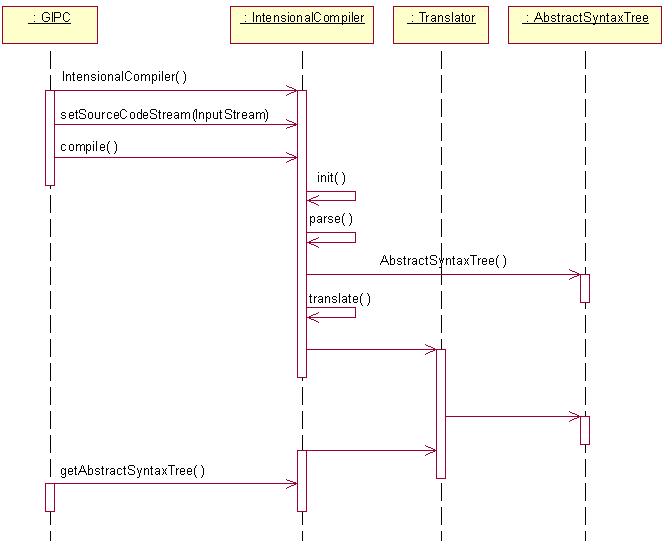}
	\caption{{\small Sequence Diagram of Intensional Compilation Process.}}
	\label{fig:intensional-sequence}
	\end{centering}
\end{figure*}

\begin{figure*}
	\begin{centering}
	\includegraphics[width=\textwidth]{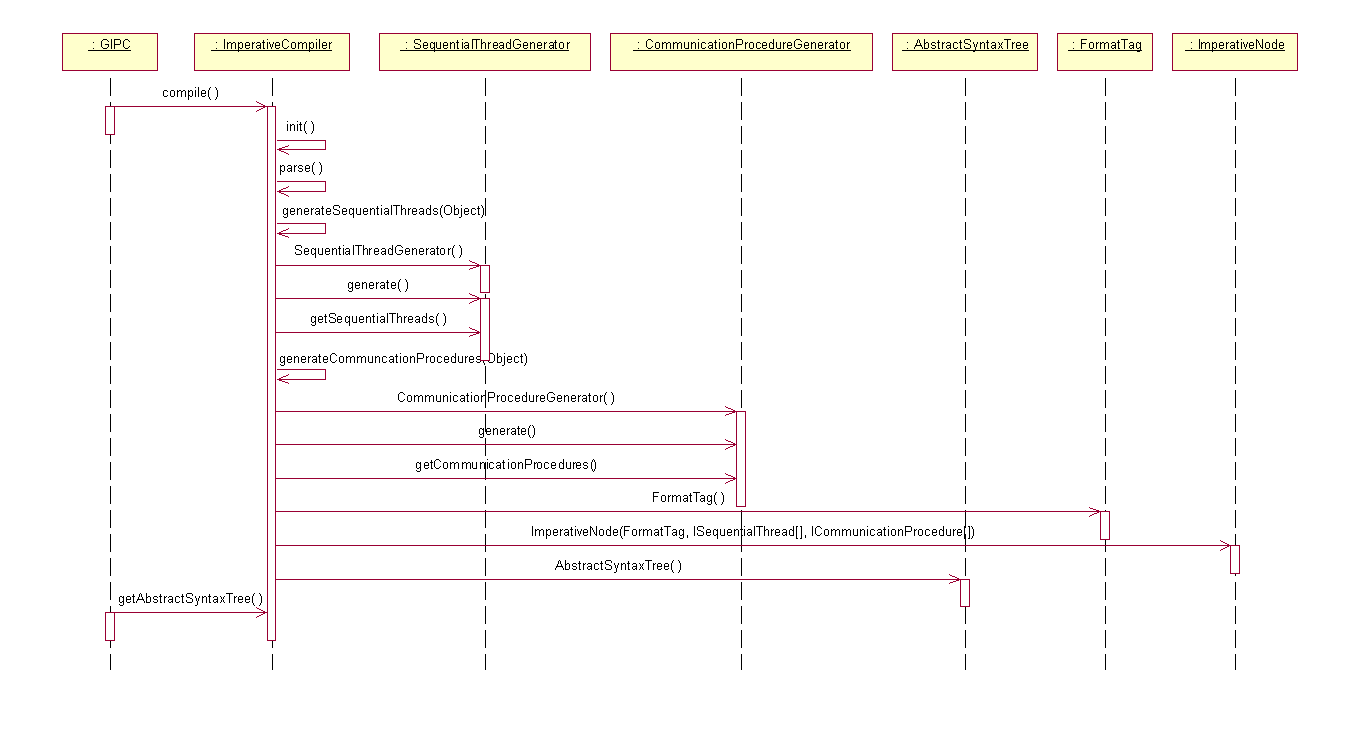}
	\caption{{\small Sequence Diagram of Imperative Compilation Process.}}
	\label{fig:imperative-sequence}
	\end{centering}
\end{figure*}

\clearpage

%
%

\subsubsection{Compiling and Linking}

\paragraph{Multiple Intensional Parts}
\index{GICF!Multiple Intensional Parts}

In a GIPSY program we may possibly have multiple intensional parts.
For example, if a GIPSY programmer gave a GIPL expression, an Indexical Lucid expression
and a couple of Java procedures in the same source GIPSY program,
what is the meaning of that setup would be?
In this case, we can say that we evaluate two independent intensional
expressions in parallel that happened to share the same imperative part.
Thus, for such a GIPSY program there will be two instances of {\gee}
running. The {\gee} is to extended to accept a forest of {\AST}s to be processed in parallel.

\paragraph{Imperative Stubs}
\index{GICF!Imperative Stubs}

When the \api{Preprocessor} completes its job, it has to create some
stubs in the intensional parts of the program for the symbols
declared outside of those parts (e.g. Java functions) so that the
appropriate intensional compiler does not complain about undefined
symbols when producing the {\AST} because the intensional compilers
are not aware of anything outside their work scope. Later on, the corresponding
stub nodes in the {\AST} are found and replaced with the real contents
at the linking stage.

\paragraph{NCP Generator as a Type Processor}
\index{GICF!Type Processor}
\index{GICF!NCP Generator}

The NCP generator will act very much like a type processor and will have
to look inside the imperative code segments analyzed/compiled by the ST
generator. This kind of type processing is needed to decide on
communication procedures (CPs) to be generated for that ST. It issues
warnings if the compiled version of the data structures to be
sent is not portable. The role of the NCP generators in the {\gipsy}
implementation is played by the imperative compilers, such as \api{JavaCompiler}.

\paragraph{GEER Generator as a Linker}
\index{GICF!GEER Generator as a Linker}
\index{GIPC!Linker}

The GEER Generator (see \api{GEERGenerator} in \xf{fig:gipsyprog-cl})
in the backend acts like a linker of all parts of
a GIPSY program. It gathers all the resources from the compiler
set, such as {\AST}s, ICs, CPs, STs, and the dictionary. Then, it replaces the
stubs in the intensional part with the nodes from the imperative {\AST}s
(STs accompanied with their respective CPs) forming a complete composite
{\AST} ready for consumption by the {\gee}. All this will be serialized as
a \texttt{GIPSYProgram} class instance. The \api{GEERGenerator}
is invoked two times -- first prior \api{SemanticAnalyzer} to assemble
a complete {\AST}, and then after semantic analysis and function elimination
to set up the finalized dictionary and program name.

%
%

\subsubsection{Semantic Analyzer}
\index{Design!Semantic Analyzer}
\index{Semantic Analyzer!Design}
\index{Integration!Semantic Analyzer}
\index{Semantic Analyzer!Integration}

\begin{figure}
	\begin{centering}
	\includegraphics[width=\textwidth]{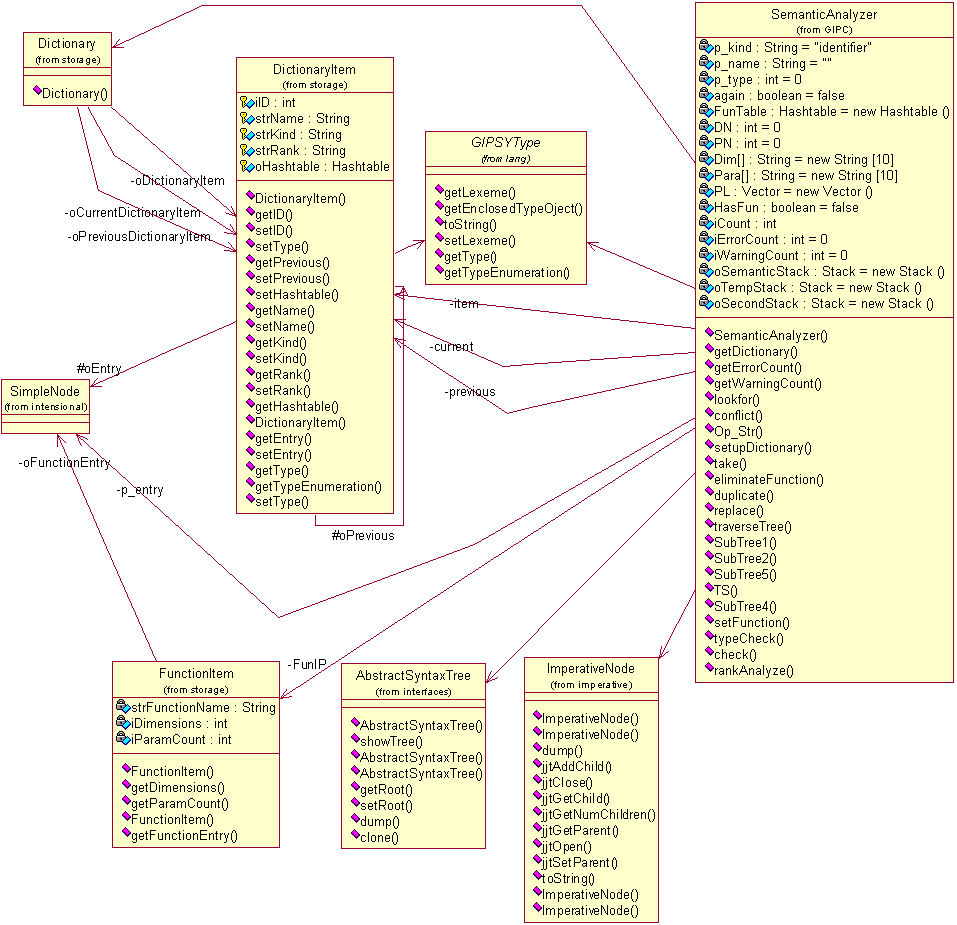}
	\caption{Semantic Analyzer.}
	\label{fig:semantics-cl}
	\end{centering}
\end{figure}

The semantic analyzer detailed design diagram is shown in \xf{fig:semantics-cl}.
Originally implemented by Aihua Wu, the class was renamed from \api{Semantic} \cite{aihuawu02}
to a more complete name of \api{SemanticAnalyzer}
and placed under the \api{GIPC} package. Relevant changes include integration
of \api{storage.Dictionary} (previously was \api{java.util.Vector}),
\api{storage.DictionaryItem} (formerly \api{Item\_in\_Dict} \cite{aihuawu02}), \api{storage.FunctionItem}
(formerly \api{Fun\_Item} \cite{aihuawu02}, serves for function description). The \api{SemanticAnalyzer} had to be taught to
recognize new GIPSY types (see \xs{sect:gipsy-types}) with base \api{GIPSYType} class for object, embed,
and array processing, \api{ImperativeNode} for sequential threads and communication
procedures, and a general \api{AbstractSyntaxTree}.

\subsubsection{Interfacing {\gipc} and {\gee} and Compiled GIPSY Program}
\label{sect:gipsy-program}
\index{GIPSY Program}
\index{GIPSY Program!Compiled}
\index{GIPSY Program!Intefacing GIPC and GEE}
\index{GIPSY Program!GEER}
\index{GEER!GIPSY Program}
\index{DPR!GIPSY Program}

Now, let us formally define the notion of a stored compiled GIPSY program, as a
{\geer} or the interface between the two major modules - {\gipc} and {\gee}.
Until this point, the {\gee} accepted from {\gipc} as the input {\AST} of
an intensional part and a dictionary of symbols. This suggests having
serialized the {\AST} and the dictionary. With the invent of
{\jlucid}, communication procedures (CPs) and sequential threads (STs)
became relevant and should belong to the GIPC-GEE interface. Thus,
a compiled GIPSY program may have several of CPs and STs serialized
along.
While STs and CPs are present within imperative AST nodes, references
to them are recorded here for quicker access and decision making by the {\gee}.
Then, as {\gee} produces demands (especially over {\rmi} or Jini, \cite{vas05})
for each intensional identifier in the dictionary an Identifier Context (IC)
class created \cite{bolu03, bolu04}. This is needed because every such identifier
represents a Lucid expression to be evaluated by the engine, and as such
should also be part of the compiled GIPSY program. The corresponding class diagram
is in \xf{fig:gipsyprog-cl}. It includes the \api{GIPSYProgram} and all its
associations with \api{GIPC}, \api{GEE}, \api{GEERGenerator}, and the storage
classes.

\begin{figure*}
	\begin{centering}
	\includegraphics[width=\textwidth]{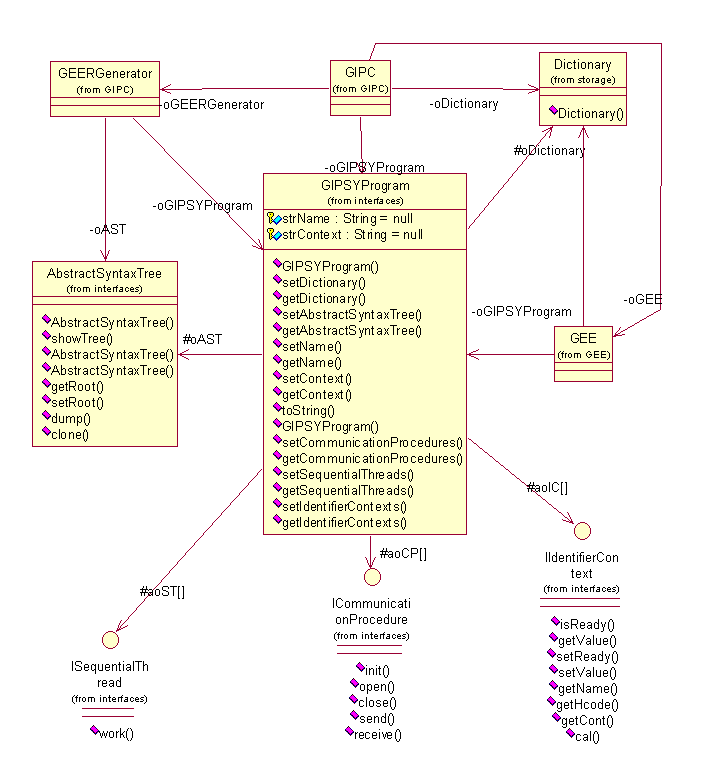}
	\caption{Class diagram describing \api{GIPSYProgram}.}
	\label{fig:gipsyprog-cl}
	\end{centering}
\end{figure*}

%
%


To summarize, the GIPC-GEE interface is the \texttt{GIPSYProgram} representing encapsulation
of the five parts:

\begin{enumerate}
\item Linked AST(s)
\item Dictionary
\item A set of STs
\item A set of CPs
\item A set of ICs.
\end{enumerate}

On the diagram in \xf{fig:gipc-preprocessor}
\api{GIPSYProgram} defines and corresponds to the {\geer}.

%
%

\clearpage
\subsection{{\jlucid}}
\label{sect:jlucid-design}
\index{JLucid!Implementation}
\index{Implementation!JLucid}

\subsubsection{Design}
\index{Design!JLucid}
\index{JLucid!Design}

\begin{figure}
	\begin{centering}
	\includegraphics[width=\textwidth]{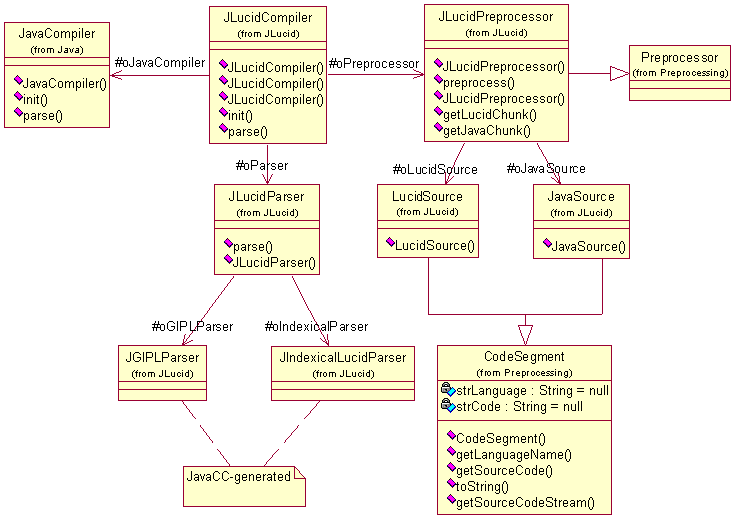}
	\caption{{\jlucid} Design.}
	\label{fig:jlucid-cl}
	\end{centering}
\end{figure}

The class diagram describing {\jlucid} is shown in \xf{fig:jlucid-cl}.
The implementation of {\jlucid} parser-wise is heavily dependent on that of
{\ilucid} as the largest chunk of the IPL work is the same. {\jlucid}
adds a preprocessor \api{JLucidPreprocessor} class that is responsible
for parsing initial source JLucid program and extract Java and Lucid
parts. The \api{JLucidParser} class is the one that manipulates \tool{javacc}-generated
parsers amended to support \api{embed()} and arrays. The sequence diagram
describing the details of the compilation sequence of {\jlucid}
is presented in \xf{fig:jlucid-sequence}.

\begin{figure}
	\begin{centering}
	\includegraphics[width=\textwidth]{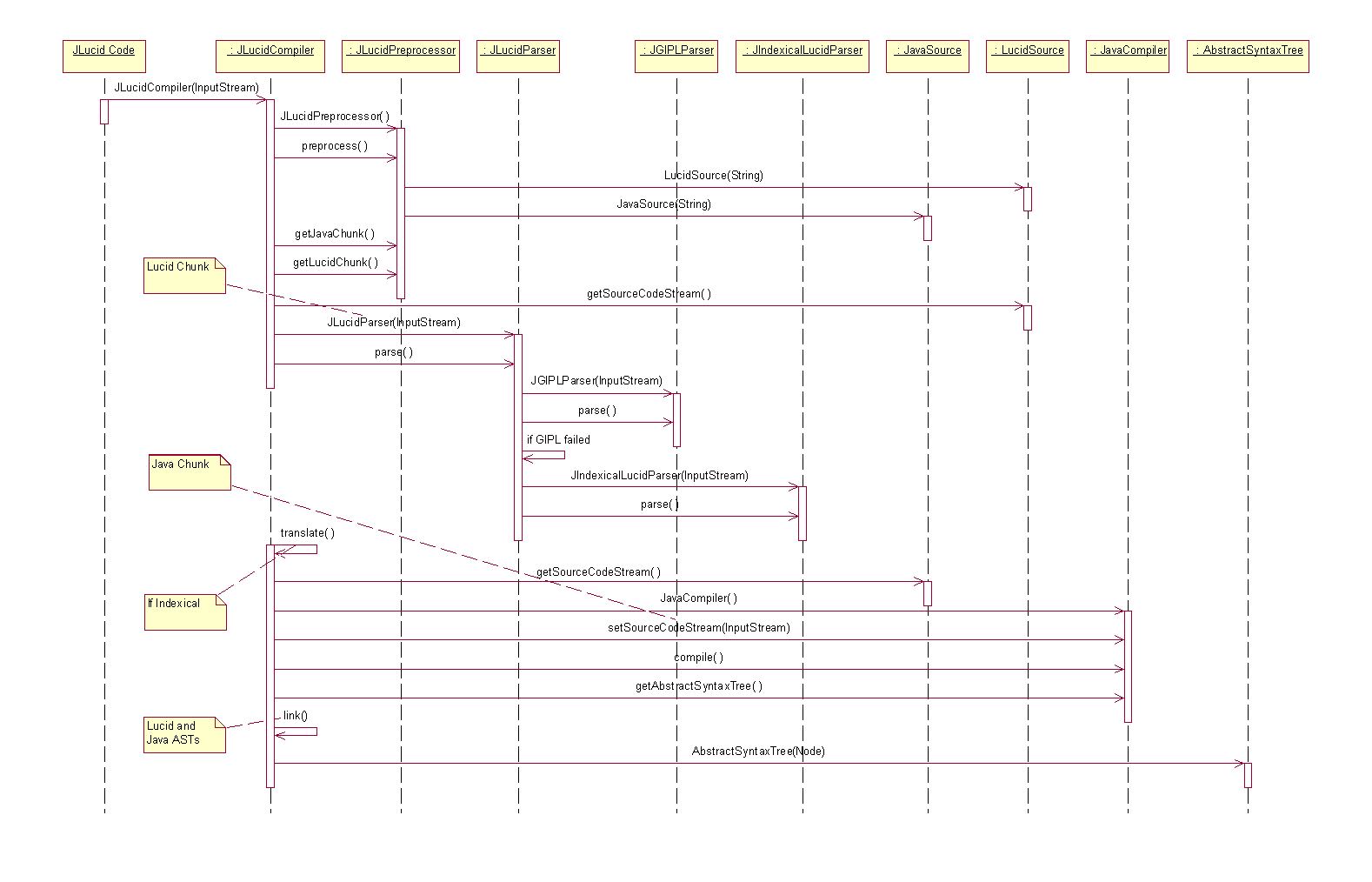}
	\caption{{\jlucid} Compilation Sequence\index{Compilation Sequence!JLucid}.}
	\label{fig:jlucid-sequence}
	\end{centering}
\end{figure}

{\jlucid} implements generation of Java sequential threads (STs) and
their communication procedures (CPs); thus, necessitating
\api{JavaSequentialThreadGenerator} and \api{JavaCommunicationGenerator}.
For uniformity, portability, and testing reasons, we also decided
to send the source code over, that can possibly be compiled on
the remote machine. All this is done by the GICF-integrated
\api{JavaCompiler}, see \xs{sect:java-integration}.

\subsubsection{Grammar Generation}
\label{sect:jlucid-scripts}
\index{JLucid!Grammar Generation}
\index{Grammar!Generation, JLucid}

As it was shown in \xc{chapt:methodology}, the {\jlucid} syntax extension
to {\gipl} and {\ilucid} is minimal. The {\javacc} grammars we use,
are stored in the \file{.jjt} files for the original two dialects.
If we decide to have very similar grammar files for {\jlucid} to
support JLucid extensions (arrays and \api{embed()}), then if the
original grammar has a bug, the fix will have to be propagated
to all the derived grammars, which will not scale from the
maintenance point of view as there will be similar small
modifications from {\olucid} and other dialects. Thus, it
was decided to only maintain the original grammars of {\gipl}
and {\ilucid} and generate the ones for the dialects with
the minimal changes, so that each dialect only maintains
the part that is relevant to its syntactic extension.

For {\jlucid} three \tool{bash} shell scripts were created to process
the original {\javacc} grammars of {\gipl} and {\ilucid} and
generate appropriate extended versions for {\jlucid}. These
include \tool{jlucid.sh} that generates {\javacc} productions
for arrays and \api{embed()}, \tool{JGIPL.sh} that alters
the original \file{GIPL.jjt} grammar to suit the needs
of {\jlucid} mostly in terms of class and package names
and the new productions. Similarly, the \tool{JIndexicalLucid.sh}
script exists for processing of the \file{IndexicalLucid.jjt} file.
The scripts are rather small and presented in the \xa{chapt:grammar-scripts}.

\subsubsection{Free Java Functions and Java Compiler}
\label{sect:java-integration}
\index{JLucid!Free Java Functions}
\index{Free Java Functions!JLucid}
\index{Free Java Functions}
\index{JLucid!Java Compiler}
\index{Java Compiler!JLucid}

As defined in \xc{chapt:methodology}, by ``free Java functions''\index{Free Java Functions}
we mean is that the corresponding Java STs don't have an enclosing Java class
as far as JLucid source code concerned. However, the enclosing class must
exist when compiling a Java program according to {\java}'s syntax and
semantics. Thus, implementation-wise we generate such a class internally that wraps
all our sequential threads, as e.g. in \xs{sect:interfaces}, and we compile that class.
This job of wrapping is delegated to the \api{JavaCompiler}, a member of
the imperative compilers framework\index{GICF}\index{Frameworks!GICF} (see \xs{sect:gicf}).
The \api{JLucidCompiler} as shown in \xf{fig:jlucid-sequence} at some point
invokes the \api{JavaCompiler}, and what the \api{JavaCompiler} does
internally is illustrated in \xf{fig:java-sequence}.

\begin{figure}
	\begin{centering}
	\includegraphics[width=\textwidth]{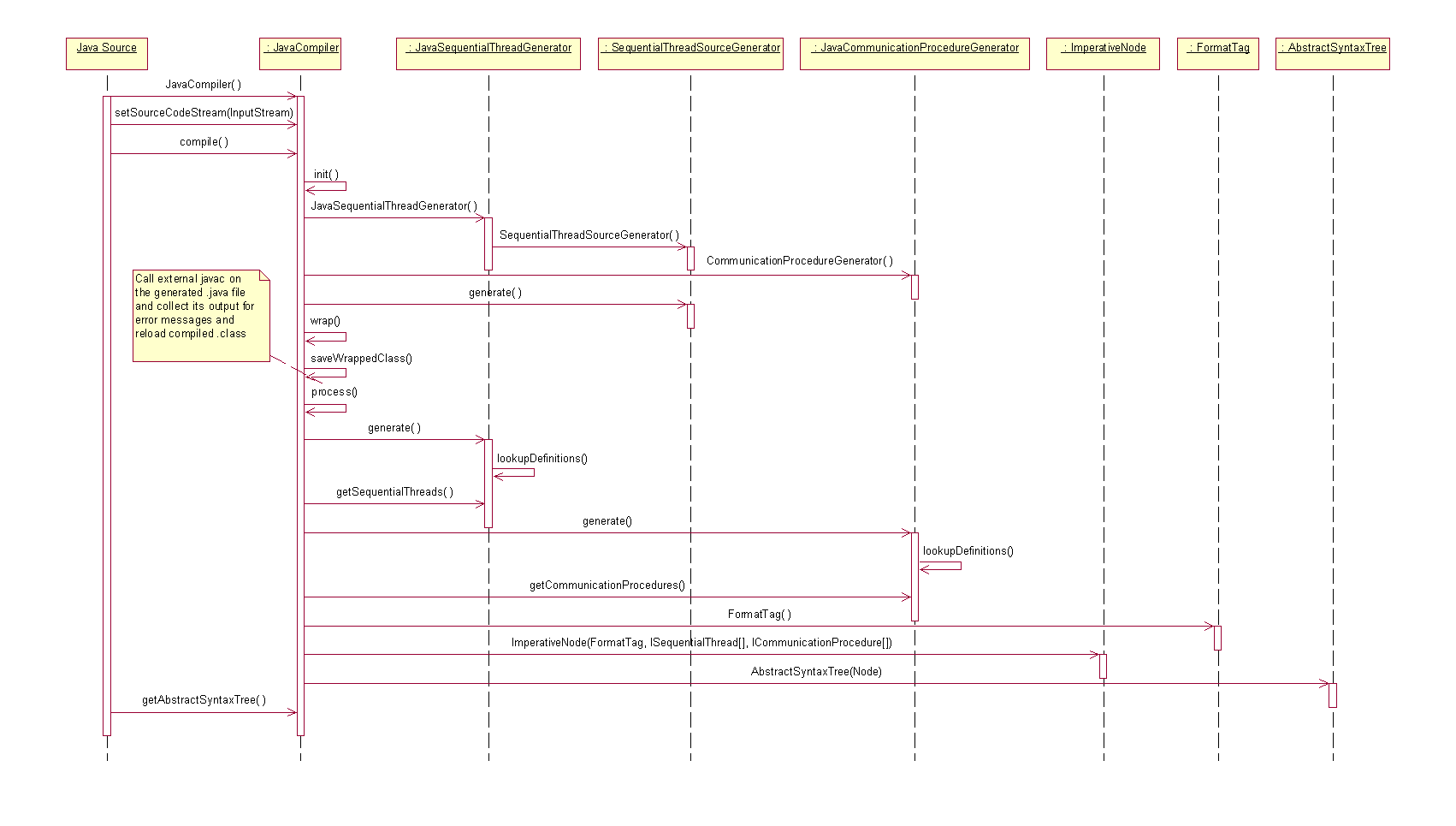}
	\caption{{\java} Compilation Sequence\index{Compilation Sequence!Java}.}
	\label{fig:java-sequence}
	\end{centering}
\end{figure}

Being an imperative compiler, the \api{JavaCompiler} is obliged to produce
the Java STs and CPs among other things. The core of this process is
the \api{wrap()} method where the actual ``wrapping'' our pseudo-free
Java functions into an internal class occurs. The generated source code
\file{.java} file is saved and is fed to the external \tool{javac}
compiler as of this implementation. If there was no compilation errors,
a corresponding \file{.class} or series of \file{.class} files (for the
case of nested classes) is generated. The generated classes are reloaded
back by the \api{JavaCompiler} and their members that are of interest to us
retrieved via the Java Reflection Framework \cite{java-reflection},
thus we obtain an array of references to the ST methods and their parameters
and assign them to our own data structures. After this process completes,
the corresponding \api{FormatTag} describing the {\java} language and
the compiler is created and all information is embedded into
the \api{ImperativeNode}, which represents a single and the only node
in the imperative \api{AbstractSyntaxTree}. Later on, this imperative node
or its pieces will replace a corresponding stub in the main intensional {\AST}.

\subsubsection{Arrays}
\index{JLucid!Arrays}
\index{Arrays!JLucid}

Implementation of arrays in {\jlucid} coincides closely with
the implementation of objects in {\olucid} in \xs{sect:olucid}.
As a part of the GIPSY Type System (see \xs{sect:gipsy-types}),
we employ the \api{GIPSYArray} (see \xf{fig:gipsy-types}) type to hold the array base
type and its members and an overall value. As proposed further,
we treat arrays internally as objects (and objects as arrays),
so \api{GIPSYArray} is an extension of \api{GIPSYObject}
that has a base type asserting the data type of the all
the elements in the arrays (as our arrays a homogenous
collection of elements). Thus, when a syntactic array token
is parsed, a corresponding instance of \api{GIPSYArray}
is created to hold the type and value information for
later processing. The \api{SemanticAnalyzer} and the
\api{Executor} are made to understand the array type
and apply similar type checking or execution rules
to a collection of values instead of a single value.

It might look like this approach will clash with the use of arrays in {\java},
i.e., when a developer wishes to use Java arrays (or if a library
already implements some functionality via Java arrays). This should not be
a problem (though will require a more thorough investigation in the future work),
when we perform type matching by the base element type, as described
in \xs{sect:gipc-preprocessor}. The \api{JavaCompiler} is responsible for the
appropriate conversion of the native-to-GIPSY type conversions, by supplying
a \api{TypeMap} such that it can also be used by the {\gee} at run-time.
Similar comments can be said of the native array types that might exist in
other imperative languages that we would be hoping to support.

\subsubsection{Implementing \api{embed()}}
\index{JLucid!embed()}
\index{embed()!implementation of}

To implement \api{embed()} we define a type \api{GIPSYEmbed}
to fetch the file pointed by the URL and hold it in there.
In {\jlucid}, a \file{.java} or \file{.class} file
(later also a \file{.jar} file) is loaded from either local
or remote location pointed by the URL as follows: if it is
a \file{.java} file, it's fetched and compiled similarly to
the generated class, but the name is static and known;
with the \file{.class} file we skip the compilation process,
but extraction of the sequential threads is the same;
for the \file{.jar} its examined with the \api{JarInputStream}
and \api{JarEntry} Java classes to extract the class information.

\subsubsection{Abstract Syntax Tree and the Dictionary}
\index{JLucid!AST}
\index{JLucid!Dictionary}

When running the JLucid compiler in stand-alone mode,
all the preprocessing and re-assembling the intensional
and imperative pieces into the combined main {\AST}
happens in here, not in the \api{GIPC}, so the JLucid
compiler returns a complete linked {\AST} with all
imperative nodes linked in place and a proper dictionary
of identifiers, both intensional and imperative. JLucid compiler,
however, reused the \api{Preprocessor} and other parts
of the new framework internally instead of re-inventing
the wheel.

The \api{JLucidPreprocessor} uses the general \api{Preprocessor}
class to do the job of chunkanizing the code segments and
preparing initial imperative stubs. This necessitated adding
the \codesegment{funcdecl} segment in the JLucid programs that
previously did not have one in \xc{chapt:methodology}, to simplify preprocessing
and generation of the dictionary. The \api{JLucidPreprocessor}
is set to reject any other code segments than \codesegment{JAVA},
\codesegment{JLUCID}, or \codesegment{funcdecl}.

If the \api{JLucidCompiler} invoked from the \api{GIPC} as
a part of general compilation process (see \xf{fig:sequence}),
the \codesegment{JAVA} segment will no longer be really processed
internally, and instead, \api{GIPC} will call \api{JavaCompiler}
externally to the \api{JLucidCompiler}, so essentially the \api{JLucidCompiler}
will be responsible only for the Lucid part (with arrays and \api{embed()}).

%
%

\clearpage
\subsection{{\olucid}}
\label{sect:olucid}
\index{Objective Lucid!Implementation}
\index{Implementation!Objective Lucid}

This section addresses problems that arise when
implementing {\olucid}. These include internal
implementation to support the dot-notation, extension
to semantic analysis to be able to manipulate object
data types (very likely user-defined), and making
it all work in the {\gicf} and General Eduction Engine ({\gee}) of the {\gipsy}
by correctly forming the abstract syntax tree ({\AST}) that
includes object data types.

%
%

\subsubsection{Design}
\index{Design!Objective Lucid}
\index{Objective Lucid!Design}

\begin{figure}
	\begin{centering}
	\includegraphics[width=\textwidth]{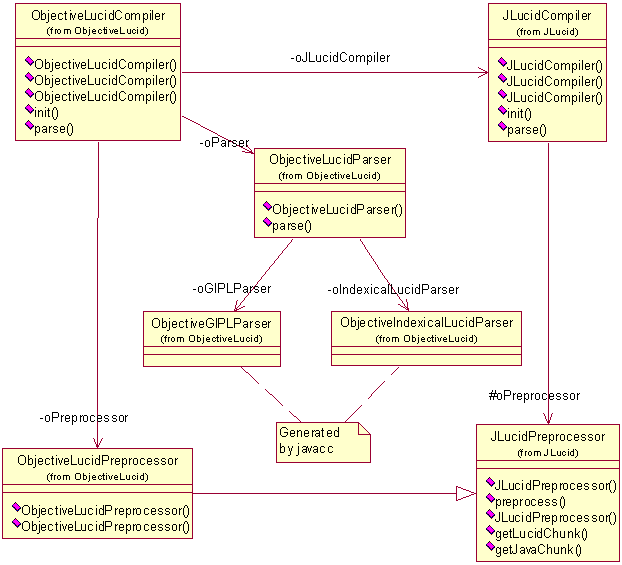}
    \caption{{\olucid} Design.}
	\label{fig:olucid-cl}
	\end{centering}
\end{figure}

\begin{figure}
	\begin{centering}
	\includegraphics[width=\textwidth]{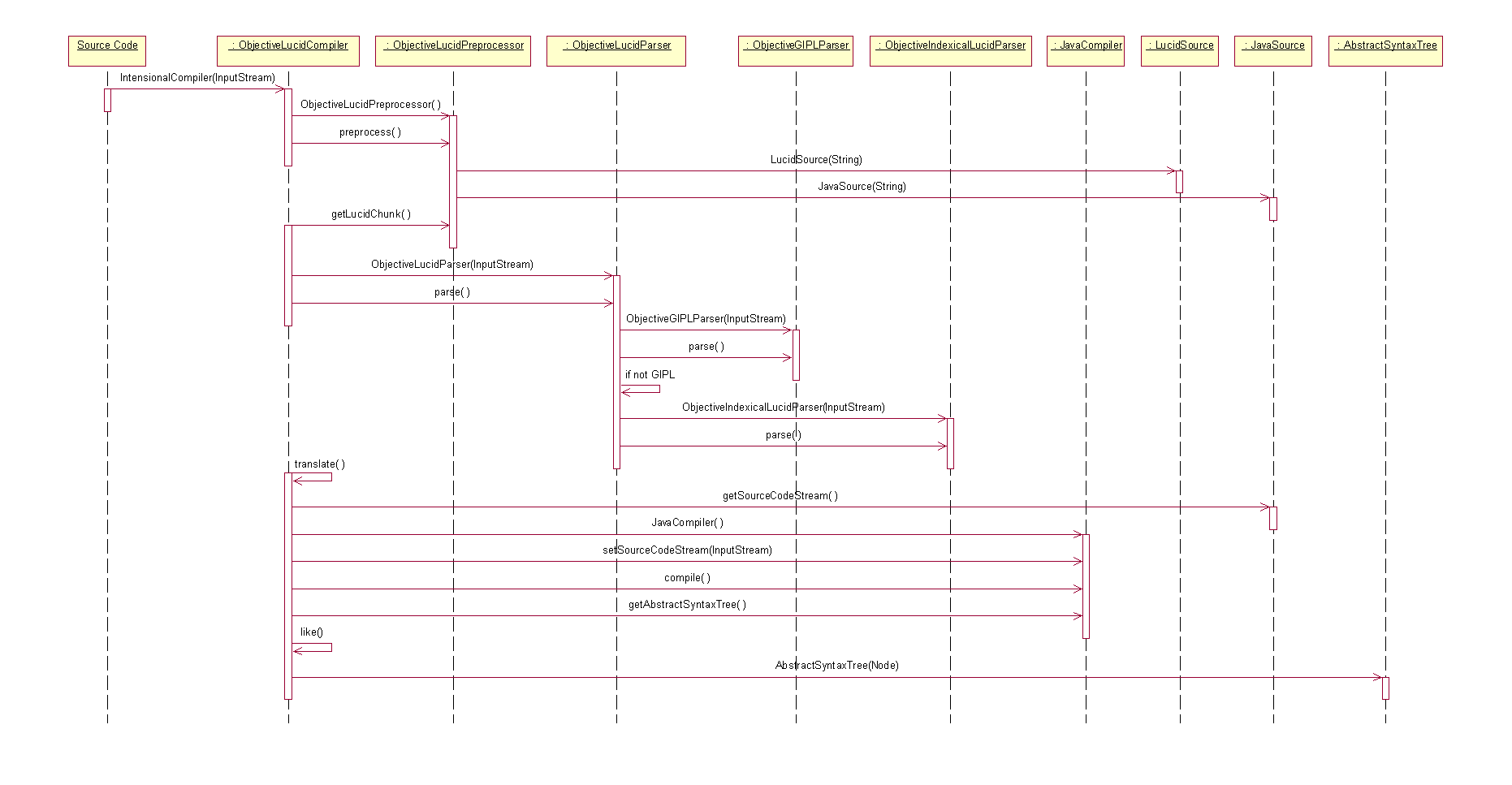}
    \caption{{\olucid} Compilation Sequence\index{Compilation Sequence!Objective Lucid}.}
	\label{fig:olucid-sequence}
	\end{centering}
\end{figure}

The class diagram describing {\olucid} is in \xf{fig:olucid-cl}.
Since the JLucid compiler already does most of the legwork, {\olucid}
simply extends it to add the dot-notation and some extra post-processing
when unrolling the objects. The corresponding compilation sequence
is shown in \xf{fig:olucid-sequence}.

\subsubsection{Grammar Generation}
\index{Objective Lucid!Grammar Generation}
\index{Grammar!Generation, Objective Lucid}

Like with {\jlucid}, the grammar files are generated for {\olucid}
using \tool{bash} shell scripts, \tool{ObjectiveGIPL.sh} and \tool{ObjectiveIndexicalLucid.sh}.
These scripts work with the grammars produced by the JLucid scripts (see \xs{sect:jlucid-scripts}) by simply
extending them with the dot-notation production and fixing up names
of classes and packages. These scripts are presented in the \xa{chapt:grammar-scripts}.

\subsubsection{Object Instantiation}
\index{Objective Lucid!Object Instantiation}

Normally, when a Lucid program refers to a Java object, it
has to instantiate it first by either calling a pseudo-free Java
function that returns an object instance or to call the constructor
directly. This instantiation has to be explicit at the beginning of the program
to avoid Java's \texttt{NullPointerException} at run-time. Internally,
the object instance is created using Java Reflection \cite{java-reflection}
by first loading and then initializing the
needed class with \texttt{Class.forName("ClassXB").newInstance()}. Referencing
static members do not require a class instance, and can be
accessed using the class name, in this case we just keep the
\texttt{Class.forName("ClassXB")}. We also keep the needed
references to the object itself and its members in the
\api{GIPSYObject} type of the GIPSY Type System.

\subsubsection{The Dot-Notation}
\index{Objective Lucid!The Dot-Notation}

Implementing the dot-notation extension of JLucid is
the easiest task of the three. In fact, the \texttt{E.id}
productions are just a syntactic sugar that can be wrapped
around already existing mechanisms of JLucid to include
Java functions as mentioned in \xs{sect:jlucid-pseudo-oop}.
The compiler simply generates a set of pseudo-free Java
functions for every object member referenced from the intensional
program. These will be easy to place into the AST just the way
JLucid does it. In other words, this is achieved by automatic
generation of implicit accessor Java
functions that had to be explicit in {\jlucid}.

\subsubsection{Abstract Syntax Tree and the Dictionary}
\index{Objective Lucid!AST}
\index{Objective Lucid!Dictionary}

The {\gipc} (General Intensional Programming Compiler) generates abstract
syntax trees ({\AST}) of all compiled GIPSY program parts, and constructs the GEER
(General Eduction Engine Resources), which is a data dictionary storing
all program identifiers, encapsulated with all ASTs generated at compile time.
Simply put, the {\geer} encapsulates all the meaning of a GIPSY program, and all
necessary resources to enable the {\gee} to execute the programs correctly.
The {\AST} and the dictionary contain the generated accessor
identifiers that are processed by the JLucid mechanisms, as
described previously. This is possible because Java's built-in
class \api{Class} can provide us with all the meta-information
about its members through enumeration that we can place in the
{\AST} and the dictionary. Little changes from the way {\jlucid}
processes that except that the object members are put into the
dictionary and acted upon as an array of homogeneous types as
described in the follow up section.

The \api{ObjectiveLucidPreprocessor} also makes use of
the general \api{Preprocessor}, but unlike \api{JLucidPreprocessor},
it also accepts the \codesegment{typedecl} segment as with objects
come user-defined types, so these have to be listed if used
by the {\lucid} part.

%
%

\subsubsection{Objects as~Arrays and Arrays as~Objects}
\label{sect:objective-arrays}
\index{Lucid!Arrays as Objects}
\index{Lucid!Objects as Arrays}

Implementation-wise, we propose to treat arrays
of {\jlucid} as a special case of objects and, the other way around,
the objects be a generalization of arrays. An array can be broken
into its elements where every element is evaluated as an expression
under the same context. Thus, evaluating:

\noindent
\fbox
{
	\scriptsize
	\begin{minipage}[t]{0.97\textwidth}
		\texttt{A[4] @ [d:4]}

		\texttt{where}

		\quad\texttt{dimension d;}

		\quad\texttt{A[\#.d] = 42 * \#.d fby.d (\#.d - 1);}

		\texttt{end;}
	\end{minipage}
	\normalsize
}\\

\noindent
is equivalent to evaluating four Indexical Lucid expressions
(possibly in parallel). Under this point of view objects
can be viewed as arrays where every atomic member is evaluated
as if it were an array element. Basically, we denormalize an object
into primitives and evaluate them. If an object encapsulates
other objects, then these are in turn denormalized and put into
the definition environment (dictionary).
In other words, if you have an array of four elements \texttt{a[4]}, the elements
are evaluated as four independent expressions. Likewise, an object
that has four data members, each of them is evaluated as an expression
under the same context.

Essentially, an array is a collection of atomic elements of the same type.
When evaluating say an array of four elements \texttt{a[4]} at some context \texttt{[d:4]},
we are, in fact, evaluating four ordinary Lucid expressions (possibly in parallel)
in the same context.
Likewise, an object is a collection of atomic elements of (possibly)
different types. In case an object encapsulates another object, that other
object can in turn be split into atoms, and so on. All atoms of an object
evaluate as independent Lucid expressions, just like array elements.

\clearpage
\noindent
Thus, from {\olucid}'s point of view, the following are equivalent:

\noindent
\fbox
{
	\scriptsize
	\begin{minipage}[t]{0.97\textwidth}
	\texttt{(a) int a[4];}
	\end{minipage}
	\normalsize
}\\

\noindent
\fbox
{
	\scriptsize
	\begin{minipage}[t]{0.97\textwidth}
	\texttt{(b) class foo}

	\qquad\texttt{\{}

	\qquad\qquad\texttt{int a1;}

	\qquad\qquad\texttt{int a2;}

	\qquad\qquad\texttt{int a3;}

	\qquad\qquad\texttt{int a4;}

	\qquad\texttt{\}}
	\end{minipage}
	\normalsize
}\\

\noindent
So, internally, we represent (a) in the definition environment as:

\noindent
\fbox
{
	\scriptsize
	\begin{minipage}[t]{0.97\textwidth}
	\texttt{a\_4     // scope identifier}

	\texttt{a\_4.a1}

	\texttt{a\_4.a2}

	\texttt{a\_4.a3}

	\texttt{a\_4.a4}
	\end{minipage}
	\normalsize
}\\

\noindent
Under the scope of array \texttt{a\_4} (a generated \myid) there are four members, and
\texttt{a\_4.a*} comprise a denormalized identifier, also generated.
And (b) will become:

\noindent
\fbox
{
	\scriptsize
	\begin{minipage}[t]{0.97\textwidth}
	\texttt{foo    // scope identifier}

	\texttt{foo.a1}

	\texttt{foo.a2}

	\texttt{foo.a3}

	\texttt{foo.a4}
	\end{minipage}
	\normalsize
}\\

\noindent
where \texttt{foo.a*} are generated variable identifiers in the definition environment.
Encapsulation will be handled in the following way:

\noindent
\fbox
{
	\scriptsize
	\begin{minipage}[t]{1.8in}
	\texttt{class bar}

	\texttt{\{}

	\qquad\texttt{int b1;}

	\qquad\texttt{int b2;}\\

	\qquad\texttt{foo oFoo = new foo();}

	\texttt{\}}
	\end{minipage}
	\normalsize
}
\hfill
\fbox
{
	\scriptsize
	\begin{minipage}[t]{1.8in}
	\texttt{bar}

	\texttt{bar.b1}

	\texttt{bar.b2}

	\texttt{bar.foo}

	\texttt{bar.foo.a1}

	\texttt{bar.foo.a2}

	\texttt{bar.foo.a3}

	\texttt{bar.foo.a4}
	\end{minipage}
	\normalsize
}\\

\noindent
To paraphrase and explain in another example, if we have three separate
Lucid expressions:

\noindent
\fbox
{
	\scriptsize
	\begin{minipage}[t]{1.5in}
	\texttt{// float}

	\texttt{a @ [d:2]}

	\texttt{where}

	\quad\texttt{dimension d;}

	\quad\texttt{a = 2.5}

	\qquad\texttt{fby.d (a + 1);}

	\texttt{end;}
	\end{minipage}
	\normalsize
}
\hfill
\fbox
{
	\scriptsize
	\begin{minipage}[t]{1.5in}
	\texttt{// integer}

	\texttt{b @ [d:2]}

	\texttt{where}

	\quad\texttt{dimension d;}

	\quad\texttt{b = 1}

	\qquad\texttt{fby.d (b + 1);}

	\texttt{end;}
	\end{minipage}
	\normalsize
}
\hfill
\fbox
{
	\scriptsize
	\begin{minipage}[t]{1.5in}
	\texttt{// ASCII Char}

	\texttt{c @ [d:2]}

	\texttt{where}

	\quad\texttt{dimension d;}

	\quad\texttt{c = 'a'}

	\qquad\texttt{fby.d (c + 1);}

	\texttt{end;}
	\end{minipage}
	\normalsize
}\\

\noindent
Now if we group \texttt{a}, \texttt{b}, and \texttt{c} as a class:

\noindent
\fbox
{
	\scriptsize
	\begin{minipage}[t]{0.97\textwidth}
	\texttt{class foo}

	\texttt{\{}

	\qquad\texttt{float a = 2.5;}

	\qquad\texttt{int   b = 1;}

	\qquad\texttt{char  c = 'a';}

	\qquad\texttt{public foo() \{\}}

	\texttt{\}}
	\end{minipage}
	\normalsize
}\\

\noindent
So when we write:

\noindent
\fbox
{
	\scriptsize
	\begin{minipage}[t]{0.97\textwidth}
	\texttt{f @ [d:2]}

	\texttt{where}

	\qquad\texttt{dimension d;}

	\qquad\texttt{f = foo() fby.d (f + 1);}

	\texttt{end;}
	\end{minipage}
	\normalsize
}\\

\noindent
we mean there start three subexpression evaluations:

\noindent
\fbox
{
	\scriptsize
	\begin{minipage}[t]{1.5in}
	\texttt{f.a @ [d:2]}

	\texttt{where}

	\quad\texttt{dimension d;}

	\quad\texttt{f.a = foo().a}

	\qquad\texttt{fby.d (f.a + 1);}

	\texttt{end;}
	\end{minipage}
	\normalsize
}
\hfill
\fbox
{
	\scriptsize
	\begin{minipage}[t]{1.5in}
	\texttt{f.b @ [d:2]}

	\texttt{where}

	\quad\texttt{dimension d;}

	\quad\texttt{f.b = foo().b}

	\qquad\texttt{fby.d (f.b + 1);}

	\texttt{end;}
	\end{minipage}
	\normalsize
}
\hfill
\fbox
{
	\scriptsize
	\begin{minipage}[t]{1.5in}
	\texttt{f.c @ [d:2]}

	\texttt{where}

	\quad\texttt{dimension d;}

	\quad\texttt{f.c = foo().c}

	\qquad\texttt{fby.d (f.c + 1);}

	\texttt{end;}
	\end{minipage}
	\normalsize
}\\

\noindent
We say these are equivalent where the \texttt{f} in all expressions refers to the same
object's instance (i.e. there are not three objects constructed, only one).
Similarly (nearly identically) we implement arrays:

\noindent
\fbox
{
	\scriptsize
	\begin{minipage}[t]{0.97\textwidth}
	\texttt{a[3] @ [d:2]}

	\texttt{where}

	\qquad\texttt{dimension d;}

	\qquad\texttt{a = [1, 2, 3] fby.d (a + 1);}

	\texttt{end;}
	\end{minipage}
	\normalsize
}\\

\noindent
The above means:

\noindent
\fbox
{
	\scriptsize
	\begin{minipage}[t]{0.97\textwidth}
	\texttt{array a}

	\texttt{\{}

	\qquad\texttt{int a1 = 1;}

	\qquad\texttt{int a2 = 2;}

	\qquad\texttt{int a3 = 3;}

	\qquad\texttt{int length = 3;}

	\texttt{\}}
	\end{minipage}
	\normalsize
}\\

\noindent
\fbox
{
	\scriptsize
	\begin{minipage}[t]{1.5in}
	\texttt{a1 @ [d:2]}\\
	\texttt{where}

	\quad\texttt{dimension d;}

	\quad\texttt{a1 = 1 fby.d (a1 + 1);}

	\texttt{end;}
	\end{minipage}
	\normalsize
}
\hfill
\fbox
{
	\scriptsize
	\begin{minipage}[t]{1.5in}
	\texttt{a2 @ [d:2]}\\
	\texttt{where}

	\quad\texttt{dimension d;}

	\quad\texttt{a2 = 1 fby.d (a2 + 1);}

	\texttt{end;}
	\end{minipage}
	\normalsize
}
\hfill
\fbox
{
	\scriptsize
	\begin{minipage}[t]{1.5in}
	\texttt{a3 @ [d:2]}\\
	\texttt{where}

	\quad\texttt{dimension d;}

	\quad\texttt{a3 = 1 fby.d (a3 + 1);}

	\texttt{end;}
	\end{minipage}
	\normalsize
}\\

\noindent
The three subexpressions run in parallel, but refer back to the same
array. Should there be a need in one of the three subexpressions to use an
array value produced by another subexpression, they generate a demand for
that value.

%
%

\section{External Design}
\label{sect:external-design}
\index{Design!External}

The external design encompasses user interface design as well as
external software interfaces. In this work, a web interface to the
{\gipsy} as well as command-line interfaces are presented as a part of
UI followed by the API of the two external libraries used, {\javacc} and {\marf}.

\subsection{User Interface}
\label{sect:ui}
\index{Design!User Interface}

\subsubsection{WebEditor -- A Web Front-End to the {\gipsy}}
\index{GIPSY!Web Front-End}
\index{WebEditor}
\index{GIPSY!WebEditor}
\index{GIPSY!Web Portal}

The user interface designed for the {\gipsy}
in the scope of this thesis includes
a Servlet-driven web interface to the GIPSY
daemon server running on our development server
for trying out GIPSY programs online.
The web interface in a form of a web page allows
a connected user to enter, compile, run, and
trace GIPSY programs.
Users are able to submit their own GIPSY
programs (in any supported Lucid dialect) or choose and modify from
existing programs from the GIPSY CVS repository (see \cite{gipsy})
and then launch the computation.
The GIPSY servlet front-end
generates demands through \api{RIPE} and returns back results along with an execution
trace to a web form.
A screenshot of this interface is illustrated in \xf{fig:webeditor-ui}.

\begin{figure}
	\begin{centering}
	\includegraphics[width=\textwidth]{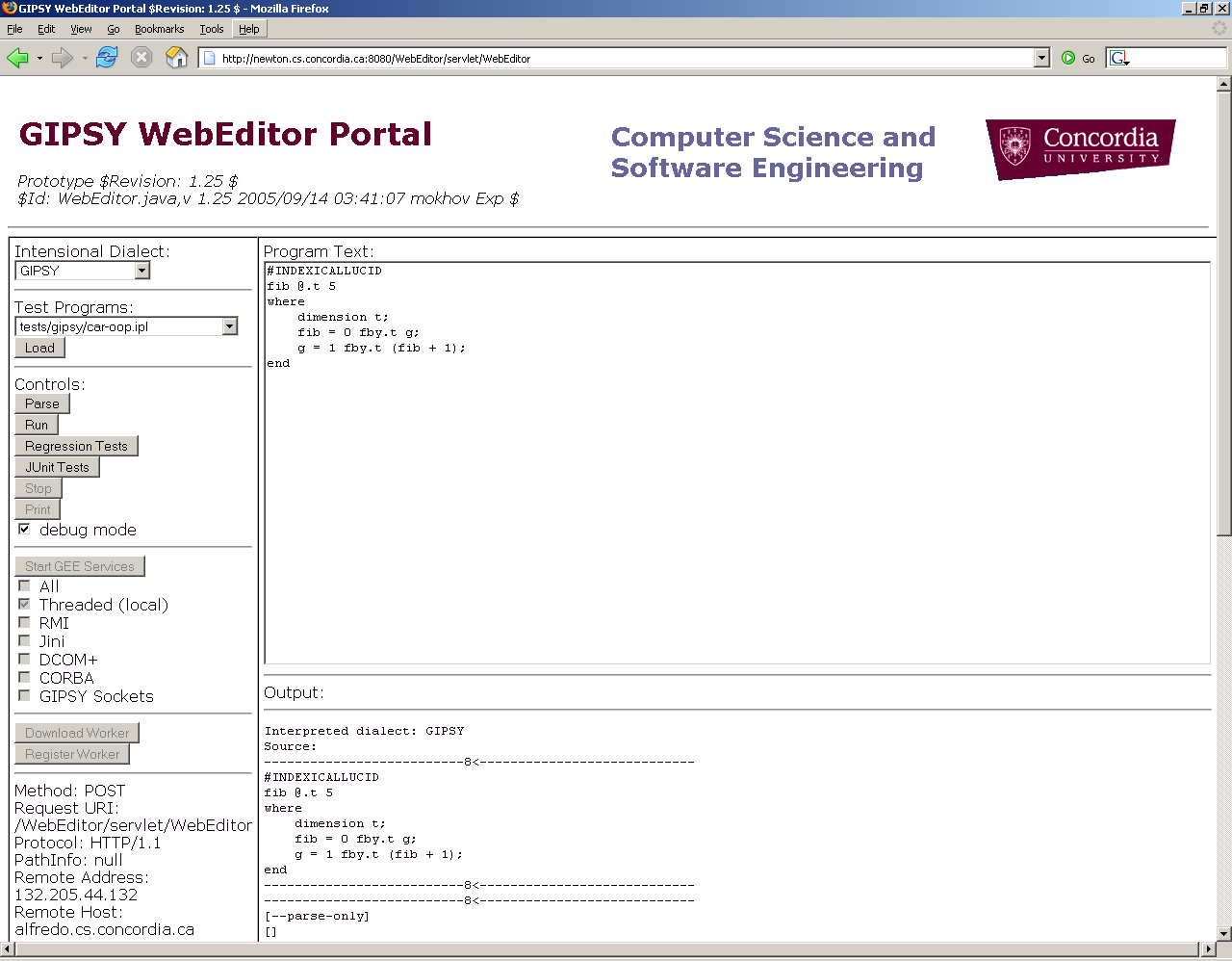}
	\caption{{\gipsy} WebEditor Interface.}
	\label{fig:webeditor-ui}
	\end{centering}
\end{figure}

\clearpage

%
%

\subsubsection{GIPSY Command-Line Interface}
\index{GIPSY!Command-Line Interface}
\index{Command-Line Interfaces!GIPSY}

Synopsis:

\begin{verbatim}
    gipsy [ OPTIONS ]
    gipsy --help | -h
\end{verbatim}

This is an all-entry point for all of {\gipsy} that bundles all
the modules. It generally passes all the options to \api{RIPE}
for further dispatching. When the server part (see \xs{sect:gipsy-server})
is complete, this will be a GIPSY daemon server.
The command line interface includes the following options:

\begin{itemize}
\item
	\option{--help} or \option{-h} displays application's usage information.

\item
	\option{--compile-only} tells to compile a GIPSY program only
	and return the result of the compilation (error or success messages)
	and the compiled program itself. This will not invoke the {\gee}
	for execution after compilation. The option is primarily for quick tests
	in development setups.

\item
	\option{--debug} tells to run in the debug/verbose mode.
\end{itemize}

It is possible to run the \api{GIPSY} by either invoking the
\file{GIPSY.class} directly, by running a corresponding \file{gipsy.jar} (see \xa{sect:gipsy-modules})
file, or using a provided wrapper script \tool{gipsy}. The latter is
the simplest one to use as it includes all the necessary options for
the JVM and searches for the executable \file{.jar} in several common
places. A good idea is to put \tool{gipsy} somewhere under one's PATH.
(A similar approach applies to the other tools mentioned in the
follow up sections, such as \tool{ripe}, \tool{gipc}, \tool{gee},
and \tool{regression}. The tools exist for both Unix and Windows
in the form of shell scripts and batch files.)

Example uses of the \api{GIPSY} application include:

\begin{itemize}
\item
	\texttt{gipsy} or \texttt{gipsy --help}
\item
	\texttt{gipsy --compile-only}
\item
	\texttt{gipsy --compile-only --debug}
\end{itemize}

Where \option{--debug} can be combined with any of these, otherwise
the options are exclusive.

%
%

\subsubsection{RIPE Command-Line Interface}
\index{RIPE!Command-Line Interface}
\index{Command-Line Interfaces!RIPE}

Synopsis:

\begin{verbatim}
    ripe [ OPTIONS ]
    ripe --help | -h
\end{verbatim}

The \api{RIPE} command-line interface right now acts mostly
to activate various own submodules (e.g. textual or DFG editors)
or dispatch requests from users to the other main modules, such
as \api{GIPC} and \api{GEE}.
The command-line interface includes the following options:

\begin{itemize}
\item
	\option{--help} or \option{-h}
	displays application's usage information.
\item
	\option{--gipc=`$<$GIPC OPTIONS$>$'}
	tells \api{RIPE} to invoke \api{GIPC} with a set of
	GIPC options (see \xs{sect:gipc-options}).
\item
	\option{--gee=`$<$GEE OPTIONS$>$'}
	tells \api{RIPE} to invoke \api{GEE} with a set of
	GEE options (see \xs{sect:gee-options}).
\item
	\option{--regression=`$<$REGRESSION OPTIONS$>$'}
	tells \api{RIPE} to invoke \api{Regression} testing with a set of
	its options (see \xs{sect:regression-options}).
\item
	\option{--dfg=`$<$DFG EDITOR OPTIONS$>$'}
	tells \api{RIPE} to start the DFG editor with
	its options. Currently, the \api{DFGEditor} Java
	class is a stub, and instead, the DFG Editor
	of Yimin Ding \cite{yimin04} is started via
	a separate program, \tool{lefty}. It is planned the \api{DFGEditor}
	class would be a wrapper for the program in the future.
	Therefore, all DFG editor options are ignored for now,
	but a provision is made for the future.
\item
	\option{--txt=`$<$TEXTUAL EDITOR OPTIONS$>$'}
	tells \api{RIPE} to start the textual editor with
	its options. Note, at the time of this writing
	\api{TextualEditor} is just a stub, and as such
	does not have any options, but a provision is made
	when it does.
\item
	\option{--debug} tells to run in the debug/verbose mode.
\end{itemize}

Example uses of the \api{RIPE} application include:

\begin{itemize}
\item
	\texttt{ripe} or \texttt{ripe --help}
\item
	\texttt{ripe --compile-only}
\item
	\texttt{ripe --compile-only --debug}
\end{itemize}

%
%

\subsubsection{GIPC Command-Line Interface}
\label{sect:gipc-options}
\index{GIPC!Command-Line Interface}
\index{Command-Line Interfaces!GIPC}

Synopsis:

\begin{verbatim}
    gipc [ OPTIONS ] [ FILENAME1.ipl [ FILENAME2.ipl ] ... ]
    gipc --help | -h
\end{verbatim}

The command line interface for \api{GIPC} inherited some options from
\api{Lucid} \cite{chunleiren02} and includes the following options:

\begin{itemize}
\item
	\option{--help} or \option{-h} displays application's usage information.

\item
	\option{[FILENAME1.ipl [FILENAME2.ipl] ...]} tells \api{GIPC} to compile
	a GIPSY program as indicated by the FILENAME. It is possible
	to have more the one input file for compilation. If this is
	the case, the same number of instances of {\gipc} threads will be initially spawned
	to compile those programs. Notice, however, this does not mean
	all the files (in case of multiple \file{.ipl} files) comprise
	one program and then linked together afterwards as in typical
	C or C++ compilers. Instead, each \file{.ipl} file is treated
	as a stand-alone independent GIPSY program.

\item
	\option{--stdin} tells \api{GIPC} to interpret the standard input as
	a source GIPSY program. This is the default if
	no FILENAME is supplied.

\item
	\option{--gipl} or \option{-G} (came from \api{Lucid} \cite{chunleiren02}
	for backwards compatibility)
	tells \api{GIPC} to interpret the source program unconditionally
	as a GIPL program (by default no assumption is made and \api{GIPC}
	attempts to treat the incoming source code as a general GIPSY program).
	It is primarily used to quickly test the GIPL compiler only, without
	extra overhead or preprocessing. It is also used by the
	\api{Regression} application for that same reason.

\item
	\option{--indexical} or \option{-S} (came from \api{Lucid} \cite{chunleiren02})
	tells \api{GIPC} to interpret the source program unconditionally
	as an Indexical Lucid program.

\item
	\option{--jlucid}
	tells \api{GIPC} to interpret the source program unconditionally
	as a JLucid program.

\item
	\option{--objective}
	tells \api{GIPC} to interpret the source program unconditionally
	as an Objective Lucid program.

\item
	\option{--translate} or \texttt{-T} (came from \api{Lucid} \cite{chunleiren02})
	enables SIPL-to-GIPL translation.
	This option is implied by default (as opposed to be optional in \api{Lucid}). It
	tells the \api{GIPC} to interpret the input program unconditionally
	as a non-GIPL program that requires operator and function
	translation. The option has no effect with \texttt{--gipl} as {\gipl}
	is the only intensional language that does not require any
	further translation.

\item
	\option{--disable-translate} turns off automatic translation
	(in case the user knows that an incoming non-GIPL program has nothing to
	translate, which is rarely the case; otherwise,
	the \api{GIPC} will bail out with an error).

\item
	\option{--warnings-as-errors} tells to treat compilation warnings
	as errors and stop compilation after displaying them.

\item
	\option{--gee} tells \api{GIPC} to run the compiled program
	immediately after compilation (if successful) by feeding
	it directly to the {\gee}. The default is that the compiled
	GIPSY program is saved into a file where the original name
	is suffixed with the \texttt{.gipsy} extension.

\item
	\option{--dfg} tells \api{GIPC} to perform DFG code generation
	as a part of the compilation process.

\item
	\option{--debug} to run in a debug/verbose mode.
\end{itemize}

Example uses of the \api{GIPC} application include:

\begin{itemize}
\item
	\texttt{gipc} or \texttt{gipc --help} or \texttt{gipc -h}
\item
	\texttt{gipc life.ipl}
\item
	\texttt{gipc --disable-translate --gee --debug life.ipl}
\item
	\texttt{gipc --gipl --debug gipl.ipl}
\item
	\texttt{gipc --jlucid --stdin}
\end{itemize}

%
%

\subsubsection{GEE Command-Line Interface}
\label{sect:gee-options}
\index{GEE!Command-Line Interface}
\index{Command-Line Interfaces!GEE}

Synopsis:

\begin{verbatim}
    gee [ OPTIONS ] [ FILENAME1.gipsy [ FILENAME2.gipsy ] ... ]
    gee --help | -h
\end{verbatim}

The command line interface includes the following options:

\begin{itemize}
\item
	\option{--help} or \option{-h} displays application's usage information.

\item
	\option{[FILENAME1.gipsy [FILENAME2.gipsy] ...]} tells \api{GEE} to run a stored version of
	a compiled GIPSY program as indicated by the FILENAME. It is possible
	to have more than one input file for execution. If this is
	the case, the same number of instances of {\gee} threads will be initially spawned
	to run those programs. The programs will run concurrently, but there should not
	be any interference or communication in their execution except they may share the output
	resource.

\item
	\option{--stdin} tells \api{GEE} to interpret the standard input as
	a compiled GIPSY program. This is the default if
	no FILENAME is supplied.

\item
	\option{--all} tells \api{GEE} to start all implemented services/servers
	locally (threaded, {\rmi}, {\jini}, {\complus}, and {\corba}).

\item
	\option{--threaded} tells \api{GEE} to start the threaded server only.

\item
	\option{--rmi} tells \api{GEE} to start the {\rmi} service.

\item
	\option{--jini} tells \api{GEE} to start the {\jini} service.

\item
	\option{--dcom} tells \api{GEE} to start the {\complus} service.

\item
	\option{--corba} tells \api{GEE} to start the {\corba} service.

\item
	\option{--debug} tells \api{GEE} to run in the debug/verbose mode.
\end{itemize}

Example uses of the \api{GEE} application include:

\begin{itemize}
\item
	\texttt{gee} or \texttt{gee --help} or \texttt{gee -h}
\item
	\texttt{gee life.gipsy}
\item
	\texttt{gee --disable-translate --threaded --debug life.gipsy}
\item
	\texttt{gee --all --debug gipl.gipsy}
\item
	\texttt{gipc --rmi --jini indexical.gipsy}
\end{itemize}

%
%

\subsubsection{Regression Testing Application Command-Line Interface}
\label{sect:regression-options}
\index{Regression Testing Application!Command-Line Interface}
\index{Command-Line Interfaces!Regression}

Synopsis:

\begin{verbatim}
    regression [ OPTIONS ]
    regression --help | -h
\end{verbatim}

The \api{Regression} application and its test suite are presented
in detail in \xs{sect:regression}. The application, based on options,
invokes either \api{GIPC} or \api{GEE} or both directly feeding
a pre-selected list of test source programs.
The command line interface includes the following options:

\begin{itemize}
\item
	\option{--help} or \option{-h} displays application's usage information.
\item
	\option{--sequential} tells to run sequential tests (default).
\item
	\option{--parallel} tells to run parallel tests.
\item
	\option{--gipl} tells to test pure GIPL programs only.
\item
	\option{--indexical} tells to test pure GIPL and Indexical programs with the Indexical Lucid compiler.
\item
	\option{--gipsy} tells to test general-style GIPSY programs with code segments.
\item
	\option{--gee} if specified, tells to run the {\gee} after compilation (default).
\item
	\option{--all} tells to do all of the above tests in one run (default).
\item
	\option{--directory} tells to pick source test files from
	a specified directory instead of pre-set directories from
	the GIPSY source tree
\item
	\option{--debug} tells to run in the debug/verbose mode.
\end{itemize}

Example uses of the \api{Regression} application include:

\begin{itemize}
\item
	\texttt{regression} or \texttt{regression --help} or \texttt{regression -h}
\item
	\texttt{regression --gipl}
\item
	\texttt{regression --parallel --indexical}
\item
	\texttt{regression --all --debug}
\item
	\texttt{regression --directory=/some/gipsy/misc/tests --all --debug}
\end{itemize}

%
%

\subsection{External Software Interfaces}
\index{Design!External Software Interfaces}
\index{External Software Interfaces}

\subsubsection{JavaCC API}
\index{External Software Interfaces!JavaCC API}

JavaCC-generated code contains a number of common classes
and interfaces, regardless of the language a parser is generated for. These have to do
with AST nodes, tokens, token types, character streams, and
alike. The most often used class out of this bundle
is \api{SimpleNode}, which is a concrete node in the AST.
These classes
have to be periodically refreshed by compiling the source grammar
when a newer version of \tool{javacc} comes out.

\begin{figure}
	\begin{centering}
	\includegraphics[width=\textwidth]{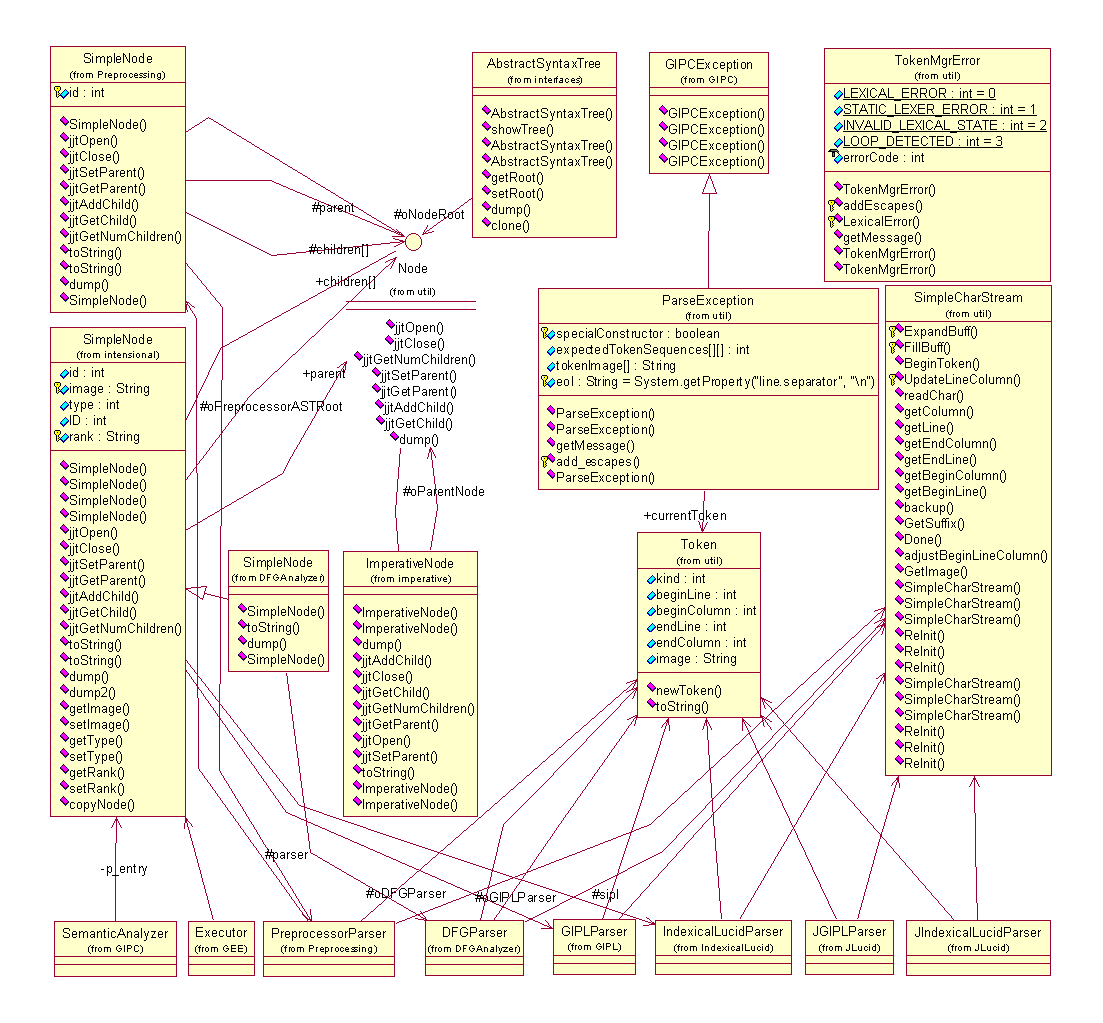}
	\caption{JavaCC- and JJTree-generated Modules Used by Several {\gipc} Modules.}
	\label{fig:javacc-util-cl}
	\end{centering}
\end{figure}

The below are JavaCC API/modules \cite{javacc} used by
the {\gipsy} and their description.
The corresponding class diagram is in \xf{fig:javacc-util-cl}.

\begin{itemize}

\item
\api{Node} is the common interface for all occurrences of \api{SimpleNode}
to implement (see below).

\item
The \api{SimpleNode} class represents a concrete node in every AST in the {\gipc}.
Once generated, this class is usually customized according to the needs
of the given parser/compiler. All concrete instances, however, implement the same \api{Node} interface
above. At the time of this writing, there are three \api{SimpleNode} occurrences
in the GIPSY source tree: the common one in \texttt{gipsy.GIPC.intensional} for
all the {\sipl}s and {\gipl}, as per original implementation presented
in \cite{chunleiren02}. It is a basis for a GIPL {\AST} aside from the related parsers
known to the \api{SemanticAnalyzer} and {\gee}'s \api{Executor}.
This implementation is wrapped-around by
\api{AbstractSyntaxTree} that the rest of the modules know.
Then, a customized subclass of it is in
\texttt{gipsy.GIPC.DFG.DFGAnalyzer} of Yimin Ding \cite{yimin04}. It was
made a subclass because a large portion of the code is identical.
Finally, the last one is in \texttt{gipsy.GIPC.Preprocessing} used by
the \api{Preprocessor}. This occurrence of \api{SimpleNode} was
kept as-is due to the significant differences and purpose with
the former two.

\item
The \api{ImperativeNode} is another implementation of the \api{Node}
interface created manually for all the imperative language compilers.
The \api{ImperativeNode} represents an {\AST} of a single node
encapsulating STs, CPs, some meta information that came from a given
imperative compiler. The reason for this is to maintain a global AST
for a GIPSY program where all nodes implement the same interface.

\item
\api{SimpleCharStream} is a common \tool{javacc} utility
that treats incoming source code stream as a set of ASCII
characters without extra UNICODE processing.

\item
\api{ParseException} is a common generated type of exception
to indicate a parse error. It was made manually to subclass
\api{GIPCException} from the GIPSY Exceptions Framework (see \xs{sect:exceptions}) for
uniform exception handling.

\item
\api{TokenMgrError} a subclass of \api{java.lang.Error}
primarily to signal lexical errors in the incoming
source code or token processing in general by a given
parser (e.g. by invoking a static parser twice).

\end{itemize}

%
%

\subsubsection{MARF Library API}
\index{External Software Interfaces!MARF Library API}

\begin{figure}
	\begin{centering}
	\includegraphics[width=\textwidth]{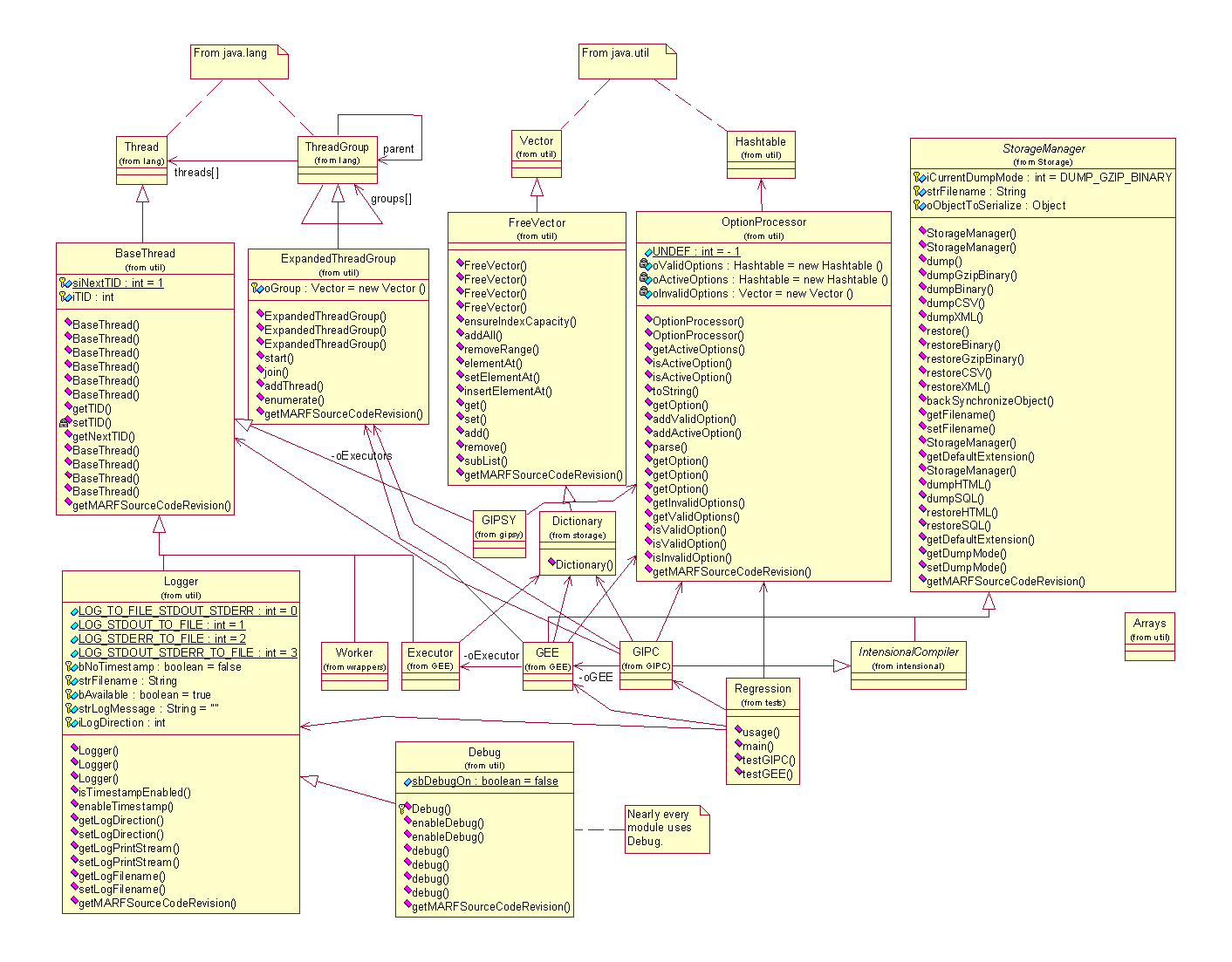}
	\caption{MARF Utility Classes used by the {\gipsy}.}
	\label{fig:marf-util-cl}
	\end{centering}
\end{figure}

\begin{figure}
	\begin{centering}
	\includegraphics[width=\textwidth]{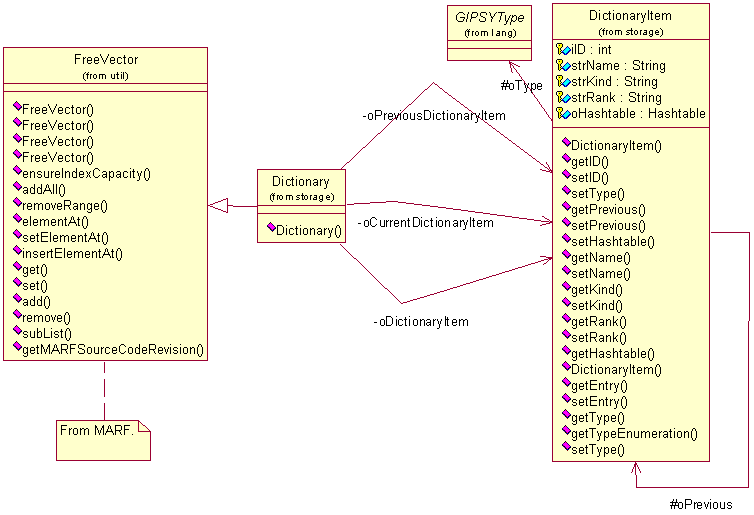}
	\caption{\api{Dictionary} and \api{DictionaryItem} API}
	\label{fig:dict-cl}
	\end{centering}
\end{figure}

\begin{figure}
	\begin{centering}
	\includegraphics[width=\textwidth]{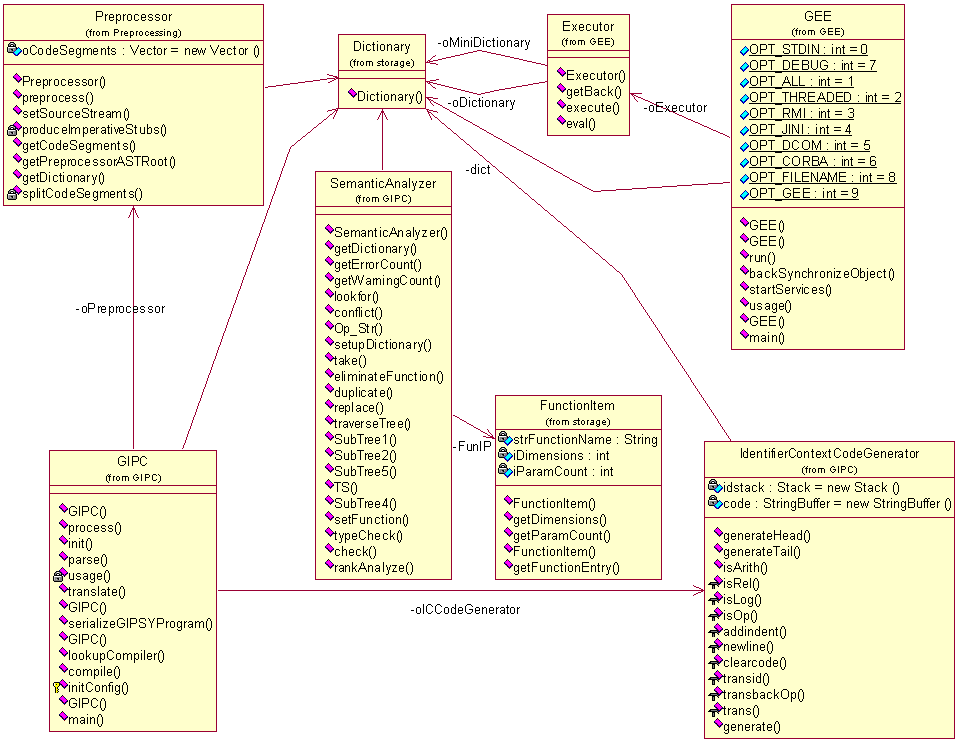}
	\caption{\api{Dictionary} Usage within the {\gipsy}}
	\label{fig:dict-usage-cl}
	\end{centering}
\end{figure}

\tool{MARF} (see \xs{sect:marf-in-tools}) has a variety of useful utility and storage-related modules
that conveniently found their place in {\gipsy}. Most of these
come from the \api{marf.util} package as well as
\api{marf.Storage}.\footnote{Later some natural language processing (NLP)
modules in \api{marf.nlp} of \tool{MARF} might also get used in the {\gipsy}
as a part of another research project.}
The below are MARF API/modules used by {\gipsy} and their description:

\begin{itemize}

\item
\api{marf.util.FreeVector} is an extension of \api{java.util.Vector}
that allows theoretically vectors of infinite length, so it is possible
to set or get an element of the vector beyond its current physical bounds.
Getting an element beyond the boundaries returns \texttt{null}, as if the
object at that index was never set. Setting an element beyond bounds
automatically grows the vector to that element. In the {\gipsy},
\api{marf.util.FreeVector} is used as a base for our \api{Dictionary}
as shown in \xf{fig:dict-cl}. \xf{fig:dict-usage-cl} shows all the
modules that are now using \api{Dictionary} instead of \api{java.util.Vector}.
The corresponding class diagram of the MARF's \api{util} API is shown in \xf{fig:marf-util-cl}.

\item
\api{marf.util.OptionProcessor} module is extensively
used by the command-line user interfaces (see \xs{sect:ui})
of \api{GIPSY}, \api{GIPC}, \api{GEE}, and \api{Regression}. A convenient way of managing command-line
options in a hash table and validating them.

\item
\api{marf.util.BaseThread} class encapsulates some useful functionality
used in threaded versions of {\gee} and {\gipc}, which Java's \api{java.lang.Thread}
does not provide:

	\begin{itemize}
	\item
		maintaining unique thread ID (TID) among multiple
		threads and reporting it (for tracing, debugging, and {\ripe}). A note is
		added here that Java 1.5.* now also provides a notion of a TID,
		but \api{marf.util.BaseThread} was written prior to that and {\gipsy}
		remains Java 1.4-compliant still. Plus, {\marf}'s way of handling
		this is more flexible.

	\item
		adapted human-readable trace information via \api{toString()}

	\item
		access to the \api{Runnable} target that was specified upon creation.

	\item
		integration with \api{marf.util.ExpandedThreadGroup}, see below.
	\end{itemize}

\item
\api{marf.util.ExpandedThreadGroup}
allows to start, stop, or other group operations that Java's
\api{java.lang.ThreadGroup} doesn't provide. \api{ExpandedThreadGroup}
is, for example, used in \api{GIPC} to create a group of compiler threads
(in {\gipsy} every compiler is a thread),
one for each language chunk, that will run concurrently. Additionally,
a group of \api{GEE}, or rather, \api{Executor} threads would run
in the case of a forest of ASTs.

\item
\api{marf.util.Arrays} groups more array-related functionality together
than the \api{java.util.Arrays} class does, for example copying
(homo- and heterogeneous types)
and converting to \api{java.util.Vector}, and provides some extras.

\item
\api{marf.Storage.StorageManager} provides basic implementation of
the (possibly compressed) object serialization, and in our case
the \api{GIPC} and \api{GEE} are storage manager with respect to
a compiled GIPSY program.

\item
\api{marf.util.Logger} is primarily used by the \api{Regression} application
to log standard output before calling \api{GIPC} or \api{GEE} to a file,
for future comparison with an expected output.

\item
\api{marf.util.Debug} is used in many places for debugging convenience allowing
to issue debug messages only if the debug mode is globally on, which is also
maintained within the class.

\end{itemize}

%
%

\subsubsection{Servlets API}
\index{External Software Interfaces!Servlets API}

The Java Servlets technology from Sun \cite{servlets} was used
to implement the \api{WebEditor} interface outlined earlier. While
the actual API specification of servlets is rather vast, the key used
components used here are listed:

\begin{itemize}

\item
The \api{HttpServlet} class is the base for all servlets, including
\api{WebEditor}.

\item
The \api{doGet()} must be overridden to respond to the GET HTTP
requests.

\item
The \api{doPost()} must be overridden to respond to the POST HTTP
requests. In our implementation, \api{doPost()} is a simply a
wrapper around \api{doGet()}, so both GET and POST requests are
handled identically.
\end{itemize}

%
%

\subsection{Architectural Design and Unit Integration}
\index{Implementation!Unit Integration}
\index{Implementation!Architectural Design}

Unit integration according to the initial design decisions of the {\gipsy}
system and setting up package hierarchy played an important role
in the success of this work. A proposed directory structure (see \xa{sect:dir-structure})
and a corresponding breakdown of the Java packages (see \xa{sect:java-packages})
hierarchy are important to the success of {\gipsy}, especially for public use.
The author of this work inherited the previous GIPSY iteration without any
structure or packaging and proposed and restructured the system to what it is now.

%
%

\subsubsection{{\gipsy}}

When integrating several components of a large system
and re\-designing some of their API, the overall system
design has to be considered. In \xf{fig:gipsy-cl} is a
high-level view of the main GIPSY modules. These
modules can be run as stand-alone Java applications
or start each other.

\begin{figure}
    \begin{centering}
    \includegraphics[width=384pt]{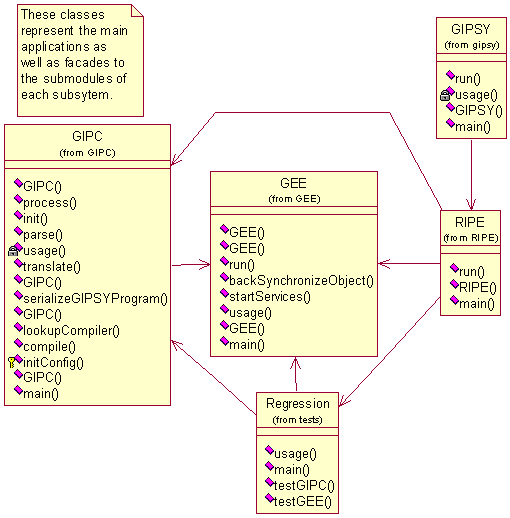}
    \caption{{\gipsy} Main Modules.}
    \label{fig:gipsy-cl}
    \end{centering}
\end{figure}

\begin{itemize}
\item
The \api{GIPSY} class on the diagram represents
a stand-alone server for a client-server type
of application, which is capable of spawning
\api{GIPC} and \api{GEE} upon client's request.
The prime goal of it is testing of intensional
programs that users can submit online and get
the result in case they don't have the development
environment set up from where they are working.

\item
The \api{GIPC} class when run as a stand-alone
application invokes all the intensional and
imperative compilers required and produces
a compiled version of a submitted GIPSY program.
Optionally, if requested, \api{GIPC} can pass
the compiled program on to \api{GEE} for execution.
The \api{GIPC} along with \api{GEE} subsumes what was previously known
as \api{Lucid} and \api{Facet} defined by Chun Lei
Ren in \cite{chunleiren02}.

\item
The \api{GEE} when run as a stand-alone application,
begins demand-driven execution of a GIPSY program
that was either compiled and stored or compiled and
passed from \api{GIPC}.

\item
The \api{Regression} class is the main driver
for the Regression Testing Suite of {\gipsy}, that
also calls these modules for regression and
unit testing.
\end{itemize}

%
%

\subsubsection{GIPSY Exceptions Framework}
\label{sect:exceptions}
\index{Frameworks!GIPSY Exceptions}
\index{GIPSY Exceptions}
\index{Exceptions}

\begin{figure}
	\begin{centering}
	\includegraphics[width=\textwidth]{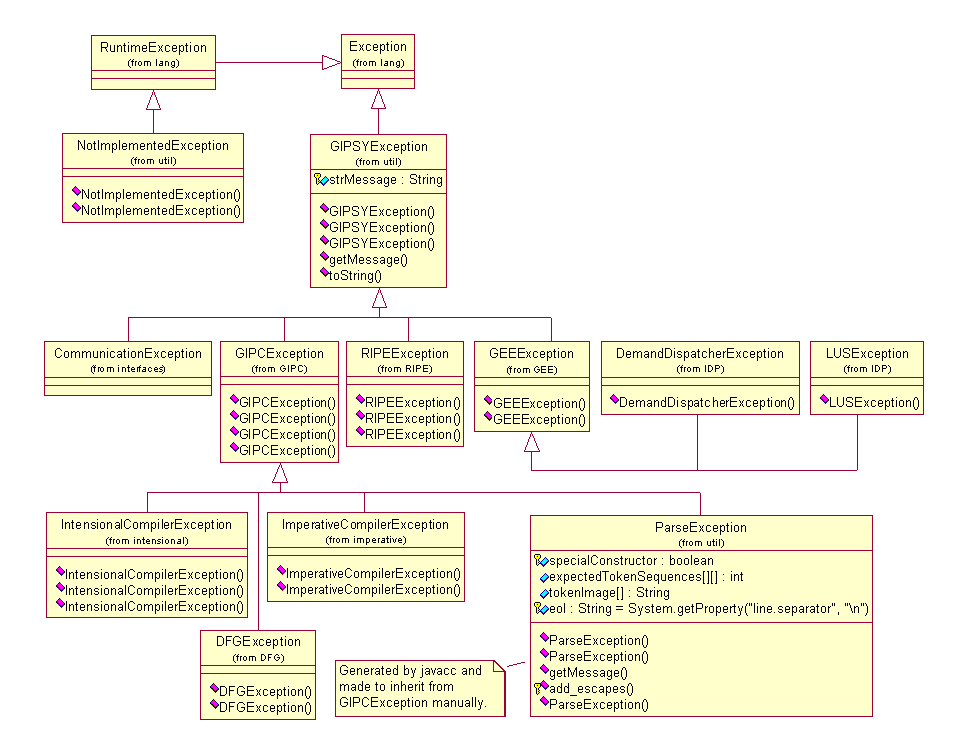}
	\caption{GIPSY Exceptions Framework\index{Frameworks!GIPSY Exceptions}.}
	\label{fig:exceptions-cl}
	\end{centering}
\end{figure}

The class diagram describing the GIPSY Exceptions Framework\index{Frameworks!GIPSY Exceptions}
is in \xf{fig:exceptions-cl}.
The main exception type is \api{GIPSYException} that provides some machinery
encapsulating other exceptions. Every major module, like {\gipc}, {\gee}, or {\ripe}
in {\gipsy} defines its own sublcass of \api{GIPSYException}. By doing this, the
applications using the modules can differentiate the exception types and handle them appropriately.
The \api{NotImplementedException} is an easy way to use to indicate some unimplemented
but important stubs, if called. It is a subclass of \api{RuntimeException} because
it can happen virtually everywhere and run-time exceptions do not need to be
declared to be thrown or caught. The \api{GIPCException}, \api{GEEException},
and \api{RIPEException} represent base exception objects for the corresponding
modules; the rest are primarily subclasses of these.

%
%

\subsubsection{{\gee} Design}
\index{Design!GEE}
\index{GEE!Design}
\index{GEE!Integration}
\index{Integration!GEE}

The general overview of {\gee} is in \xf{fig:gee-cl}.
The several modules under the \api{gipsy.GEE} package
carry out a complex GIPSY program execution task.
The \api{GEE} is the facade and the main starting point
for all of {\gee}. \api{GEE} may act as either an application
on its own or be invoked by the \api{GIPC}. For the stand-alone
execution a user has to supply a filename of a valid compiled
\api{GIPSYProgram}. This program is loaded and \api{GEE} starts
the \api{Executor} thread to actually execute it. Before \api{Executor}
begins the \api{GEE} may optionally start the available demand
propagation services, such as local (just threads), {\rmi},
Jini-based and the like. The \api{Executor} while executing
the program generates demands for the identifiers listed in the
program and then performs the final calculation based on the
results received. The \api{Executor} was formerly known as
\api{XLucidInterpreter} and the Java version of which was implemented
by Bo Lu in \cite{bolu04} and reworked to handle sequential threads,
arrays, objects, and other than integer and float data types.

\begin{figure}
	\begin{centering}
	\includegraphics[width=\textwidth]{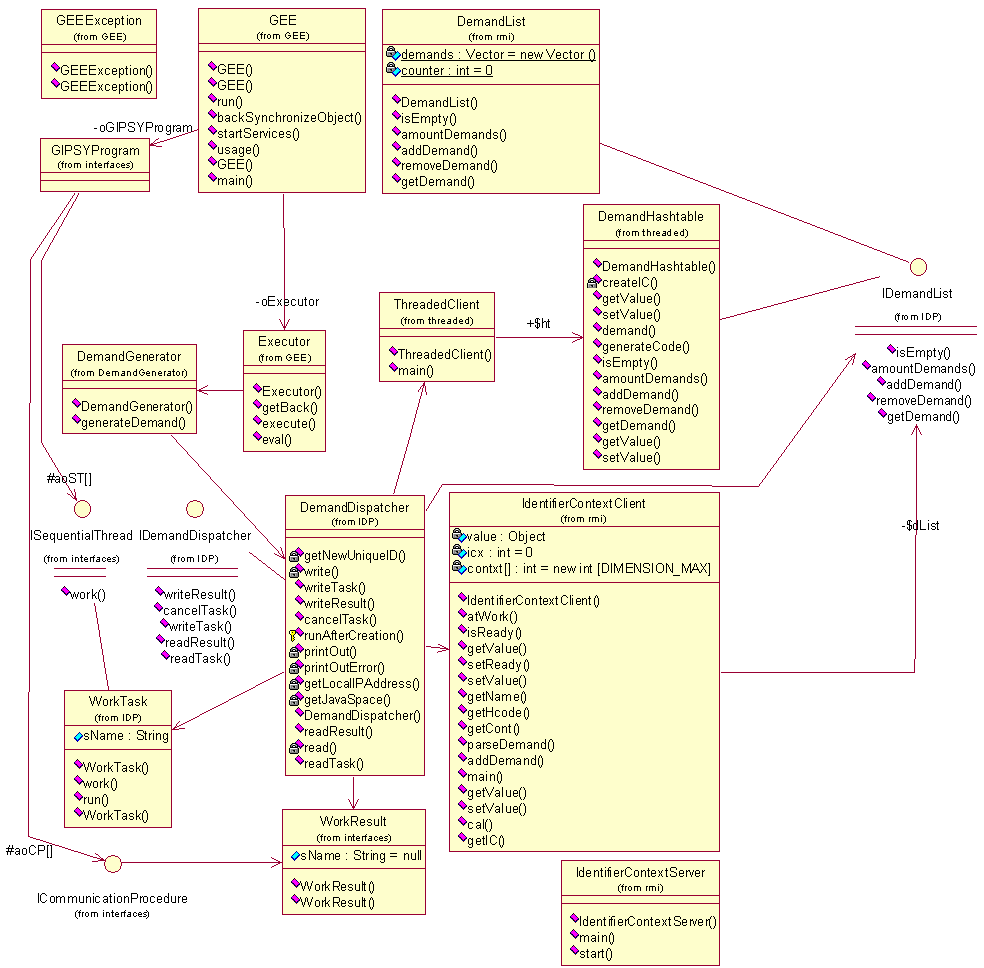}
	\caption{{\gee} Design.}
	\label{fig:gee-cl}
	\end{centering}
\end{figure}

%
%

\paragraph{Demand Dispatcher}
\index{Integration!Demand Dispatcher}
\index{Integration!Jini}
\index{Demand Dispatcher!Integration}
\index{Jini!Integration}

In \xf{fig:demand-dispatcher-cl} is a high-level overview of the
\api{DemandGenerator} and related classes. Most of the demand propagation in {\jini} and
JavaSpaces is implemented by Emil Vassev in \cite{vas05}. The integration part
included making sure the \api{IDemandList} interface is consistently
used by the \api{DemandGenerator} along with the \api{DemandDispatcherAgent} to be
compliant to the rest of the {\gee}. The \api{IDemandList} interface
was originally designed by Bo Lu in \cite{bolu04} and redesigned by the author
of this thesis to be implemented by the {\rmi} and threaded versions of {\gee}
and was formerly known as \api{DemandList}. Next, the temporary class \api{WorkTask}
was made to implement the \api{ISequentialThread} interface according to the overall
{\gipsy} design for sequential threads. This class is marked as deprecated (and later on will be removed)
as every sequential thread class is generated by the \api{SequentialThreadGenerator}
and is different from one GIPSY program to another. Finally, the \api{LUSException} (service look up exception)
and \api{DemandDispatcherException} were made to be a part of the GIPSY Exceptions
Framework \xs{sect:exceptions} by inheriting from the \api{GEEException}. For further
implementation details of the \api{DemandDispatcher} please refer to Emil's work \cite{vas05}.

\begin{figure}
	\begin{centering}
	\includegraphics[width=\textwidth]{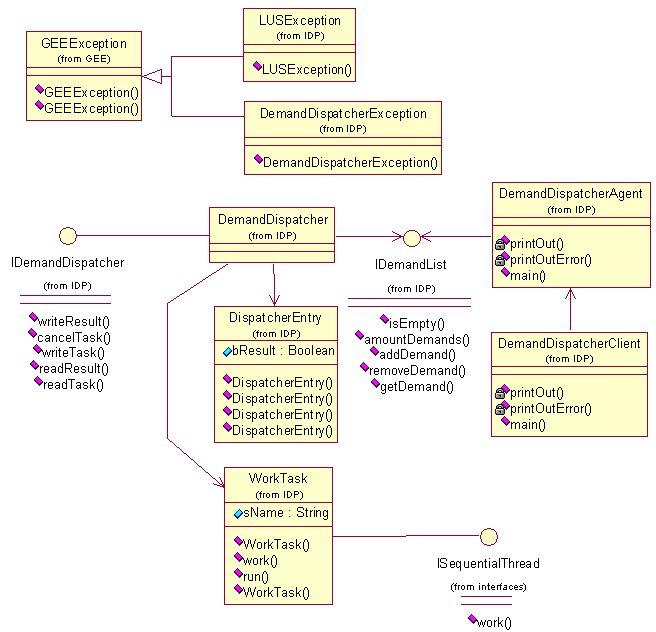}
	\caption{The Demand Dispatcher Integrated and Implemented based on {\jini}.}
	\label{fig:demand-dispatcher-cl}
	\end{centering}
\end{figure}

\paragraph{Intensional Value Warehouse and Garbage Collection}
\index{Integration!Intensional Value Warehouse}
\index{Integration!Garbage Collection}

\begin{figure}
	\begin{centering}
	\includegraphics[width=\textwidth]{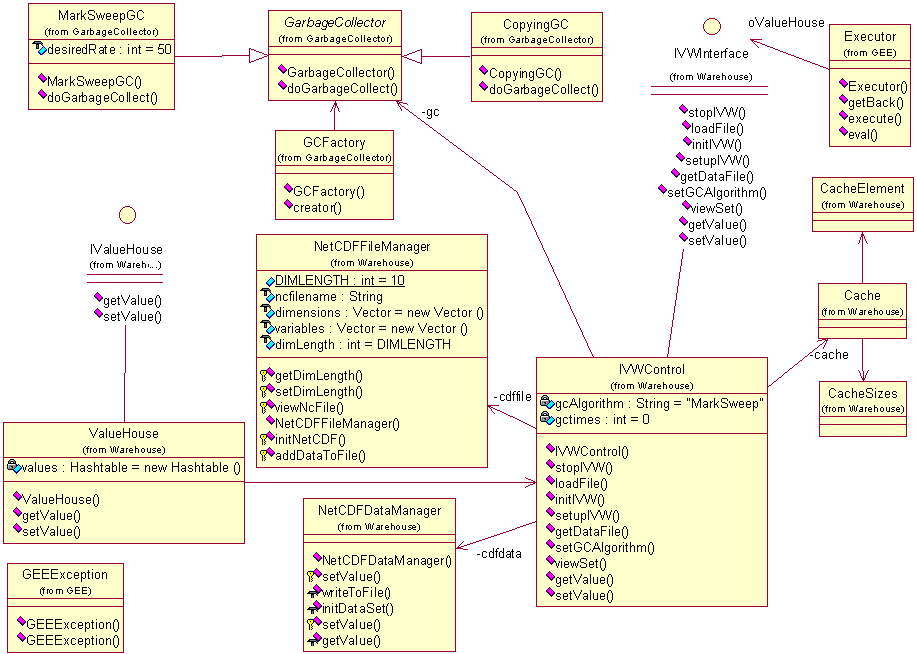}
	\caption{Integration of the Intensional Value Warehouse and Garbage Collection.}
	\label{fig:ivwgc}
	\end{centering}
\end{figure}

Intensional Value Warehouse and Garbage Collection were implemented
by Lei Tao in \cite{leitao04}. After integration, his contributions
became to look like as shown in \xf{fig:ivwgc}. The \api{IValueHouse}
and its extension \api{IVWInterface} are the ones used by the \api{Executor}
to communicate to a concrete implementation of a warehouse, allowing
adding/changing warehouse implementations easily without affecting
the \api{Executor}.
All the exception handling is based on the \api{GEEException}.

%
%

\subsubsection{{\ripe} Design}
\index{Frameworks!RIPE}

\begin{figure}
	\begin{centering}
	\includegraphics[width=\textwidth]{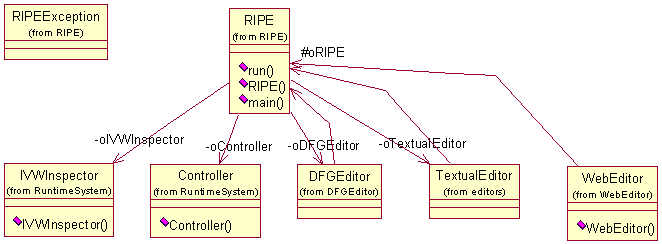}
	\caption{{\ripe} Design.}
	\label{fig:ripe-cl}
	\end{centering}
\end{figure}

The class diagram describing {\ripe} is in \xf{fig:ripe-cl}.
The \api{RIPE} class represents a facade to the rest of the
{\ripe} modules. It is semi-implemented, as many things are
not clear on this side of the project yet. The only part
of {\ripe} that was advanced well so far by Yimin Ding in
\cite{yimin04} is the Data-Flow-Graph (DFG) editor, which
is not implemented in {\java}. The \api{DFGEditor} {\java}
class is meant to be main {\java} program acting like
a bridge between {\java} and the LEFTY language, but did
not get implemented yet. The rest of the modules are planned
stubs.

%
%

\subsubsection{Data Flow Graphs Integration}
\index{Integration!DFG}
\index{DFG!Integration}

\begin{figure}
	\begin{centering}
	\includegraphics[width=\textwidth]{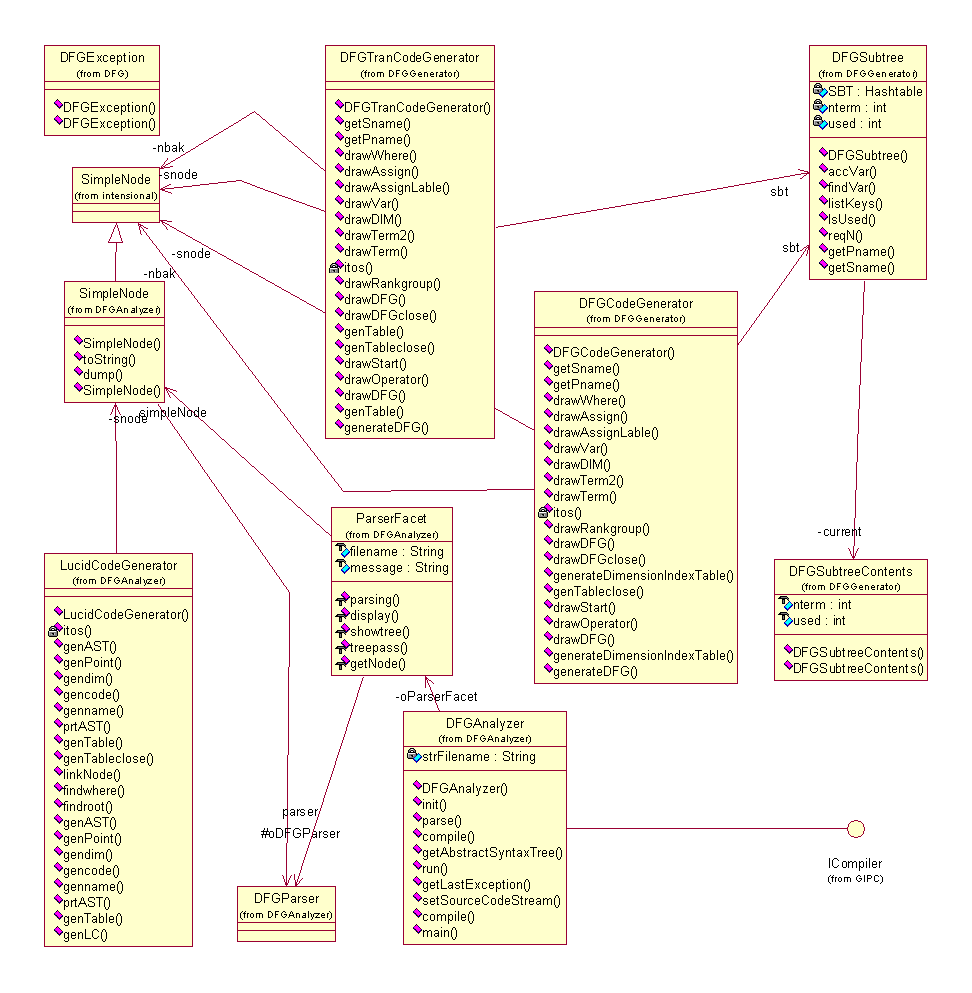}
	\caption{DFG Integration Design.}
	\label{fig:dfg-cl}
	\end{centering}
\end{figure}

The integration of Yimin Ding's \cite{yimin04} DFG-related work
is presented in \xf{fig:dfg-cl}. The \api{DFGAnalyzer} was augmented
to implement the \api{ICompiler} interface as it follows the same structure
as the rest of our compilers, which compiles a Lucid code from DFG. The \api{DFGException}
class, a subclass of \api{GIPCException} has been created to indicate
an error situation in the DFG processing. \api{DFGAnalyzer}'s
\api{SimpleNode} was updated to inherit from \api{GIPC.intesional.SimpleNode}
due to vast functionality overlap. The two analyzer and generator modules
have been placed under the \api{GIPC.DFG.DFGAnalyzer} and \api{GIPC.DFG.DFGGenerator}
packages.

%
%

\clearpage
\section{Summary}

This chapter presented most of the development effort went
into integration, design, and implementation of {\jlucid},
{\olucid}, and {\gicf}. User interfaces (both web and command line)
has been outlined. Regression Test Suite has been introduced.
The follow up chapter presents a variety of testing approaches
went into the {\gipsy} to prove successful integration of the old and
implementation of new modules.

To summarize, {\olucid}, as opposed to {\glu} \cite{glu1, glu2} and {\jlucid}, provides
access to the object members and is real object-oriented
hybrid language. While {\jlucid} may indirectly manipulate
objects through pseudo-free functions, the actual objects are still
a ``black box'' to it.

The {\gicf} and {\iplcf} gave an ability for an easier integration of
intensional and imperative languages in the {\gipsy}. The below are
the steps one needs to perform to add a new compiler to the {\gipsy}:

\begin{itemize}
\item
	create a package where the language compiler
	will reside (usually under \\\file{imperative/LANGUAGE}
	or \file{intensional/SIPL/LANGUAGE}.

\item
	add a compiler class that extends either one of \api{IntensionalCompiler},
	\\\api{ImperativeCompiler}, or implements one of their superinterfaces

\item
	the code segment and fully qualified class name should be added
	to either \api{EImperativeLanguages} or \api{EIntensionalLanguages}

\item
	optionally implement a custom version of a preprocessor if it is a hybrid
	language

\item
	implement translation rules to GIPL if it is a {\sipl} if it is
	an intensional language

\item
	implement proper ST/CP generation for an imperative language
	according to that language's semantics and typing instructions

\item
	implement type mapping table upon the need if it is an imperative
	language
\end{itemize}

The above might still sound complex, but it is much more easier
and flexible than before. Additionally, some of the steps can
be abstracted and simplified, but it is impossible to eliminate
manual work altogether.



\chapter{Testing}
\label{chapt:testing}
\index{Testing}

This chapter addresses the testing aspect of this thesis for the
following two main reasons: integration and refactoring of a variety
of the {\gipsy} modules including {\gicf} and the development and
operation of the two new Lucid dialects developed in this work,
namely {\jlucid} and {\olucid}. Notice, this testing is far from
comprehensive and does not include testing of the execution
performance of any of the programs and many compilation aspects
are still to be resolved as of this writing (and be resolved
in the final version). This is, however,
a starting point of setting up the GIPSY testing infrastructure
for the projects to come to do mandatory systematic tests, which
are now a necessity given the size of the system, a centralized source tree,
and the number of subprojects developed simultaneously.

\section{Regression Testing}
\label{sect:regression}
\index{Testing!Regression}

\subsection{Introduction}
\index{Regression!Introduction}
\index{Regression!Testing}
\index{Testing!Regression}

The regression testing is a comprehensive set of tests for the implementation
and integration of the {\gipsy} modules.
They test most of the operations and capabilities of the {\gipsy}.
The test cases primarily are various intensional programs (hybrid or not)
that exercise the main modules, such as {\gipc} and {\gee} as well
as their submodules with the major focus on {\gipc}.

\subsection{Regression Testing Suite}
\label{sect:regression-testing}
\index{Regression Testing Suite}

The regression tests can be run against already pre-compiled \file{gipsy.jar}, or
by using a temporary installation within the
source tree using the \api{Regression} application.
Next, there are a ``sequential'' and ``parallel'' modes to run the tests.
In the sequential mode tests run
in strict sequence, whereas in the parallel mode multiple threads are started
to run groups of tests in parallel.


\subsubsection{Unit Testing with JUnit}
\label{sect:unit-testing}
\index{Testing!Unit}

The core of the \api{Regression} application is based on the
{\junit} framework\index{Frameworks!JUnit} introduced
in \xs{sect:junit-intro}. \api{Regression} represents
a \api{TestSuite}, that contains \api{ParallelTestCase}
and \api{SequentialTestCase}, a subclasses of \api{TestCase}.
Both types of tests are customizable based on the options
supplied to the \api{Regression} application (see \xs{sect:regression-options}).
{\junit} helps to tell us what errors happened and in which
modules and the reason of the failures dynamically at run-time.

\subsubsection{Unit Testing with \tool{diff}}
\label{sect:diff-testing}
\index{Testing!Diff}

It becomes cumbersome to use {\junit} for all possible cases,
in a large system, where often we are generally interested in the output
behaviour changes only. Here the {\unix} tool \tool{diff} helps us.
A collection of hand-checked outputs are said to be ``expected'',
one ore more file for each test case.
Then, when the next time the test is run, a current directory
is created with the current outputs, and the current and expected
output directories are compared with the \tool{diff} to show
the differences in the output produced by the modules.
This is all achieved by
the \tool{regression} script.

\subsubsection{Tests}

The actual test cases in the form of {\gipl}, {\ilucid},
{\olucid}, {\jlucid}, and {\gipsy} programs, are located
under the corresponding \file{src/tests/*} directories in the source
tree in the form of \file{*.ipl} files. These comprise
most of the examples presented earlier in this work as
well as developed in \cite{paquetThesis}, \cite{chunleiren02},
\cite{aihuawu02}, and \cite{bolu04}. The regression tests for the 
{\dfg} generation (\cite{yimin04}), Intensional
Value Warehouse and Garbage Collector \cite{leitao04} and
Demand Migration System ({\dms}) \cite{vas05} are not
present as of this implementation.


\section{Portability Testing}
\label{sect:portability-testing}
\index{Testing!Portability}

{\gipsy} has been tested and is known as expected (regression tests pass) to run
on \rhl{9}\index{Testing!Red Had Linux 9}, \fcore{2}\index{Testing!Fedora Core 2},
\macos{X}\index{Testing!MacOS X}, \solaris{9}\index{Testing!Solaris 9}, \win{98SE/2000/XP}\index{Testing!Windows 98SE/2000/XP}
systems under JDK 1.4 and 1.5. The corresponding hardware architectures
were Intel or Intel-compatible processors (Pentium II, III, and IV with 233 MHz to 1.4 GHz)
and G3 and G4 processors from Apple and IBM.
For the WebEditor interface, Tomcat 5 on \macos{X} were tested, but
it is believed to run on other platforms the Jakarta Tomcat runs on.


\clearpage
\section{Solving Problems}
\label{sect:problems}
\index{Problems!Solving}

This section is targeting some common problems of
synchronization in parallel and distributed environment
and how they are solved using the {\gipsy} system relieving
the programmer from the need of explicitly synchronize
the objects. They also illustrate the use of arrays and
embedded {\java}, and Java objects. These programs are among
many other test cases from the Regression Tests Suite.

%
%

\subsection{Prefix Sum}
\index{Prefix Sum}
\index{Examples!Prefix Sum}

\sourcefloat
    {\begin{verbatim}
pseudocode (for thread 'j')

'shared'  a 'future' 'int' 'array' [1..logP, 1..P] := undefined;
'private' sum 'int' := j,
          hop 'int' := 1;

'do' level = 1, logP  --->
    'if' j <= P - hop ---> a[level, j] := sum        'fi'
    'if' j >      hop ---> sum +:= a[level, j - hop] 'fi'
     hop := 2 * hop
'od'
\end{verbatim}
}
    {fig:prefix-sum-pseudo}
    {Pseudocode of a thread $j$ for the Prefix Sum Problem.}

\sourcefloat
    {\begin{verbatim}
/*
 * PREFIX SUM in GIPL-style JLucid program.
 * Numbers are from 1 to 8.
 * S[I] will contain prefix sum for number 'i'
 */

#JLUCID

// Array of prefix sums
S[8] @d 8
where
    dimension d;

    S[I] = if(#d = 0)
                then 1
                else (2 * S[I] - 1) @d (#d - 1)
           fi;

    // Index the array varies within.
    I @i 8
    where
        dimension i;
        I = if(#i = 0) 1 else (I - 1) @d (#i - 1);
    end;
end;
\end{verbatim}
}
    {fig:prefix-sum}
    {The Prefix Sum Problem in {\jlucid} in GIPL Style.}

\sourcefloat
    {\begin{verbatim}
/*
 * PREFIX SUM in Indexical Lucid-style JLucid
 */

#JLUCID

S[8] @d 8
where
    dimension d;

    S[I] = 1 fby.d (2 * S[I] - 1);

    I @i 8
    where
        dimension i;
        I = 1 fby.i (I - 1);
    end;
end;
\end{verbatim}
}
    {fig:prefix-sum-indexical}
    {The Prefix Sum Problem in {\jlucid} in Indexical Lucid Style.}

The pseudocode of for a thread $j$ is in \xf{fig:prefix-sum-pseudo} \cite{probstPrefixSum}.
The \xf{fig:prefix-sum} shows the program translated into {\lucid}.
The \xf{fig:prefix-sum-indexical} shows the program translated into
Indexical Lucid for numbers from 1 to 8.
Below is an equivalent implementation
of the problem (targeting only TLP) in Java; compare the program's
line count and complexity to that of {\jlucid}:

\source{\begin{verbatim}
// Modified from Dr. Probst's Cyclic.java
public class PrefixSum
{
    public static final int P     = 8; // number of workers
    public static final int logP  = 3; // number of rows in logP x P matrix

    // For permutation of workers
    private static int[] col = {3, 6, 5, 7, 4, 2, 1, 0};

    // These two mimic a 2D array of future variables
    public static int[][]       a       = new int      [logP][P];
    public static Semaphore[][] futures = new Semaphore[logP][P];

    // The resulting sums are to be placed here.
    public static int[]         sums    = new int[P];

    public static void main(String[] argv)
    {
        Worker w[] = new Worker[P];

        for(int j = 0; j < futures.length; j++ )
            for(int k = 0; k < futures[j].length; k++)
                futures[j][k] = new Semaphore(0);

        for(int j = 0; j < P; j++)
        {
            w[col[j]] = new Worker(col[j] + 1);
            w[col[j]].start();
        }

        for(int j = 0; j < P; j++)
        {
            try
            {
                w[j].join();
            }
            catch(InterruptedException e)
            {
            }
        }

        for(int j = 0; j < P; j++)
            System.out.println ("Prefix Sum of " + (j + 1) + " = " + sums[j]);

        System.out.println ("System terminates normally.");
    }
}

class Semaphore
{
    private int value;

    Semaphore(int value1)
    {
        value = value1;
    }

    public synchronized void Wait()
    {
        try
        {
            while(value <= 0)
            {
                wait();
            }

            value--;
        }
        catch (InterruptedException e)
        {
        }
    }

    public synchronized void Signal()
    {
        ++value;
        notify();
    }
}

class Worker extends Thread
{
    private int j;
    private int sum;
    private int hop = 1;

    public Worker(int col)
    {
        sum = j = col;
    }

    public void run()
    {
        System.out.println("Worker " + j + " begins execution.");
        yield();

        for(int level = 0; level < PrefixSum.logP; level++)
        {
            if(j <= PrefixSum.P - hop)
            {
                System.out.println
                (
                    "Worker " + j +
                    " defines a[" + level + "," + (j-1) +"]."
                );

                PrefixSum.a[level][j - 1] = sum;
                PrefixSum.futures[level][j - 1].Signal();
            }

            if(j > hop)
            {
                PrefixSum.futures[level][j - 1 - hop].Wait();

                System.out.println
                (
                    "Worker " + j +
                    " uses a[" + level + "," + (j - 1 - hop) + "]."
                );

                sum += PrefixSum.a[level][j - 1 - hop];
            }

            hop = 2 * hop;
        }

        PrefixSum.sums[j - 1] = sum;
        System.out.println ("Worker " + j + " terminates.");
    }
}
\end{verbatim}
}

%
%

\clearpage
\subsection{Dining Philosophers}
\index{Dining Philosophers}
\index{Examples!Dining Philosophers}
\index{Problems!Dining Philosophers}

Below is a JLucid implementation of
the Dining Philosophers problem \cite{dijkstra65, dijkstra71, gingras90}.
We have arrays of 8 philosophers
and 8 forks, each represented as integers. A philosopher is either
thinking (1) or eating (2); likewise for forks, taken or not.
A philosopher may eat when they have exactly two forks, not less,
if the forks are available. If none available, the philosopher
waits (implicit, guaranteed by the {\gee}). The special variable $I$
serves as an intensional index for our arrays.

\source{\begin{verbatim}
/**
 * Dining Philosophers Problem
 * in JLucid
 *
 * @author Serguei Mokhov, mokhov@cs.concordia.ca
 * @version $Revision: 1.10 $ $Date: 2005/03/02 02:57:31 $
 */

#funcdecl

int getIninitalRandomState();
boolean chew(int);
boolean brainstormIdea(int);


#JLUCID

/*
 * Assume 8 philosophers and two states.
 * States: 2 - eating, 1 - thinking
 * Forks are either available or not; hence, 2 states as well.
 */
PHILOSOPHERS[8] @states 2
where
    dimension states;

    // Initialize all forks
    FORKS[8] @availability 2
    where
        dimension availability;

        FORKS[I] = getIninitalRandomState();

        I @d 8
        where
            dimension d;
            I = 1 fby.d (I - 1);
        end;
    end;

    /*
     * Run the actual algorithm.
     * NOTE: in this implementation the computation
     * never terminates (normally). It is an infinite loop.
     */
    PHILOSOPHERS[I] =
        if(#states == 1) then
            eat(I) @states 2

            eat(I) =
                getForks(I) && chew(I);

            getForks(I) = g(l, r)
            where
                l = FORK[I] @availability 1;
                r = FORK[I] @availability 1;
            end;
        else
            think(I) @states 1

            think(I) =
                putForks(I) && brainstormIdea(I);

            putForks(I) = p(l, r)
            where
                l = FORK[I] @availability 2;
                r = FORK[I] @availability 2;
            end;
        fi;

    I @d 8
    where
        dimension d;
        I = 1 fby.d (I - 1);
    end;
end;

#JAVA

int getIninitalRandomState()
{
    // Either 1 or 2
    return new Random().nextInt(2) + 1;
}

boolean chew(int i)
{
    try
    {
        System.out.println("Philo " + i + " is chewing smth tasty now.");
        sleep(new Random().nextInt(i * 1200));
        System.out.println("Philo " + i + " finished chewing.");
        return true;
    }
    catch(InterruptedException e)
    {
        return false;
    }
}

boolean brainstormIdea(int i)
{
    try
    {
        System.out.println("Philo " + i + " is heavily thinking now.");
        sleep(new Random().nextInt(i * 1200));
        System.out.println("Philo " + i + " finished thinking.");
        return true;
    }
    catch(InterruptedException e)
    {
        return false;
    }
}
\end{verbatim}
}

%
%

\clearpage
\subsection{Fast Fourier Transform}
\index{JLucid!Examples -- FFT}
\index{FFT}
\index{Fast Fourier Transform}
\index{Examples!FFT}
\index{Problems!FFT}

This is an example on how one would compute Fast Fourier Transform (FFT)
in the {\gipsy} for an array of double values.
This is straightforward in {\lucid} because it's deterministic with
plenty of parallelism. A JLucid program implementing FFT is in \xs{sect:fft-jlucid}. 
The algorithm is based on the Java algorithm implemented in {\marf}\index{MARF!FFT}
\cite{marf,numericalrecipes,dspdimension}, a code fragment of which is
in \xs{sect:fft-marf}, 
originally written by Stephen Sinclair.
The JLucid version omits the imaginary part of the transform,
but it would not be hard to add it.


\subsubsection{Fast Fourier Transform\index{FFT} in {\jlucid}.}
\label{sect:fft-jlucid}

\source{\begin{verbatim}
/*
 * FFT implementation in JLucid.
 * Serguei Mokhov
 * $Id: fft.ipl,v 1.2 2005/08/13 01:37:23 mokhov Exp $
 */

#funcdecl
double sin(double);
double pi();

#JAVA

double sin(double pdValue)
{
    return Math.sin(pdValue);
}

double pi()
{
    return Math.PI;
}

#JLUCID

A
where
    // A is an array of 9 FFT values with a
    // normal FFT applied to the array below.

    A = fft([1, 2, 3, 4, 6, 7, 8, 9], 9, 1);

    fft(inputValues, length, sign) = fftValues
        where
            fftValues = apply(length, reverse(length, inputValues), sign);

            apply(len, coeffs, direction) = coeffs @.s (N - 1)
                where
                    dimension s;
                    
                    N = 2 * len;
                    mmax = (2 fby.s istep) upon(mmax < N);

                    coeffs[J / 2] = coeffs[I / 2] - tempr;
                    coeffs[I / 2] = coeffs[I / 2] + tempr;

                    where
                        istep = mmax fby.s (istep) * 2;
                        
                        M @.m mmax
                        where
                            dimension m;

                            M = (0 fby.m (M + 2)) upon (M < mmax);

                            tempr = wr * coeffs[J / 2] - wi * coeffs[J / 2];

                            J = I + mmax;

                            wr = 1.0 fby.m ((wtemp = wr) * wpr - wi * wpi + wr);
                            wi = 0.0 fby.m (wi * wpr + wtemp * wpi + wi);
    
                            where
                                dimension i;
                                I = (M fby.i (I + istep)) upon (I < N);
                                theta = (direction * 2 * pi()) / mmax;
                                wtemp = sin(0.5 * theta);
                                wpr   = -2.0 * wtemp * wtemp;
                                wpi   = sin(theta);
                            end;
                        end;
                    end;
                end;
    
            // Binary reversion
            reverse(l, vals) = out @.i length
                where
                    dimension i;
                    out[t] = vals[#.i] @ (#.i + 1) @.bit maxbits(length);
                    where
                        dimension bit;

                        t = 0 fby.bit ((t * 2) | (n & 1));
                        n = #i fby.bit (n / 2);
                    end;
                end;
        
            // Determine max number of bits
            maxbits(len) = (mbits - 1) @.m 16
                where
                    dimension m;

                    mbits = ( 0 fby.m (mbits + 1) ) upon (mbits < 16 && n != 0);
                    n = len fby.m (n / 2);
                end;
        end;
end;

// EOF
\end{verbatim}
}


\subsubsection{Fast Fourier Transform\index{FFT} code fragment in {\java} from {\marf}\index{MARF!FFT}.}
\label{sect:fft-marf}

\source{\begin{verbatim}
...
        /**
         * <p>FFT algorithm, translated from "Numerical Recipes in C++" that
         * implements the Fast Fourier Transform, which performs a discrete Fourier transform
         * in O(n*log(n)).</p>
         *
         * @param InputReal InputReal is real part of input array
         * @param InputImag InputImag is imaginary part of input array
         * @param OutputReal OutputReal is real part of output array
         * @param OutputImag OutputImag is imaginary part of output array
         * @param direction Direction is 1 for normal FFT, -1 for inverse FFT
         * @throws MathException if the sizes or direction are wrong
         */
        public static final void doFFT
        (
            final double[] InputReal,
            double[] InputImag,
            double[] OutputReal,
            double[] OutputImag,
            int direction
        )
        throws MathException
        {
            // Ensure input length is a power of two
            int length = InputReal.length;

            if((length < 1) | ((length & (length - 1)) != 0))
                throw new MathException("Length of input (" + length + ") is not a power of 2.");

            if((direction != 1) && (direction != -1))
                throw new MathException("Bad direction specified.  Should be 1 or -1.");

            if(OutputReal.length < InputReal.length)
                throw new MathException("Output length (" + OutputReal.length + ") < Input length (" + InputReal.length + ")");

            // Determine max number of bits
            int maxbits, n = length;

            for(maxbits = 0; maxbits < 16; maxbits++)
            {
                if(n == 0) break;
                n /= 2;
            }

            maxbits -= 1;

            // Binary reversion & interlace result real/imaginary
            int i, t, bit;

            for(i = 0; i < length; i++)
            {
                t = 0;
                n = i;

                for(bit = 0; bit < maxbits; bit++)
                {
                    t = (t * 2) | (n & 1);
                    n /= 2;
                }

                OutputReal[t] = InputReal[i];
                OutputImag[t] = InputImag[i];
            }

            // put it all back together (Danielson-Lanczos butterfly)
            int mmax = 2, istep, j, m;                                // counters
            double theta, wtemp, wpr, wr, wpi, wi, tempr, tempi;    // trigonometric recurrences

            n = length * 2;

            while(mmax < n)
            {
                istep = mmax * 2;
                theta = (direction * 2 * Math.PI) / mmax;
                wtemp = Math.sin(0.5 * theta);
                wpr   = -2.0 * wtemp * wtemp;
                wpi   = Math.sin(theta);
                wr    = 1.0;
                wi    = 0.0;

                for(m = 0; m < mmax; m += 2)
                {
                    for(i = m; i < n; i += istep)
                    {
                        j = i + mmax;
                        tempr = wr * OutputReal[j / 2] - wi * OutputImag[j / 2];
                        tempi = wr * OutputImag[j / 2] + wi * OutputReal[j / 2];

                        OutputReal[j / 2] = OutputReal[i / 2] - tempr;
                        OutputImag[j / 2] = OutputImag[i / 2] - tempi;

                        OutputReal[i / 2] += tempr;
                        OutputImag[i / 2] += tempi;
                    }

                    wr = (wtemp = wr) * wpr - wi * wpi + wr;
                    wi = wi * wpr + wtemp * wpi + wi;
                }

                mmax = istep;
            }
        }
...
\end{verbatim}
}

%
%

\clearpage
\subsection{Moving Car}
\index{Objective Lucid!Examples -- Moving Car}
\index{Examples!Moving Car}
\index{Problems!Moving Car}

A less contrived example of an {\olucid} program is presented in \xf{fig:car-oop}.
This is an example where a \api{Car} object changes with time.
Eliminating $S$, and ignoring the print call, we have have:

\sourcefloat
	{\begin{verbatim}
#typedecl
Car;

#JAVA
public class Car
{
    public int x = 0;

    public float speed;
    public float speeddrop;
    public float fuel;
    public float fueldrainrate;

    public Car()
    {
        // Assume initially car was already moving.
        speed = 100.0; fuel = 40.5;
        fueldrainrate = 0.018; speeddrop = 0.1;
    }

    // Move by a number of steps assuming constant speed
    // and decelerate when ran out of fuel.
    public Car move(int steps)
    {
        if(fuel > 0)
        {
            fuel -= fueldrainrate * speed * steps;
            x += steps;
        }
        else if(speed > 0)
        {
            x += steps;
            speed -= speeddrop * steps;
        }

        return this;
    }

    public void printCarState()
    {
        System.out.println
        (
            "Speed: " + speed + ", fuel: " + fuel +
            ", drain: " + fueldrainrate + ", x: " + x +
            ", speeddrop: " + speeddrop
        );
    }
}

#OBJECTIVELUCID

(C @.time 15).printCarState()
where
    C = Car() fby.time S;
    S = C.move(#time);
end;
\end{verbatim}
}
	{fig:car-oop}
	{\small Objective Lucid example of a Car object that changes in time.}

\noindent
\fbox
{
	\begin{minipage}[t]{0.97\textwidth}
		\texttt{C @.time 15 where}
		
		\qquad\texttt{C = Car() fby.time C.move(\#.time)}
	\end{minipage}
}\\

\noindent
Using the definition of \lucidop{fby} gives:

\noindent
\fbox
{
	\begin{minipage}[t]{0.97\textwidth}
		\texttt{C @.time 15}
		
		\qquad\texttt{= (Car() fby.time C.move(\#.time)) @.time 15}
		
		\qquad\texttt{= if 15 <= 0 then Car() else (C.move(\#.time)) @.time (15 - 1)}
		
		\qquad\texttt{= C.move(14)}
	\end{minipage}
}\\

\noindent
Our intention is that \lucidop{fby} will give the sequence:

\noindent
\fbox
{
	\begin{minipage}[t]{0.97\textwidth}
		\texttt{Car()  Car.move(1)  Car.move(2)  ...  Car.move(15)}
	\end{minipage}
}\\

This will work as follows. When one generates a demand for \texttt{C.move(15)}
it's not satisfied until \texttt{C.move(14)} is until \texttt{C.move(13)}
is ... until \texttt{C.move(1)} is until \texttt{Car()}, so
it recurses back and finally the \texttt{Car()} object instance gets constructed,
and then the demands flow from 1 to 15 and the instance already exists.

The car also does not accelerate indefinitely. It moves until it has enough fuel,
else it returns the car object with its members unmodified. The drop of speed is
also in place when
fuel is depleted.

To further illustrate this idea let's take the existing example
of a simpler problem of natural numbers presented in \xf{fig:nat2}
and convert it into {\olucid} as in \xf{fig:Nat42}. First, we will present the eduction
tree of the natural numbers problem (see \xf{fig:exegraphn}, a corrected version
of the one produced by Paquet in \cite{paquetThesis}) and
then transmute it into the eduction tree of the execution of the equivalent
Objective Lucid propgram, as shown in \xf{fig:exegraphobj}. The program
in \xf{fig:Nat42} exhibits the same properties as the Car example, so
the eduction tree will be similar but will take more space. The important
aspect here is to illustrate the difference between demands for STs and their lazy
execution (which is italisized, e.g. {\it N.inc()}); thus, the actual invocation of a ST method happens at a later time
after the demand is made so we avoid not having called constructor prior
execution of an instance method. In the eduction trees the normal arrows
correspond to demands made for expressions and the bullet arrows correspond
to the result of evaluation of the demands, which are also \textbf{\textit{bold and italic}}.
In the Objective Lucid eduction tree object instance is denoted as
{\it ClassName:MemberName:value} and the \{d:X\} presents the context of evaluation.
The result of evaluation of the Objective Lucid variant is said to be {\it true}
because, as previously defined, \api{void} methods are mapped to return {\it true}
and the last expression bit that is evaluated here is the \texttt{print()}
method call of the instance of a \api{Nat32} class, which returns \api{void}.

\begin{figure}
	\begin{centering}
	\includegraphics[totalheight=\textheight]{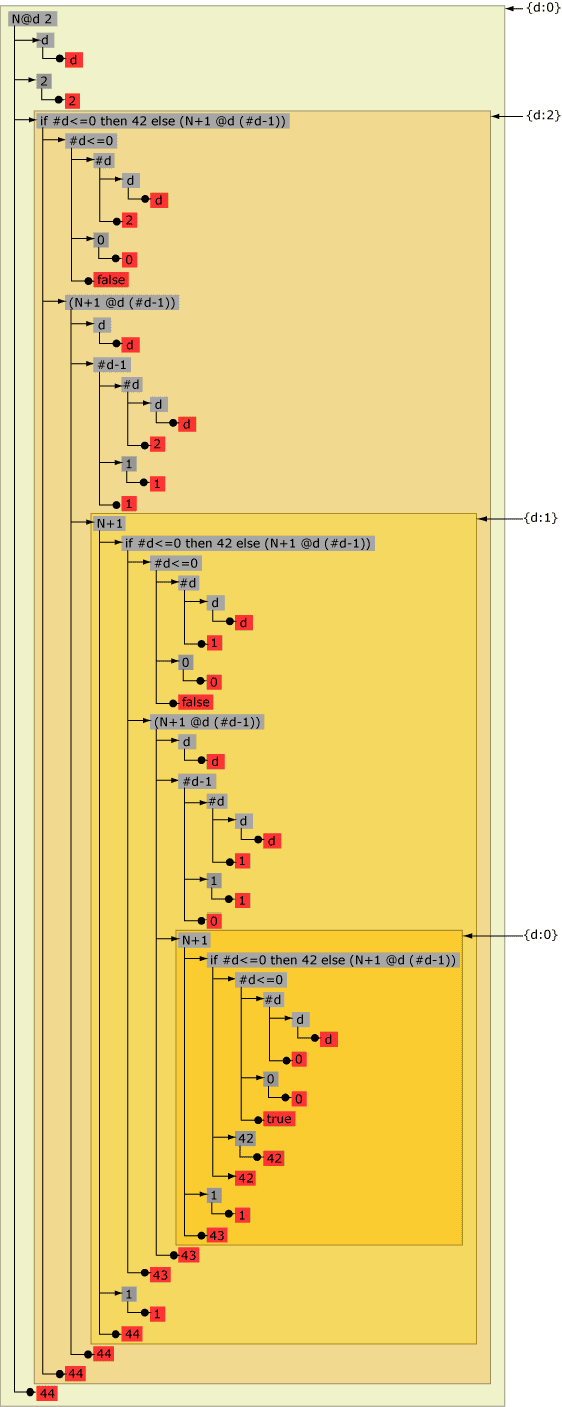}
	\caption{Eduction Tree for the Natural Numbers Problem.}
	\label{fig:exegraphn}
	\end{centering}
\end{figure}

\sourcefloat
	{\begin{verbatim}
#typedecl
Nat42;

#JAVA
class Nat42
{
    private int n;

    public Nat42()
    {
        n = 42;
    }

    public Nat42 inc()
    {
        n++;
        return this;
    }

    public void print()
    {
        System.out.println("n = " + n);
    }
}

#OBJECTIVELUCID

(N @.d 2).print[d]()
where
    dimension d;
    N = Nat42[d]() fby.d N.inc[d]();
end
\end{verbatim}
}
	{fig:Nat42}
	{\small The Natural Numbers Problem in {\olucid}.}

\begin{figure}
	\begin{centering}
	\includegraphics[totalheight=\textheight]{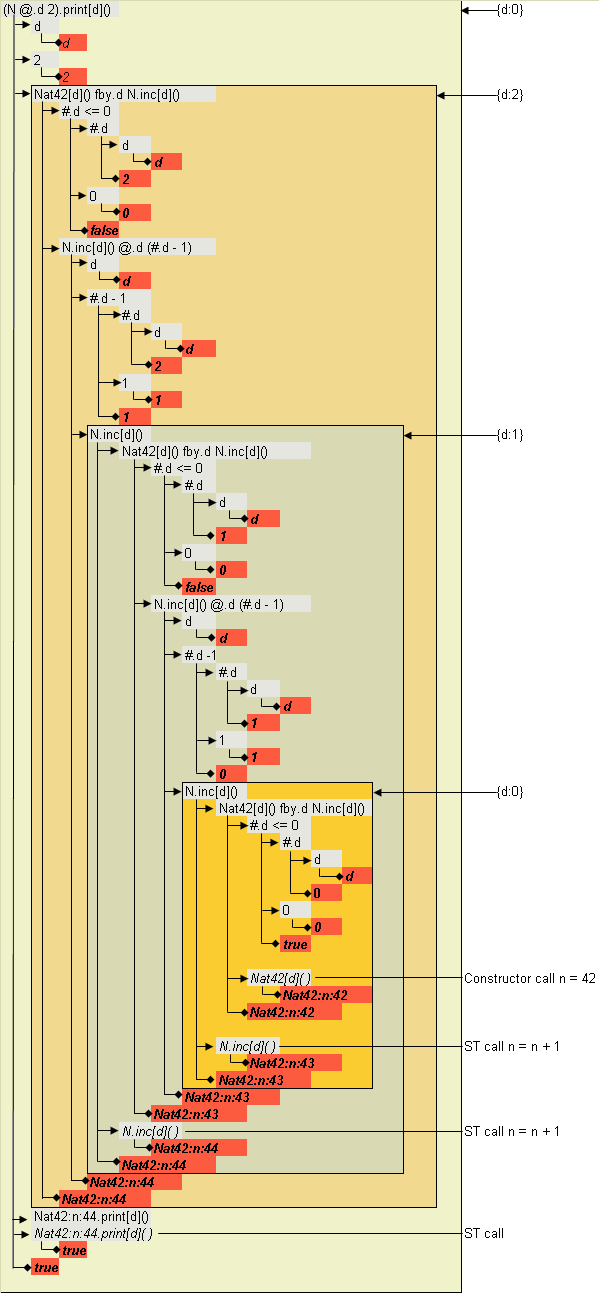}
	\caption{Eduction Tree for the Natural Numbers Problem in {\olucid}.}
	\label{fig:exegraphobj}
	\end{centering}
\end{figure}

\clearpage

%
%

\subsection{Game of Life}
\index{Examples!Game of Life}
\index{Problems!Game of Life}

The Game of Life \cite{conwaylife} would make a good benchmark for
the {\gipl}. Life takes place on a 2D grid and evolves in time, so it's a 3D problem.
The value of a cell at time $T+1$ depends on the value of the cell and
its 8 neighbours at time $T$.  Thus, there is a high branching factor and
the IVW will get plenty of exercise.
Peter Grogono wrote a version in {\haskell}, which is functional and lazy but is not
concurrent and does not have an IVW. The author of this work made a version
in {\ilucid}.
In \xf{fig:life-haskell} is the top-level function.
The Game of Life program is included in the test suite as a good elaborate test case, but
this work does not address any of the performance and efficiency issues
related to the execution and wareshousing, so no measurements have been
done two compare the efficiency of the program with and without the
warehouse nor with the Haskell program.

\sourcefloat
    {\begin{verbatim}
life = evolve T initial (conway life) where
    initial = F(\i ->
        if val Y i == 0 && 0 <= val X i && val X i < 5 then 1 else 0)

    conway v = F(\i ->
        let neighbours v =
            ev v (n i) + ev v (ne i) + ev v (e i) + ev v (se i) +
            ev v (s i) + ev v (sw i) + ev v (w i) + ev v (nw i) in
            b2i(neighbours v == 3 || ev v i == 1 && neighbours v == 2))

    evolve d s e = F(\i ->
        if val d i == 0 then ev s i else ev e (prev d i))

    b2i b = if b then 1 else 0

    n i = F(...)\end{verbatim}
}
    {fig:life-haskell}
    {The Life in {\haskell}.}

\sourcefloat
    {\begin{verbatim}
#INDEXICALLUCID

life = evolve(T, initial(T), conway(life, T))
where
    dimension T;

    evolve(d, u, v) = u fby.d v;

    initial(d) = 
        if(Y == 0 && 0 <= X && X < 5) then 1 else 0
        where
            X = 0 fby.d X + 1;
            Y = 0 fby.d Y + 1;
        end;
        
    conway(d, v) = b2i(neighbours == 3 || (v == 1 && neighbours == 2))
    where
        neighbours = n(d) + ne(d) + e(d) + se(d) + s(d) + sw(d) + w(d) + nw(d);
        where
            n(d)  = v @.(d - 5);
            ne(d) = v @.(d - 4);
            e(d)  = v @.(d + 1);
            se(d) = v @.(d + 6);
            s(d)  = v @.(d + 5);
            sw(d) = v @.(d + 4);
            w(d)  = v @.(d - 1);
            nw(d) = v @.(d - 6);
        end;
        
        b2i(b) = if(b) then 1 else 0;
    end;
end;
\end{verbatim}
}
    {fig:life}
    {The Life in {\ilucid}.}

\noindent
Explanations:

\begin{itemize}
\item
$\mathit{evolve(d,u,v)}$ allows a value to evolve in the dimension $d$.  The
first value of the stream is given by $u$ and subsequent values by $v$.

\item
$\mathit{initial(d)}$ defines the initial configuration
(five ones in the row 0, zeroes everywhere else in the matrix 5-by-5).

\item
$\mathit{conway(d, v)}$ computes the successor of state $v$.  The functions $n$, $ne$, $e$,
$se$, $s$, $sw$, $w$, and $nw$ are ``navigators'' that find values of neighbours.

\item
$\mathit{b2i(d)}$ converts a Boolean to integer to decide the new value of
an entity.
\end{itemize}

\clearpage

%
%

\section{Summary}

There were many tests developed and exercised for the {\gipsy}.
This section attempted to show the reader the most representative
ones and how the Regression Tests Suite works in the {\gipsy} for the most
modules of {\gipc} and {\gee} and how {\junit} is applied to make it possible and maintainable.
Now, every new module added to the {\gipsy} system will have to have
a corresponding unit and/or regression test (or several tests) exercising
most of the features of this module added.



\chapter{Conclusion}
\label{chapt:conclusion}

To conclude, it is believed {\gipsy} is well off the ground and is steadily
getting ready for its first large public release to the research
community. It is becoming a lot more usable not only
by a small circle of GIPSY developers, but also by
scientists and researchers from other research groups.
Preliminary testing (see \xc{chapt:testing}) and results (\xs{sect:results})
give confidence in the success of an
important step for the {\gipsy} in the are
of flexible hybrid intensional-imperative
programming.
To summarize, the newly introduced
features for the innovative intensional research
platform {\gipsy} are a valuable
asset allowing us to release GIPSY to the
masses and a new release will be made at
the SourceForge.net at \url{http://sf.net/projects/sfgipsy}
circa the end of December 2005 - January 2006.


\section{Results}
\label{sect:results}
\index{Results}

\subsection{Experiments}

The experiments conducted on the {\gipsy}
research platform were primarily design, development,
and testing of hybrid programming paradigms
by fusing together intensional and imperative
languages. For test experiments please
refer to \xc{chapt:testing}.

\subsection{Interpretation of Results}

After extensive testing of the design
and implementation of ideas presented
in \xc{chapt:methodology} we can see
an enhanced, more flexible
{\gipsy} system taking off the ground.
Most of regression tests pass for the
developed sample programs with known
errors and failures.


\section{Discussions and Limitations}
\label{sect:limitations}

\subsection{Lack of Hybrid Intensional-Imperative Semantics Proofs}

The semantics for the GIPSY Type System was not defined
and the one of {\jlucid} and {\olucid} was not
formally proven to be correct.

\subsection{Genuine Imperative Compilers}

The most serious limitation of the
current implementation of the hybrid
paradigm is that there are no genuine
imperative GIPSY compilers. The Java
wrapper compiler classes merely resort
to the external tools from the library
of enumerated tools. This makes overall
error checking and reporting cumbersome.
Additionally,
this slows down the compilation process.

\subsection{Cross-Language Data Type Mapping}

When implementing other imperative language
compilers than {\java}, or a genuine compiler
for {\java}, a special mapping has to be
explicitly established in the form of \api{TypeMap}.
We can avoid this for {\C}/{\cpp} with the JNI \cite{jni},
but not for other popular languages.

\subsection{Dimension Index Overflow}

While this limitation is not directly related to the main
topics of this thesis, it has to be mentioned.
In the current implementation of the dimension type in all
{\lucid} variants is done as a simple Java integer, and as such,
is finite. Thus, incorrectly written Lucid programs or programs
that may require high dimension values may overflow the dimension
index rendering execution of the program incorrect. This limitation
is not handled by the {\gee} nor constrained in the operational
semantics of {\lucid}.

\subsection{Hybrid-DFG Integration}

This thesis does not address
placement, rendering, and integration
of the hybrid {\AST} nodes into {\dfg}s.

\subsection{Dealing With Side Effects and Abrupt Termination}

As of this implementation, {\gee} has very limited control over what's
happening inside the STs in terms side effects, exceptions,
non-termination, etc. in the Java (or other imperative language) code causing
it to exit prematurely or to hang. Likewise, we cannot do warehousing
of non-immutable STs due to the side effects, i.e. when the same arguments are given
to an ST may yield a different result at different invocations.
This is serious aspect, which is related to the development of any
future semantics of the hybrid programming languages and deserves
a separate publication.

\subsection{Imperative Function Overloading}

It is an error to write the following:

\noindent
\fbox
{
	\begin{minipage}[t]{0.97\textwidth}
		\texttt{\#funcdecl}

		\texttt{int foo(int);}

		\texttt{int foo(float);}

		\texttt{...}
	\end{minipage}
}\\

\noindent
but it shouldn't be. This is an error in the
sense that only the last declaration is retained
due to the way function identifiers are handled,
so no function overloading at this moment is
officially supported.
The issue of dealing with the semantics of a type system
in which this is possible, especially if we support multiple
imperative PLs, where each may have potentially its own type system
or even paradigm is complex. However, this feature is nice to have and some practical
aspects can be implemented, which will be a research topic on its own.

\subsection{Cross-Imperative Language Calls}

Normally, an ST written say in \codesegment{JAVA} cannot
call another ST in say \codesegment{C}. This limitation
is that only the intensional part can make calls to
the imperative functions. This eliminates the need
to keep the type mappings between all possible combinations of
the imperative languages and semantics associated with this.

However, depending on the language, procedures written
in the same language can possibly communicate by calling
each other. E.g. in {\java}, defining free members and
passing state between free functions is possible as nothing
is done to prevent this.

\noindent
\fbox
{
	\begin{minipage}[t]{0.97\textwidth}
		\texttt{\#JAVA}

		\texttt{int i;}\\

		\texttt{int foo() \{}

		\qquad\texttt{return i + 1;}

		\texttt{\}}\\

		\texttt{int bar() \{}

		\qquad\texttt{i++;}

		\qquad\texttt{return foo();}

		\texttt{\}}
	\end{minipage}
}\\

This is based on the knowledge about the internal implementation i.e.
the ``\texttt{int i;}'' bit will also be wrapped in the class, so it'd
be legal to have it from the {\java}'s point of view;
however, is considered to be a kludge and non-portable feature.
To be on the safer side, the STs like that should be written assuming
no knowledge of internal state for communication is available.

\subsection{Security}

{\jlucid}, {\olucid}, and {\gicf} opened up doors for very
flexible use of external languages and resources as a part
of intensional computation. Unfortunately, there are security
considerations to deal with when embedding a vulnerable unsigned
code from possibly untrusted remote location and then propagate
it to all the workers participating in computation can result
resulting either gaining some unwanted  privileges to
the attackers or DDoS.



\chapter{Future Work}
\label{chapt:future-work}

The future work to take on will focus in the following
areas to either address the limitations outlined in \xs{sect:limitations} or to introduce
new features, not necessarily all related to the topics of this
thesis.

\begin{itemize}
\item
Integration of the Demand Migration System (DMS) \cite{vas05}.
\item
Formal semantic verification from {\ilucid} through {\olucid}.
\item
Placement of hybrid nodes into DFGs.
\item
Security.
\item
Trial C compiler with JNI.
\item
Fully Explore Array Properties.
\item
Genuine imperative compilers in {\gicf}.
\item
Introduction functional language compilers.
\item
Visualization and control of communication patterns and load balancing.
\item
Target Host Compilation.
\item
Java wrapper for the DFG Editor of Yimin Ding.
\end{itemize}

\section{Formal Verification of Semantic Rules and the GIPSY Type System}

One needs to formally conduct verification proofs of the semantic rules
from {\ilucid} to {\olucid} in PVS\index{PVS} or Isabelle\index{Isabelle}, so this project
can be undertaken in the near future and the work on it has
already began. Specifically, a relation to the semantic of objects
and Java's operational semantics has to be made.
Likewise, the semantics of the newly introduced
GIPSY type system has to be formally defined.

\section{Dealing with Data Flow Graphs in Hybrid Programming}

This thesis did not deal with the way on how to augment
\api{DFGAnalyzer} and \api{DFGGenerator} to support
hybrid GIPSY programs. This can be addressed
by adding an unexpandable imperative DFG node to the graph.
To make it more useful, i.e. expandable and so it's
possible to generate the GIPSY code off it or reverse it,
would require having the genuine compilers as in \xs{sect:genuine-compilers}
for imperative languages, which is far from trivial.

\section{Security}
\index{GIPSY!Security}

Security is a substantial concern in distributed computing.
The great flexibility provided by embedded Java in {\jlucid} (and later
in {\olucid}) can be misused and be a source of security breaches
or DDoS attacks (e.g., due to explicit oversynchronization using Java's
synchronization primitives explicitly). Thus, the follow-up work in this direction
would include malicious code detection in embedding and distributing
as well as explicit synchronization points so that there are no
deadlocks and DDoS potential is reduced. This concern touches the compiler ({\gipc}), the
Generator-Worker architecture, the GIPSY Server, and the GIPSY
Screen Saver components of the {\gipsy} system.

\section{Implementation of the C Compiler in GICF}

An methodology of implementing a C compiler, and therefore, C CPs and STs
has been devised, but never implemented, so in the future a C compiler
will be implemented as a part of {\gicf} with the JNI\index{JNI} \cite{jni}.

\section{Fully Explore Array Properties}

The arrays in {\jlucid}, {\olucid}, and their generalization in {\gicf}
requries further exploration and formalization and mapping of the GIPSY
arrays to their native equivalents.

\section{Genuine Imperative and Functional Language Compilers}
\label{sect:genuine-compilers}

Future work in this area is to focus on writing our
genuine compilers for the mentioned imperative languages
and extending support for more imperative and functional
languages (e.g. {\lisp}, {\scheme}, or {\haskell}) and make it as much automated as possible.

\section{Visualization and Control of Communication Patterns and Load Balancing}

It is proposed to have a ``3D editor'' within {\ripe}'s \api{DemandMonitor}
that will render in 3D space the current communication patterns of a GIPSY
program in execution or replay it back and allow the user visually to
redistribute demands if they go off balance between workers. A kind of
virtual 3D remote control with a mini expert system, an input from which can be used to teach
the planning, caching, and load-balancing algorithms to perform efficiently
next time a similar GIPSY application is run.

\section{Target Host Compilation}

This has to do with enabling the {\gee} to deliver
the ST source code around and compile it on the target
host instead of sending a pre-compiled version of the STs.
This is an experimental feature can be useful and dangerous
and requires a lot of research.

%
%

\section{The GIPSY Screen Saver}
\index{GIPSY!Screen Saver}

This is a sample implementation of a worker, outlined in \xs{sect:worker},
would represent an application for a PC as a way to contribute to
a GIPSY program execution. Three sample implementations of screen saver workers
exist one for Windows, one for Linux and one for MacOS X.

%
%

\section{The GIPSY Server}
\label{sect:gipsy-server}
\index{GIPSY!Server}

A so-called ``GIPSY server'' will be implemented to be able to
serve intensional or otherwise requests primarily through the HTTP protocol,
thus acting like a mini-GIPSY intensional web server. It would accept request
from remote clients via HTTP or local clients via command line and be the 
starting point of computation (an intensional computation resource) available
to all those who have no resources to set up GIPSY. This is not duplicate any
of the {\dms} \cite{vas05} nor it is a part of {\ripe}, as it is primarily non-interactive
and runs on the background.



%
%


\addcontentsline{toc}{chapter}{Bibliography}
\label{chapt:bibliography}
\bibliography{mcompsci-mokhov-thesis}
\bibliographystyle{alpha}



\addcontentsline{toc}{chapter}{Appendix}
\appendix

%
%

\chapter{Definitions and Abbreviations}
\label{appdx:defs-and-abbrs}

\section{Abbreviations}
\label{appdx:abbrs}

\begin{itemize}

\item
AST - Abstract Syntax Tree\index{AST}

\item
COM - Component Object Model

\item
{\corba} - Common Object Requester Broker Architecture

\item
CLP - Cluster-Level Parallelism

\item
CP - Communication Procedure, \xs{sect:cp}

\item
CVS\index{CVS} - Concurrent Versions System

\item
DCOM - Distributed COM

\item
DDoS - Distributed Denial of Service (attack).

\item
FFT - Fast Fourier Transform\index{FFT}

\item
FTP - File Transfer Protocol\index{FTP}

\item
{\dpr} - Demand Propagation Resource, \xs{sect:dpr}, \cite{gipsy, gipsy2005}

\item
{\gee} - General Eduction Engine

\item
{\geer} - GEE Resources, \xs{sect:gipsy-program}

\item
{\gipc} - General Intensional Program Compiler, \xf{fig:gipsy-general}, \cite{gipsy, gipsy2005}

\item
{\gipl} - General Intensional Programming Language, \cite{paquetThesis, gipsy, gipsy2005}

\item
{\gipsy} - General Intensional Programming System, \cite{gipsy, gipsy2005}

\item
{\glu} - Granular Lucid, \cite{glu1, glu2, paquetThesis}

\item
HTTP - Hyper-Text Transfer Protocol\index{HTTP}

\item
IDP - Intensional Demand Propagator, \xs{sect:worker}, \cite{gipsy, gipsy2005}

\item
IDS - Intensional Data-dependency Structure

\item
IP - Intensional Programming

\item
IPL - Intensional Programming Language
(e.g. {\gipl}, {\glu}, {\lucid}, {\ilucid}, {\jlucid}, {\tlucid}, {\olucid},
{\onyx} \cite{grogonoonyx2004})

\item
IVW - Intensional Value Warehouse, \xs{sect:worker}, \cite{gipsy, gipsy2005}

\item
JDK - Java Developer's Kit

\item
JNI\index{JNI} - Java Native Interface

\item
JRE\index{JRE} - Java Runtime Environment

\item
JSSE - Java Secure Socket Extension\index{JSSE}

\item
{\marf} - Modular Audio Recognition Framework \cite{marf}

\item
MPI - Message Passing Interface\index{MPI}

\item
NCP - Native Communication Procedure

\item
NST - Native Sequential Thread

\item
NUMA - Non-Uniform Memory Access\index{NUMA}

\item
PVM - Parallel Virtual Memory System

\item
RFE - Ripe Function Executor, \xs{sect:worker}, \cite{gipsy, gipsy2005}

\item
{\rmi} - Remote Method Invocation

\item
{\rpc} - Remote Procedure Call

\item
{\sipl} - Specific IPL
(e.g. {\ilucid}, {\jlucid}, {\tlucid}, {\olucid}, {\onyx})

\item
SLP - Stream-Level Parallelism\index{SLP}

\item
ST - Sequential Thread, \xs{sect:st}

\item
TLP - Thread-Level Parallelism\index{TLP}

\item
TTS - Time To Solution\index{TTS}

\item
UMA - Uniform Memory Access\index{UMA}

\item
URI - Unified Resource Indentifier\index{URI}

\item
URL - Unified Resource Locatior\index{URL}

\end{itemize}


%
%

\chapter{Sequential Thread and Communication Procedure Interfaces}
\label{chapt:stcp}

In this section the actual definitions of the CP and ST interfaces,
an example of a generated wrapper class and a \api{Worker} are
presented.

%
%

\section{Sequential Thread Interface}
\label{sect:st-iface}
\index{Sequential Thread!Interface}
\index{Interfaces!Sequential Thread}

See \xf{fig:st-iface}.

\sourcefloat
	{\begin{verbatim}
package gipsy.interfaces;

import java.io.Serializable;
import java.lang.reflect.Method;


/**
 * <p>Sequential Thread represents a piece work to be done.
 * Has to extend Serializable for RMI, CORBA, COM+, Jini to work.
 * Runnable needed to run it in a separate thread.</p>
 *
 * $Id: ISequentialThread.java,v 1.13 2005/09/12 01:24:38 mokhov Exp $
 *
 * @version $Revision: 1.13 $
 * @author Serguei Mokhov, mokhov@cs.concordia.ca
 * @since Inception
 */
public interface ISequentialThread
extends Runnable, Serializable
{
    /**
     * Work-piece to be done.
     * @return WorkResult container
     */
    public WorkResult work();
    
    public WorkResult getWorkResult();
    public void setMethod(Method poSTMethod);
}

// EOF
\end{verbatim}
}
	{fig:st-iface}
	{Sequential Thread Interface.}

%
%

\section{Communication Procedure Interface}
\label{sect:cp-iface}
\index{Communication Procedure!Interface}
\index{Interfaces!Communication Procedure}

See \xf{fig:cp-iface}.

\sourcefloat
	{\begin{verbatim}
package gipsy.interfaces;
import gipsy.lang.GIPSYType;
import java.io.Serializable;

/**
 * <p>CommunicationProcedure represents the means of delivery of sequential threads.</p>
 * $Id: ICommunicationProcedure.java,v 1.11 2005/10/11 08:34:11 mokhov Exp $
 * @version $Revision: 1.11 $
 * @author Serguei Mokhov, mokhov@cs.concordia.ca
 * @since Inception
 * @see gipsy.interfaces.SequentialThread
 */
public interface ICommunicationProcedure
extends Serializable
{
    public GIPSYType getReturnType();
    public GIPSYType getParamType(final int piParamNumber);
    public GIPSYType[] getParamTypes();
    public void setReturnType(GIPSYType poType);
    public void setParamType(final int piParamNumber, GIPSYType poType);
    public void setParamTypes(GIPSYType[] paoTypes);
    public GIPSYType getParamType(String pstrLexeme);
    public GIPSYType getParamType(String pstrLexeme, String pstrID);
    public int getParamListSize();
    /**
     * Perform any initialization actions required.
     * @return status object of the result of send operation.
     * @throws CommunicationException in case of error
     */
    public CommunicationStatus init()
    throws CommunicationException;
    /**
     * Open a connection; whatever that means for a given protocol.
     * @return status object of the result of send operation.
     * @throws CommunicationException in case of error
     */
    public CommunicationStatus open()
    throws CommunicationException;
    /**
     * Close a connection; whatever that means for a given protocol.
     * @return status object of the result of send operation.
     * @throws CommunicationException in case of error
     */
    public CommunicationStatus close()
    throws CommunicationException;
    /**
     * Defines the means of sending data. Should be overridden by
     * a concrete implementation, such as JINI, COM, CORBA, etc.
     * @return status object of the result of send operation.
     * @throws CommunicationException in case of error
     */
    public CommunicationStatus send()
    throws CommunicationException;
    /**
     * Defines the means of receiving data. Should be overridden by
     * a concrete implementation, such as JINI, COM, CORBA, etc.
     * @return status object of the result of receive operation.
     * @throws CommunicationException in case of error
     */
    public CommunicationStatus receive()
    throws CommunicationException;
}
\end{verbatim}
}
	{fig:cp-iface}
	{Communication Procedure Interface.}

%
%

\section{Generated Sequential Thread Wrapper Class}
\label{sect:genst-long}
\index{Sequential Thread!Wrapper}

This is a more complete version of the generated wrapper
class for the code in \xf{fig:nat2java}.

\clearpage
\source{\begin{verbatim}
import java.util.Hashtable;
import java.util.Vector;

public class <filename>_<machine_name>_<timestamp>
implements gipsy.interfaces.ISequentialThread
{
    private OriginalSourceCodeInfo oOriginalSourceCodeInfo;

    /** 
     * Inner class with original source code information 
     */ 
    public class OriginalSourceCodeInfo
    {
        /** 
         * For debugging / monitoring; generated statically 
         */ 
        private String strOriginalSource = 
            "int getN(int piDimension)" +
            "{" +
            "    if(piDimension <= 0)" +
            "        return get42();" +
            "    else" +
            "        return getN(piDimension - 1) + 1;" +
            "}" +
            "" +
            "int get42()" +
            "{" +
            "    return 42;" +
            "}";

        /**  
         * Mapping to original source code position for error reporting 
         */ 
        private Hashtable oLineNumbers = new Hashtable();

        /** 
         * Body is filled in by the preprocessor statically 
         */ 
        public OriginalSourceCodeInfo()
        {
            Vector int_getN_int_piDimension = new Vector();

            // Start line and Length in lines
            int_getN_int_piDimension.add(new Integer(3));
            int_getN_int_piDimension.add(new Integer(7));

            this.oLineNumbers.put 
            (
                "int getN(int piDimension)",
                int_getN_int_piDimension
            );

            Vector int_get42 = new Vector();
            int_get42.add(new Integer(11));
            int_get42.add(new Integer(4));

            this.oLineNumbers.put
            (
                "int get42()",
                int_get42
            );
        }

        public Hashtable getLineNumbersHash()
        {
            return this.oLineNumbers;
        }

        public int getLineNumberForFunction(String pstrFunctionSignature)
        {
        }

        public int getFunctionSourceLength(String pstrFunctionSignature)
        {
        }

        public String toString()
        {
        }
    }

    /** 
     * Constructor 
     */ 
    public <filename>_<machine_name>_<timestamp>()
    {
        this.oOriginalSourceCodeInfo = new OriginalSourceCodeInfo();
    }

    public String toString()
    {
        return this.oOriginalSourceCodeInfo.toString();
    }

    /*
     * Implementation of the SequentialThread interface
     */

    // Body generated by the compiler
    public void run()
    {
        Payload oPayload = new Payload();
        oPayload.add("d", new Integer(42));

        work(oPayload);
    }

    // Body generated by the compiler statically
    public WorkResult work(Payload poPayload)
    {
        WorkResult oWorkresult = new WorkResult();
        oWorkresult.add(getN(poPayload.getVaueOf("d")));
        return oWorkResult;
    }

    /*
     * ------------
     * The below are generated off the source file nat2java.ipl
     * ------------
     */

    public static int getN(int piDimension)
    {
        if(piDimension <= 0)
            return get42();
        else
            return getN(piDimension - 1) + 1;
    }

    public static int get42()
    {
        return 42;
    }
}
\end{verbatim}
}
\clearpage

%
%

\section{Sample Worker's Implementation}
\label{sect:sample-worker}
\index{Worker!Implementation}

\source{\begin{verbatim}
package gipsy.wrappers;

//import gipsy.interfaces.SequentialThread;
import gipsy.interfaces.ICommunicationProcedure;
import gipsy.util.*;

import marf.util.BaseThread;

/**
 * Worker Class Definition
 * 
 * $Revision: 1.11 $ by $Author: mokhov $ on $Date: 2004/11/06 00:50:09 $
 *
 * @version $Revision: 1.11 $
 * @author Serguei Mokhov
 */
public class Worker extends BaseThread
{
    /**
     * Aggregation of sequential threads.
     */
    private Thread[] aoSequentialThreads = null;

    /**
     * Set of available communication procedures for different protocols.
     */
    private ICommunicationProcedure[] aoCommuncationProcedures = null;

    /**
     * Default settings.
     */
    public Worker()
    {
        super();
    }

    /**
     * Generate a demand.
     */
    public void demand()
    {
    }

    /**
     * Receive a result on a demand.
     */
    public void receive()
    {
    }

    /**
     * Perform computation.
     */
    public void work() throws GIPSYException
    {
        try
        {
            for(int i = 0; i < this.aoSequentialThreads.length; i++)
                this.aoSequentialThreads[i].start();
        }
        catch(NullPointerException e)
        {
            throw new GIPSYException
            (
                "Worker TID=" + getTID() +
                " did not have any sequential threads to work on."
            );
        }
    }

    /**
     * Stops worker thread.
     */
    public void stopWorker()
    {
    }

    /**
     * From Runnable interface, for TLP
     */
    public void run()
    {
        try
        {
            work();
        }
        catch(GIPSYException e)
        {
            System.err.println(e);
        }
    }
}

// EOF
\end{verbatim}
}


\chapter{Architectural Module Layout}
\label{chapt:arch-layout}

%
%




%
%

\section{GIPSY Java Packages and Directory Structure}
\label{sect:java-packages}
\index{Implementation!GIPSY Java Packages}
\index{Architecture!GIPSY Java Packages}
\index{Layout!GIPSY Java Packages}
\label{sect:dir-structure}
\index{Implementation!Directory Structure}
\index{Architecture!Directory Structure}
\index{Layout!Directory Structure}

Normally, a directory structure of a Java project corresponds
to the package naming; thus, the packages are named and
declared after the directories.
By the means of Java packages,
all the
classes within the project and external applications ``know''
how to identify and import the classes they intend to use.
A fully-qualified
class name includes all the packages starting from the ``root''
(the top-level directory of the hierarchy) all the way up to
the class itself, separated by a dot. The GIPSY Java packages
breakdown as of this writing corresponds to the \xf{fig:packages}.

\begin{figure}
	\begin{centering}
	\includegraphics[totalheight=8.5in,width=7in]{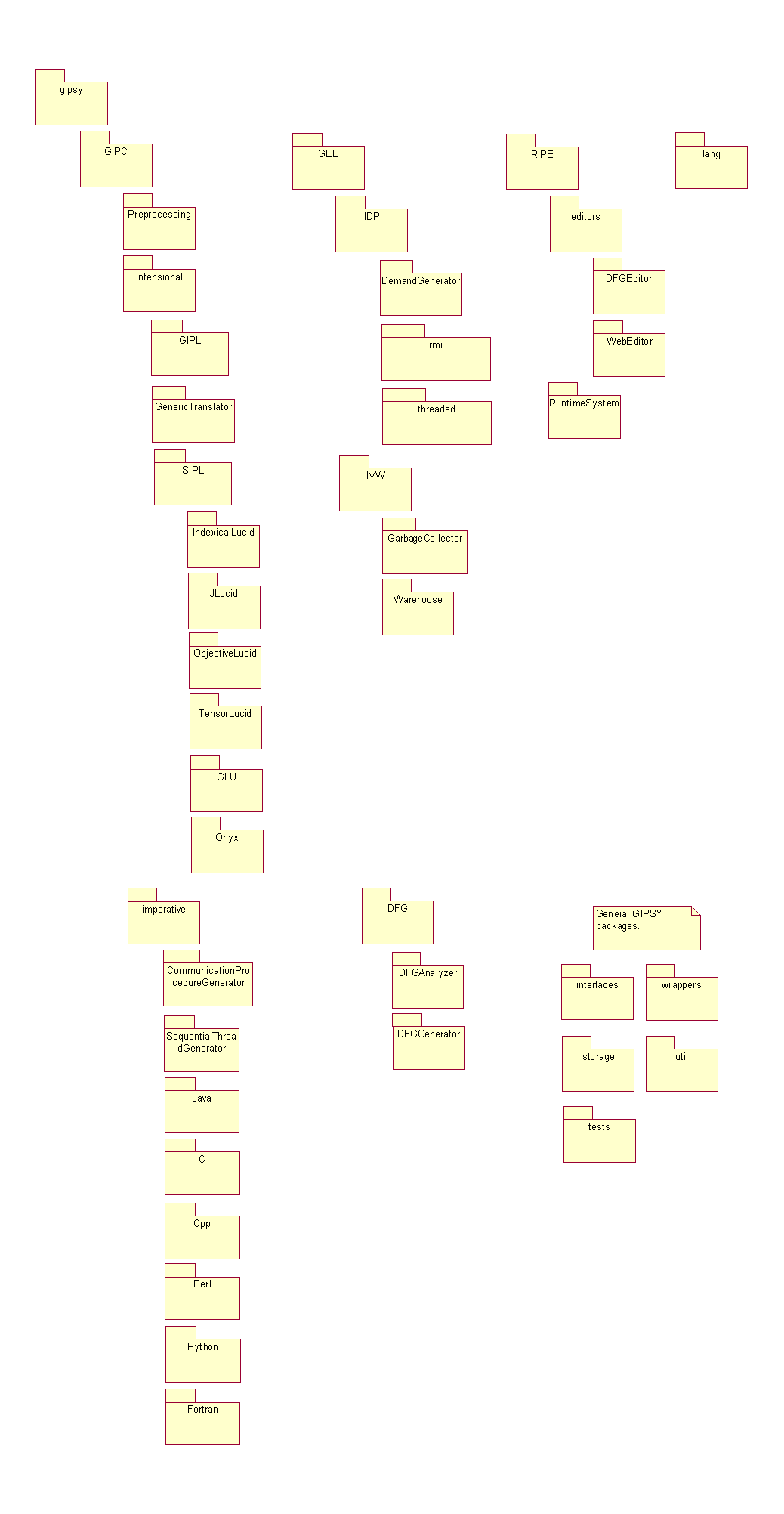}
	\caption{{\gipsy} Java Packages Hierarchy.}
	\label{fig:packages}
	\end{centering}
\end{figure}

The logical breakdown was performed in accordance with the
original conceptual design primarily produced by Joey Paquet
and further by Aihua Wu and Bo Lu, has been the primary source
of the hierarchy plus any exceptions and extensions that
various team members come up with or have been forced to
during implementation were taken into account.

The basic structure is as follows. The top root hierarchy is
logically the \api{gipsy} package. The major non-utility packages
under it, which come from the conceptual design, are \api{GIPC}, \api{GEE},
and \api{RIPE}. The major utility packages under \api{gipsy} that
are not present in the conceptual design are: \api{interfaces}
for most intermodule communication; \api{wrappers} for object
wrapping; \api{storage} for the serializable interface classes;
\api{util} for most common exceptions and utility modules (e.g. fast linked
list \cite{yimin04}); and \api{tests} for the Unit and Regression Testing Suites.

Under the \api{GIPC} package the major modules (to be discussed later in
this chapter) include \api{Preprocessing} for general GIPSY program preprocessing,
\api{intensional} and \api{imperative} language compilers and their necessary
followers (\api{GenericTranslator} for the former and \api{CommunicationProcedureGenerator}
and \api{SequentialThreadGenerator} for the latter). Then the \api{DFG} package for
Lucid-to-data-flow-graph and back generation.

The \api{GEE}'s main packages includes \api{IDP} for demand propagation and \api{IVW}
for caching and garbage collection.

Under \api{RIPE} we have interactive run-time editing and monitoring modules
that include textual editor, DFG editor, and the web-based editor.

%
%

\section{GIPSY Modules Packaging}
\label{sect:gipsy-modules}
\index{Implementation!GIPSY Modules Packaging}
\index{Architecture!GIPSY Modules Packaging}
\index{Layout!GIPSY Modules Packaging}

{\gipsy}'s major and minor modules are packaged
into a set of runnable \file{.jar} files and
distributed with wrapper scripts to be either
used as ordinary command line tools as a part
of GIPSY Development Kit or the \api{WebEditor}
web application. Different \file{.jar} files include
a subset of all GIPSY modules depending on the
need, e.g. {\gipc} includes GIPC-related classes
plus {\gee} as we allow to optionally invoke
{\gee} after successful compilation. {\ripe}, except itself,
needs both {\gipc} and {\gee}, whereas {\gee} does not at all
require presence of any other module. Thus, the GIPSY binary
distribution is broken down into five major \file{.jar} files
(notice, that these files do not include any external libraries GIPSY
references):

\begin{itemize}
\item \file{gipsy.jar} simply includes almost all of {\gipsy}.

\item \file{gipc.jar} should be used/distributed as a part of
so-called ``GIPSY Development Kit'' if someone wants to be able to compile
intensional programs and optionally run them.

\item \file{gee.jar} represents {\gipsy}'s non-interactive
run-time environment, the {\gee}. This can be distributed alone
to the hosts that only wish to run pre-compiled GIPSY programs
and have no development environment set up.

\item \file{ripe.jar} includes most of the interactive programming
environment of the {\gipsy} along with {\gipc} and {\gee}.

\item \file{Regression.jar} includes the Regression Testing application
plus all of {\gipc} and {\gee} as the most exercised modules for testing
as of this writing.
\end{itemize}

The \xt{tab:jar-modules} shows correspondence between the variety of modules and their
containment within a \file{.jar} file.

\begin{table}
\caption{{\small Correspondence of the GIPSY \file{.jar} files and the modules.}}
\begin{minipage}[b]{\textwidth}
\begin{center}
\begin{tabular}{|c|c|c|c|c|c|} \hline
Module / Jar   & \file{gipsy.jar} & \file{ripe.jar} & \file{gipc.jar} & \file{gee.jar} & \file{Regression.jar} \\ \hline\hline
GIPSY          & $\star$          &                 &                 &                &                       \\ \hline
GIPC           & $\star$          & $\star$         & $\star$         &                & $\star$               \\ \hline
RIPE           & $\star$          & $\star$         &                 &                &                       \\ \hline
GEE            & $\star$          & $\star$         & $\star$         & $\star$        & $\star$               \\ \hline
DFG/GIPC       & $\star$          & $\star$         & $\star$         &                & $\star$               \\ \hline
DFGEditor      & $\star$          & $\star$         &                 &                &                       \\ \hline
Regression     &                  &                 &                 &                & $\star$               \\ \hline
Interfaces     & $\star$          & $\star$         & $\star$         & $\star$        & $\star$               \\ \hline
WebEditor      &                  &                 &                 &                &                       \\ \hline
gipsy.lang     & $\star$          & $\star$         & $\star$         & $\star$        & $\star$               \\ \hline
gipsy.wrappers & $\star$          & $\star$         & $\star$         & $\star$        & $\star$               \\ \hline
gipsy.util     & $\star$          & $\star$         & $\star$         & $\star$        & $\star$               \\ \hline
gipsy.storage  & $\star$          & $\star$         & $\star$         & $\star$        & $\star$               \\ \hline
\end{tabular}
\end{center}
\end{minipage}
\label{tab:jar-modules}
\end{table}


\chapter{Grammar Generation Scripts for {\jlucid} and {\olucid}}
\label{chapt:grammar-scripts}

\section{\tool{jlucid.sh}}
\source{\begin{verbatim}
#!/bin/bash

strDate=`date`

cat <<GRAMMAR_TAIL
/*
 * Generated by jlucid.sh on $strDate
 */

/**
 * @since $strDate
 */
void embed() #EMBED : {}
{
    //<EMBED> <LPAREN> url() E() ( <COMMA> E() )* <RPAREN> <SEMICOLON>
    <EMBED> <LPAREN> url() <COMMA> <STRING_LITERAL> ( <COMMA> E() )* <RPAREN> <SEMICOLON>
}

/**
 * @since $strDate
 */
void array() #ARRAY : {}
{
    <LBRACKET> E() ( <COMMA> E() )* <RBRACKET>
}

/**
 * URL -> CHARACTER_LITERAL | STRING_LITERAL.
 * @since $strDate
 */
void url() #URL :
{
    Token oToken;
}
{
    (
          oToken = <CHARACTER_LITERAL>
        | oToken = <STRING_LITERAL>
    )
    {
        jjtThis.setImage(oToken.image);
    }
}

// EOF
GRAMMAR_TAIL

# EOF
\end{verbatim}
}

\section{\tool{JGIPL.sh}}
\source{\begin{verbatim}
#!/bin/bash

cat ../../GIPL/GIPL.jjt | \
    # Filter out unneeded stuff
    grep -v '// EOF' | \
    #grep -v 'import gipsy.GIPC.intensional.SimpleNode' | \
    # Fix package
    sed 's/intensional\.GIPL/intensional\.SIPL\.JLucid/g' | \
    # JLucid GIPL
    sed 's/GIPL/JGIPL/' | \
    sed 's/\/\/{EXTEND-E}/\/\/{EXTEND-E}\n\t\t| embed()/' | \
    sed 's/\/\/{EXTEND-FACTOR}/\/\/{EXTEND-FACTOR}\n\t| array()/' | \
    sed 's/<WHERE: "where">/<WHERE: "where">\n\t| <EMBED: "embed">/g' \
    > JGIPL.jjt

./jlucid.sh >> JGIPL.jjt

# EOF
\end{verbatim}
}

\section{\tool{JIndexicalLucid.sh}}
\source{\begin{verbatim}
#!/bin/bash

cat ../../SIPL/IndexicalLucid/IndexicalLucid.jjt | \
    # Filter out unneeded stuff
    grep -v '// EOF' | \
    #grep -v 'import gipsy.GIPC.intensional.SimpleNode' | \
    # Fix package
    sed 's/intensional\.SIPL\.IndexicalLucid/intensional\.SIPL\.JLucid/g' | \
    # JLucid Indexical
    sed 's/IndexicalLucid/JIndexicalLucid/' | \
    sed 's/\/\/{EXTEND-E}/\/\/{EXTEND-E}\n\t\t| embed()/' | \
    sed 's/\/\/{EXTEND-FACTOR}/\/\/{EXTEND-FACTOR}\n\t| array()/' | \
    sed 's/<WHERE: "where">/<WHERE: "where">\n\t| <EMBED: "embed">/g' \
    > JIndexicalLucid.jjt

./jlucid.sh >> JIndexicalLucid.jjt

# EOF
\end{verbatim}
}

\section{\tool{ObjectiveGIPL.sh}}
\source{\begin{verbatim}
#!/bin/bash

cat JGIPL.jjt | \
    # Filter out unneeded stuff
    grep -v '// EOF' | \
    # Fix package
    sed 's/intensional\.SIPL\.JLucid/intensional\.SIPL\.ObjectiveLucid/g' | \
    # ObjectiveLucid GIPL
    sed 's/JGIPL/ObjectiveGIPL/' | \
    sed 's/\/\/{EXTEND-E1}/\/\/{EXTEND-E1}\n\t\t\t| ( <DOT> ID() ) #OBJREF E1()/' \
    > ObjectiveGIPL.jjt

# EOF
\end{verbatim}
}

\section{\tool{ObjectiveIndexicalLucid.sh}}
\source{\begin{verbatim}
#!/bin/bash

cat JIndexicalLucid.jjt | \
    # Filter out unneeded stuff
    grep -v '// EOF' | \
    # Fix package
    sed 's/intensional\.SIPL\.JLucid/intensional\.SIPL\.ObjectiveLucid/g' | \
    # ObjectiveLucid Indexical
    sed 's/JIndexicalLucid/ObjectiveIndexicalLucid/' | \
    sed 's/\/\/{EXTEND-E1}/\/\/{EXTEND-E1}\n\t\t\t| ( <DOT> ID() ) #OBJREF E1()/' \
    > ObjectiveIndexicalLucid.jjt

# EOF
\end{verbatim}
}



\addcontentsline{toc}{chapter}{Index}
\printindex


\end{document}